\documentclass[12pt,twoside]{article}
\usepackage{amsfonts}
\usepackage{graphicx}

\newtheorem{theorem}{Theorem}[section]
\newtheorem{lemma}[theorem]{Lemma}
\newtheorem{proposition}[theorem]{Proposition}
\newtheorem{corollary}[theorem]{Corollary}
\newtheorem{definition}[theorem]{Definition}
\newtheorem{rmrk}[theorem]{Remark}

\newcommand{\R} {{\mathbb R}}

\newcommand{\Z} {{\mathbb Z}}
\newcommand{\N} {{\mathbb N}}
\newcommand{\sir} {\mathcal{R}}
\newcommand{\qed} {\hfill {\small Q.E.D.} \par\medskip}
\newcommand{\skippar} {\par\medskip}
\newcommand{\ds} {\displaystyle}
\newcommand{\proof} {\noindent \textsc{Proof.} }
\newcommand{\proofof}[1] {\noindent \textsc{Proof of {#1}.} }
\newcommand{\article}[3] {\textsc{{#1}}, {\itshape {#2}}, {{#3}}.}
\newcommand{\book}[3] {\textsc{{#1}}, {\itshape {#2}}, {{#3}}.}
\newcommand{\vol} {\textbf}
\newcommand{\eps} {\varepsilon}
\newcommand{\rset}[2] {\left\{ #1 \: \left| \: #2 \right. \! \right\} }

\renewcommand{\tilde} {\widetilde}
\renewcommand{\o} {orbit}

\newcommand{\Int}[1] {int{#1}}
\newcommand{\fn} {function}
\newcommand{\bi} {billiard}
\newcommand{\me} {measure}
\newcommand{\tr} {trajector}

\newcommand{\erg} {ergodic}
\newcommand{\ex} {existence}
\newcommand{\ac} {absolute continuity}
\newcommand{\sy} {system}
\newcommand{\hyp} {hyperbolic}
\newcommand{\ta} {Q}		
\newcommand{\f} {f}		
\newcommand{\ph} {\varphi}	
\newcommand{\si} {\mathcal{S}}	
\newcommand{\z} {z}		
\newcommand{\w} {w}		
\newcommand{\uw} {\mathcal{U}}	
\newcommand{\ma} {T}		
\newcommand{\ps} {\mathcal{M}}	
\newcommand{\ts} {\mathrm{T}}	
\newcommand{\co} {\mathcal{C}}	
\newcommand{\du} {d^u}		
\newcommand{\bu} {B^u}		
\newcommand{\wu} {W^u}		
\newcommand{\wsu} {W^{s(u)}}	
\newcommand{\len} {\ell}	
\newcommand{\qu} {P}		

\renewcommand{\l} {r}		

\newcommand{\sect}[1] {\section{{#1}} \setcounter{equation}{0}}

\newcommand{\fig}[3] {
\medskip\smallskip
\begin{figure}[ht]
	\centering
	\includegraphics[width=#2]{#1.eps}
	\begin{minipage}[t]{0.75\linewidth} 
		\caption{\baselineskip=14pt {#3}}
		\protect\label{#1}
	\end{minipage}
\end{figure}
\medskip
}

\newenvironment{remark}
{\begin{rmrk} \em}
{\end{rmrk}}

\begin{document}

\title{Semi-dispersing billiards with an infinite cusp}

\author{
	Marco Lenci \\
	Institute for Mathematical Sciences \\
	SUNY at Stony Brook \\
	Stony Brook, NY \ 11794-3660 \\ 
	U.S.A. \\
	\footnotesize{E-mail: \texttt{lenci@math.sunysb.edu}}
}

\date{October 2001}

\maketitle

\begin{abstract}
	Let $f: [0, +\infty) \longrightarrow (0, +\infty)$ be a
	sufficiently smooth convex function, vanishing at
	infinity. Consider the planar domain $Q$ delimited by the 
	positive $x$-semiaxis, the positive $y$-semiaxis, and the graph 
	of $f$.
	
	Under certain conditions on $f$, we prove that the billiard
	flow in $Q$ has a hyperbolic structure and, for some examples,
	that it is also ergodic. This is done using the cross section
	corresponding to collisions with the dispersing part of the
	boundary. The relevant invariant measure for this Poincar\'e
	section is infinite, whence the need to surpass the existing
	results, designed for finite-measure dynamical systems.

	\bigskip\noindent
	Mathematics Subject Classification: 37D50, 37D25.
\end{abstract}

\sect{Historical introduction}
\label{sec-intro}

There is a long tradition in the study of \hyp\ \bi s, especially \bi
s in the plane. This was initiated by Sinai as early as 1963
\cite{s1}, in connection with the Boltzmann Hypothesis in statistical
mechanics. In his celebrated 1970 paper \cite{s2}, Sinai proved the
first cornerstone theorem of the field: the $K$-property of a \bi\ in
a 2-torus endowed with a finite number of convex scatterers (of
positive curvature). The result was polished in a later joint work
with Bunimovich \cite{bs} and extended to a larger class of
\emph{dispersing} \bi s. This terminology was introduced precisely in
that paper and designates \bi\ tables whose boundaries are composed of
finitely many convex pieces, when seen from the interior. In
particular, the new theorem allowed for positive-angle corners at the
boundary.

In the following years this area of research made progress in several
directions. For instance, Gallavotti and Ornstein \cite{go} proved
that the Sinai \bi\ is also Bernoulli; Bunimovich \cite{b} found an
example of a \hyp\ and chaotic \bi\ with flat and \emph{focusing} (as
opposed to dispersing) boundary---the famous Bunimovich stadium. Also,
Vaserstein's 1979 paper \cite{v} provided the missing ingredient for
the proof of the \erg\ properties of a \bi\ with \emph{cusps}, i.e.,
zero-angle corners. (In short, the hurdle was that, for cuspidal \bi s,
the amount of phase-space contraction/expansion after $n$ collisions
is not uniformly large as a \fn\ of the phase-space point. In jargon,
this is a \emph{non-uniformly \hyp} system.)

However, it was not until the 1980's that this field was made into an
organized theory, the theory of \emph{singular \hyp\ dynamical \sy
s}. We will not even try to summarize the hefty literature concerned
with it, but will only mention the works that are most relevant to the
present paper. Among these, a prominent place is deserved by the book of
Katok and Strelcyn \cite{ks}. One of the most valuable results they
proved there is that a very general class of \bi s has a \emph{\hyp\
structure}. By this we mean that local stable and unstable manifolds
exist almost everywhere and these local foliations are absolutely
continuous w.r.t.\ the invariant \me. In practice, they adapted Pesin's
theory for smooth \hyp\ \sy s \cite{p} to \sy s with singularities.
Actually, again following Pesin, they also showed that such \sy s admit
a very general structure theorem. This claims that the phase space
splits into a finite or countable number of \erg\ components over which
the map has the $K$-property (more or less---in fact, a power of the map
does).

This last theorem, as nice as it is, is not of big practical
applicability, since it does not give any characterization of the
\erg\ components. This is needed, for example, to show that there is
only one of them. Researchers have henceforth tried to prove
\emph{local \erg ity} theorems, theorems that guarantee that a point
with enough properties (usually called a \emph{sufficient} point) has
a neighborhood contained in one \erg\ component. Any such result is
nowadays called a \emph{fundamental theorem} for \bi s (the
terminology comes from \cite{bs}), and there is a variety of them,
formulated in more or less general ways. Sinai and Chernov in 1987
\cite{sc} wrote a version especially tailored for \sy s of several
2-dimensional discs or 3-dimensional balls (these correspond to higher
dimensional semi-dispersing \bi s). This result was later improved and
extended by Kr\'amli, Sim\'anyi and Sz\'asz \cite{kss}. Among the more
general formulations of the fundamental theorem we have \cite{c1} and
\cite{lw}.

The latter reference, by Liverani and Wojtkowski, has the nice feature
that it uses \emph{invariant cones}. This elegant tool has been very
effective in \hyp\ dynamics. Wojtkowski in 1986 wrote a beautiful
paper \cite{w} on how to use geometrical optics to define invariant
cones for planar \bi s. (Here, however, we will use different cones,
whose construction dates back to \cite{s2}.) This work was improved by
Markarian \cite{m1} in 1988.

In the last decade researchers have focused on a yet harder problem,
of great importance for the physics that these \bi s are supposed to
resemble: the decay of correlations. We will be even more superficial
on this widely studied question, since it does not concern this note.
But it is interesting to recall that it took the community more than
one decade to get convinced that a large class of \hyp\ \bi s has
exponential decay of correlations, and even longer to prove it. This
was very recently done by Young \cite{y}, and her techniques better
adapted to \bi s by Chernov \cite{c2}. The class of \sy s for which
exponential decay holds includes at least all \bi s treated in
\cite{bs}, but does not include dispersing \bi s with a cusp, whose
decay of correlations is polynomial \cite{cm}.

\skippar

In the present article we study the case of a semi-dispersing \bi\
with a \emph{non-compact} cusp. Given a \fn\ $\f$ defined on $[0,
+\infty)$ with values in the positive numbers, sufficiently smooth,
convex and vanishing at $+\infty$, we consider the table $\ta$
delimited by the positive $x$-semiaxis, the positive $y$-semiaxis, and
the graph of $f$, as in Fig.~\ref{figa}. For many such \bi s we will
construct a \hyp\ structure, and for some of them we will also prove
\erg ity.

\fig{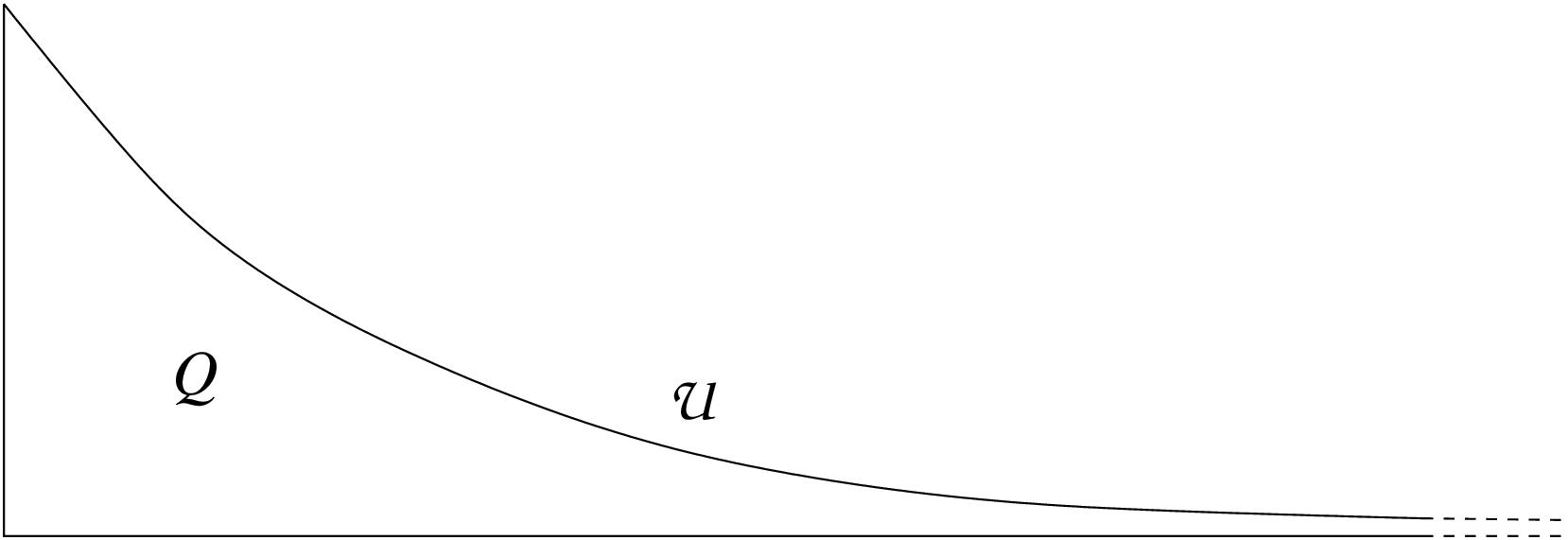} {4in} {An example of a billiard table $\ta$.} 

Geometrically speaking, one cannot fail to notice the similarities
between this \sy\ and the geodesic flow on a non-compact modular surface
in the \hyp\ half-plane. The \erg\ properties of this flow were proved
by Hopf, for both compact and non-compact surfaces \cite{cfs}.  However,
dynamically speaking, the modular surface is not a good analogue for our
\bi, even leaving aside the smoothness of the flow. In fact, in the
\hyp\ half-plane, the curvature of the metrics acts as a \emph{constant}
source of \hyp ity. In other words, it does not matter if the particle
spends a given amount of time in the ``bulk'' of the surface or in the
cusp. In the case of the \bi, instead, once the particle is deep into
the cusp, one does not know \emph{a priori} whether the increased number
of rebounds per unit time compensates the flattening of the dispersing
boundary and the decrease of the mean free path.

As important as this, is the question of the finiteness of the area.
Contrary to the modular surface case, we do not assume $\ta$ to have
finite area. In fact, our model represents one of the very few treated
examples of singular \hyp\ dynamical \sy s with \emph{infinite}
invariant \me. This claim calls for an explanation. There is a serious
amount of literature concerning infinite-\me\ \sy s, even among
dispersing \bi s \cite{lm}, or \bi s with a cusp \cite{l, k, le}
(\cite{ddl1, ddl2} are polygonal caricatures of the \sy\ at hand).
Seldom, though, are they studied from the point of view of
\hyp ity and \erg\ theory. Perhaps the only close relative to the
present model is the Lorentz gas in two dimensions. Conze has recently
proved its recurrence \cite{co}, which yields \erg ity, according to a
previous result by Sim\'anyi \cite{si}.

But even the Lorentz gas is not a very good analogue, since, thanks to
recurrence, one really studies the return map to a finite-\me\
Poincar\'e section. Despite our efforts, this did not seem to be a
fruitful approach for the infinite-cusp \bi. What we do is consider
the return map to the dispersing part of $\partial \ta$. This choice
corresponds to an infinite-\me\ cross section, independently of the
integrability of $\f$.

As a matter of fact, although the underlying ideas used here are the
same as in the literature previously cited, hardly anything works
directly. Particularly disappointing is the fact that we cannot use
\cite{ks} to construct a \hyp\ structure. Establishing this becomes
our real ``fundamental theorem''! We achieve it by introducing two
techniques: the \emph{unstable distance} (Section \ref{sec-sing}) and
the concept of \emph{fuzzy boundary} for a sub-cross-section (Section
\ref{sec-lsum}). (I have recently learned that a notion of unstable
distance was already present in \cite{m2}.)

Apart from a number of technical results, the other serious task that we
have to take on is to devise our own version of the local \erg ity
theorem. We do so at the end of this note, extending the theorem of
\cite{lw} to infinite-\me\ \sy s.

\skippar

To summarize the rest of the paper: Section \ref{sec-prelim} contains
the basic definitions and their immediate consequences; in Section
\ref{sec-cone} we introduce the invariant cone bundle; Section
\ref{sec-sing} describes the neighborhoods of the singularities
w.r.t.\ the unstable distance; in Section \ref{sec-lsum} we prove the
\ex\ of the local stable and unstable manifolds and in Section
\ref{sec-ac} their \ac; Section \ref{sec-loc-erg} is devoted to the
local \erg ity theorem, the \erg ity standing as a corollary; in
Section \ref{sec-erg-ta} we check that a certain family of \bi s
(namely those generated by $\f(x) = C x^{-p}$, with $p>0$) are in fact
\erg.  Finally, some of the less important lemmas are proved in the
Appendix.

Of course, all these results require assumptions on the \fn\ $\f$. For
the convenience of the reader, we list them all at the beginning of
Section \ref{sec-prelim} and reference them appropriately in the
formulation of each theorem.  At any rate, we can anticipate that two
classes of examples, $\f(x) = C e^{-kx}$ ($k>0$) and $\f(x) = C
x^{-p}$ ($p>0$), verify all the general theorems; that is, their
corresponding \bi s have a \hyp\ structure.  The latter family, as
already mentioned, is also \erg.  This is particularly interesting
since, within this family, we find tables of infinite area. For these
\sy s, one cannot hope to avoid the hurdles associated to
infinite-\me\ dynamical \sy s by studying a 3-dimensional flow,
instead of a 2-dimensional map. (Not that this approach would be
effortless, anyway.)

\sect{Mathematical preliminaries}
\label{sec-prelim}

A \bi\ is a dynamical \sy\ defined by the free motion of a material
point inside a domain, subject to the Fresnel law of reflection at the
boundary, i.e., the angle of reflection is equal to the angle of
incidence. Here we are concerned with domains (otherwise referred to
as \emph{tables}) of the form: $\ta := \rset{ (x,y) \in \R_{0}^{+}
\times \R_{0}^{+} } {0 \le y \le \f(x) }$ (see Fig.~\ref{figa}), where
$\f: \R_{0}^{+} \longrightarrow \R^{+}$ is three times differentiable,
bounded and convex. Thus the table is semi-dispersing. The angle at
the vertex $V := (0,\f(0))$ is $\pi/2 + \arctan \f'(0^{+})$ and is
allowed to be zero; so the \bi\ might have a compact cusp, together
with the non-compact cusp at $x = +\infty$.

For many geometrical proofs, throughout the paper, it will be convenient
to introduce two more domains in the plane, $\ta_2 := \rset{ (x,y) \in
\R_{0}^{+} \times \R } { |y| \le \f(x) }$, a two-fold copy of $\ta$,
and $\ta_4 := \rset{ (x,y) \in \R \times \R } { |y| \le \f(|x|) \, }$,
a four-fold copy (see Figs.~\ref{figr} and \ref{figb} later on).

Before moving on, we list all the assumptions on $\f$ that we use in
the rest of this paper, specifying which result requires which
hypothesis. To this purpose, we give an extra definition, that will
be introduced with the aid of Fig.~\ref{figr}.

\fig{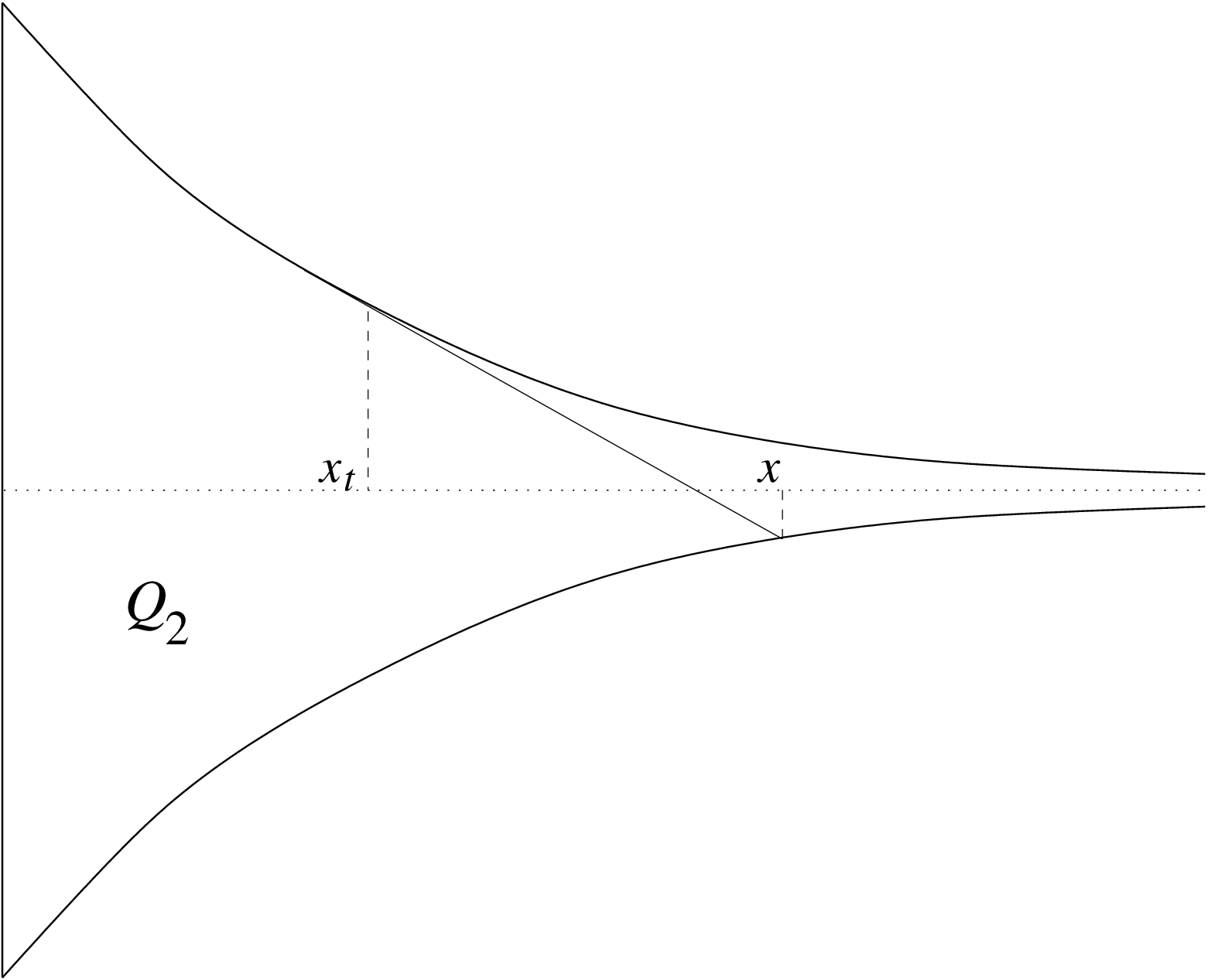} {3.5in} {The definition of $x_t$.} 

In $\ta_2$, for $x > 0$, consider the straight line passing through
$(x,-\f(x))$ and tangent to $\partial \ta$ (i.e., to the part of
$\partial \ta_2$ that lies in the first quadrant). Denote by $x_t =
x_t(x)$ the abscissa of the tangency point. This is uniquely
determined by the equation 
\begin{equation}
	\frac{\f(x) + \f(x_t)} {x - x_t} = -\f'(x_t).
	\label{cond-xt}
\end{equation}
In the sequel, for $f,g \ge 0$, we use the notation $f(x) \ll g(x)$ to
indicate that there is constant $C$ such that $f(x) \le C\, g(x)$, as
$x \to +\infty$; likewise for $\gg$. Also, $f(x) \sim g(x)$ means
that, when $x \to +\infty$, $f(x)/g(x)$ is bounded away from 0 and
$+\infty$. Later on, where there is no danger of confusion, we use the
same symbols for different asymptotics, such as $\eps \to 0^{+}$, and
so on.

\skippar

\noindent
\textsc{Assumptions for the existence of local stable and unstable
manifolds:}
$$
	\f''(x) \to 0
	\eqno{\mathrm{(A1)}}
$$
$$
	|\f'(x_t)| \ll |\f'(x)|;
	\eqno{\mathrm{(A2)}}
$$
$$
	\frac{\f(x) \f''(x)} {(\f'(x))^2} \gg 1;
	\eqno{\mathrm{(A3)}}
$$
$$
	\frac{|\f'''(x)|} {\f''(x)} \ll 1.
	\eqno{\mathrm{(A4)}}
$$
\textsc{Assumption for the absolute continuity of local stable and
unstable manifolds:}
$$
	|\f'(x)| \gg (\f(x))^\xi, \mbox{ for some } \xi>0.
	\eqno{\mathrm{(A5)}}
$$
\textsc{Examples of \erg ity:}
$$
	\f(x) = C x^{-p}, \mbox{ with } C,p>0,
	\eqno{\mathrm{(E1)}}
$$

\begin{remark}
	We have worked out the \erg ity result for a class of
	examples, rather than giving general conditions, because such
	conditions would be rather unhandy to write down and to verify
	(see Section \ref{sec-erg-ta}). On the other hand, it is a
	safe conjecture to say that \erg ity must hold for many other
	examples as well, including $\f(x) = C e^{-kx}$.
\end{remark}

We notice that (A1) and (A4) imply, via a De l'H\^opital-like
argument, that
\begin{equation}
	\frac{|\f'(x)|} {\f(x)} \ll \frac{\f''(x)} {|\f'(x)|} \ll 1,
	\label{a9}
\end{equation}
since $\f''$ (hence $\f$ and $\f'$) vanish at infinity.  It is evident
that none of the above assumptions depend on multiplicative constants
in front of $\f$. We now check that $\f(x) = x^{-p}$, with $p>0$, and
$\f(x) =e^{-kx}$, with $k>0$, verify (A1)-(A5). 

The other conditions being trivially satisfied, we only need to worry
about (A2). For $\f(x) = x^{-p}$, after a simple algebraic
manipulation, (\ref{cond-xt}) reads
\begin{equation}
	\frac{ \left( \ds \frac{x_t}{x} \right)^p + 1} { 1 -
	\ds \frac{x_t}{x}} = p \left( \frac{x_t}{x} \right)^{-1},
	\label{verif-a2-1}
\end{equation}
which gives $x_t/x = const$, whence (A2). For $\f(x) = e^{-kx}$,
(\ref{cond-xt}) amounts to
\begin{equation}
	\frac{ e^{-k(x-x_t)} + 1} {x-x_t} = k;
	\label{verif-a2-2}
\end{equation}
hence $x - x_t = const$, yielding (A2) for this case, too. 

\skippar

A state in our system is completely specified by the position $(x,y)$ and
velocity $v$ of the point. Since the kinetic energy is a constant, one
can assume that $|v|=1$. Hence the natural phase space for the flow is
$\ta \times S^1$. In the terminology of \cite{s2}, points in this space
are called \emph{line elements}.  The relevant Liouville \me\ here is
the Lebesgue measure, which is therefore left invariant by the flow.

It is customary, especially if one is only interested in the \erg ity,
to work with a cross-section. For \bi s, one usually chooses the
cross-section corresponding to rebounds of the point against $\partial
\ta$. In our case, the geometry of the \bi\ (see $\ta_4$) suggests
that we restrict to rebounds against the dispersing part of $\partial
\ta$, which we denote by $\uw$. More precisely, we consider only unit
vectors based in $\uw$ and pointing towards the interior of $\ta$.  We
parameterize these line elements by $\z := (\l, \ph)$, where $\l \in
[0, +\infty)$ is the arc length variable along $\uw$, and $\ph \in
[0,\pi]$ \me s the angle between $\partial/\partial \l$ and the
velocity vector (Fig.~\ref{figb}).

\fig{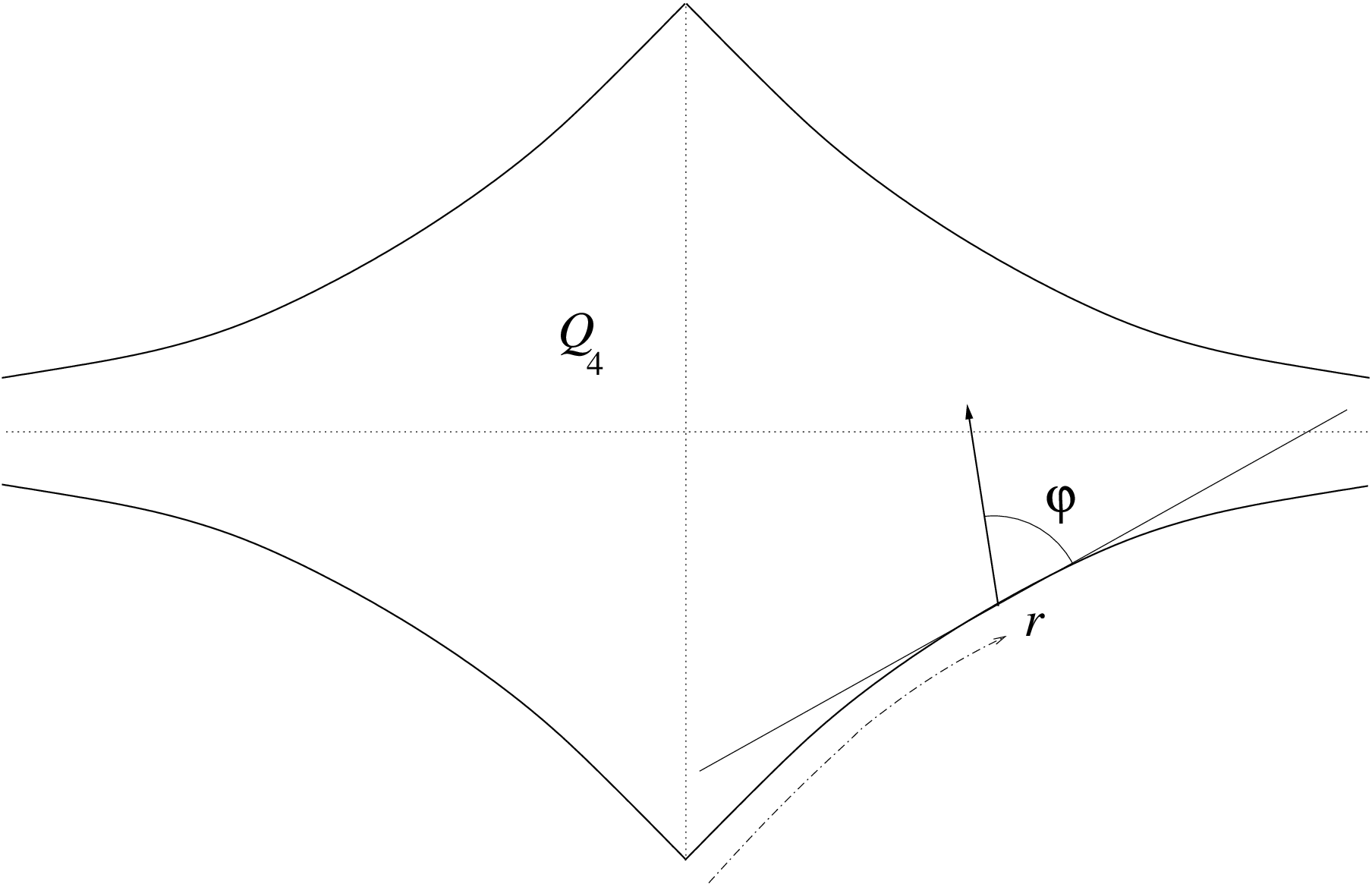} {4in} {The definition of $\l$ and $\ph$.}

So the manifold over which we define our dynamical system is $\ps :=
(0,+\infty) \times (0,\pi)$ (it will be clear in the sequel why it is
a good idea to exclude the boundary of $\ps$). The \bi\ flow defines
on $\ps$ a Poincar\'e return map $\ma$ which, according to an easy
classical result, preserves the \me\ $d\mu(\l,\ph) = \sin \ph \, d\l
d\ph$.

\fig{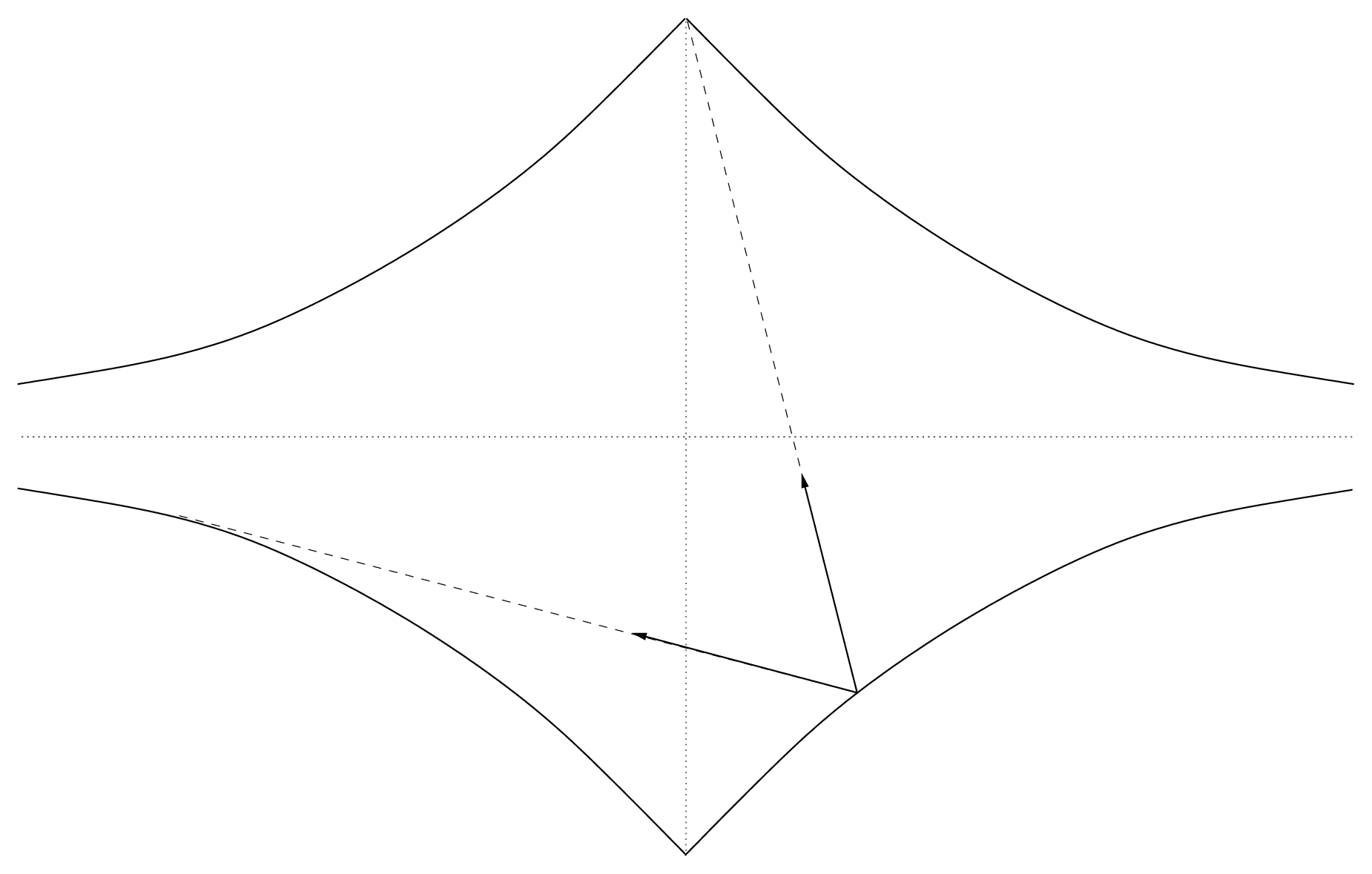} {4in} {The geometrical meaning of the singularities.}

We do not define $\ma$ at those points of $\ps$ that would end up in $V$
or hit $\uw$ tangentially (see Fig.~\ref{figc}); that is, we morally
exclude the set ``$\ma^{-1} \partial \ps$''. In fact, these points make
up the discontinuity set of $\ma$. We will see later that they are often
singularities as well---every time they correspond to a tangency. They
are arranged in two curves, depicted in Fig.~\ref{figd}. $\si^{1+}$
corresponds to tangencies to $\partial \ta_4$ in the fourth quadrant
(i.e., in $\ta$, tangencies to $\uw$ after a rebound on the vertical
side); this curve is a regular as $\f$. As for $\si^{2+}$, its first
part corresponds to line elements pointing into $V$ (in $\ta$, after a
rebound on the horizontal side); as $\l$ increases, these become
elements tangent to $\partial \ta$. The border between these two
behaviors is the only non-regular point of $\si^{2+}$. We denote
$\si^{+} := \si^{1+} \cup \si^{2+}$.

\fig{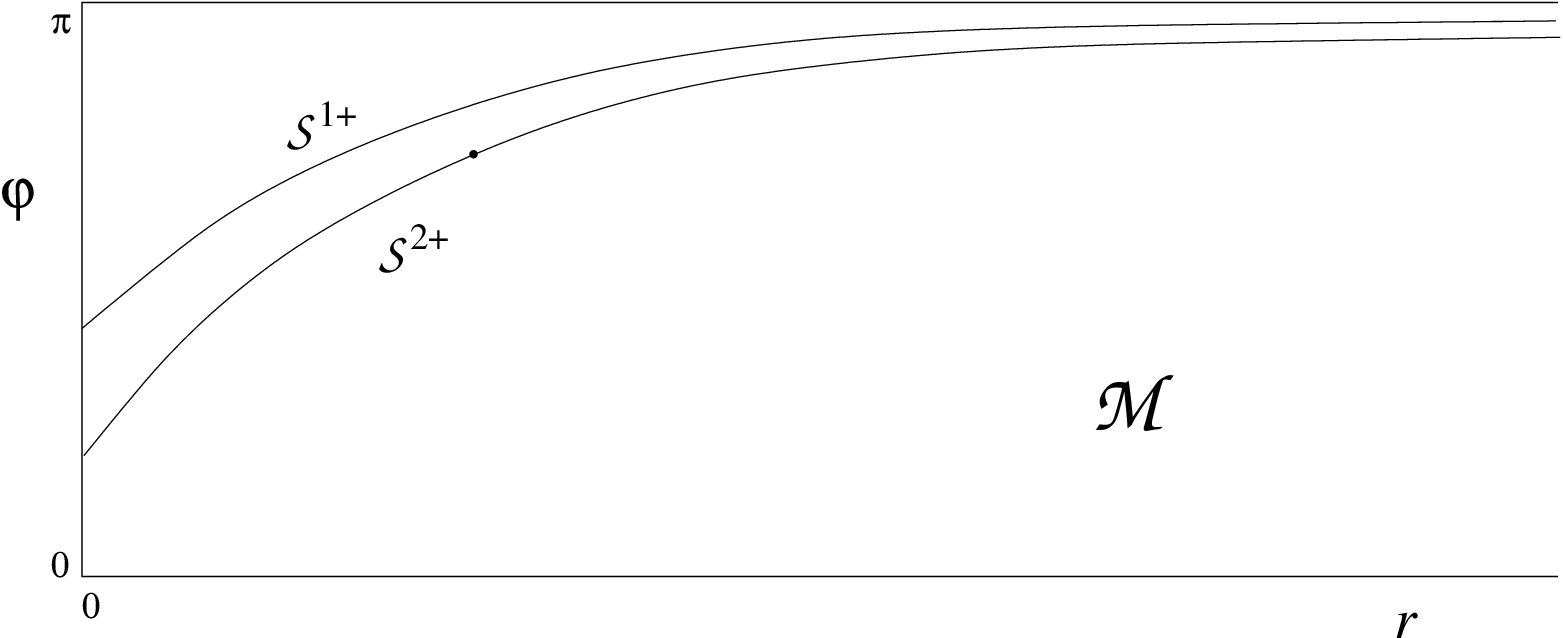} {5in} {The singularity lines $\si^{1+}$ and $\si^{2+}$ in
$\ps$.}

Analogously we name $\si^{-} := \si^{1-} \cup \si^{2-}$, where the
$\si^{i-}$ are the singularity lines of $\ma^{-1}$, obtained from
$\si^{i+}$ through the application of the time-reversal operator:
$(\l,\ph) \mapsto (\l,\pi-\ph)$. Whenever the superscript $\pm$ is
dropped we will always mean $\si^{+}$.

For all points in $\ps \setminus \si^{+}$, the differential of $\ma$
is known \cite[\S 14]{lw}, and not too hard to compute, anyway. If $
(\l_1,\ph_1) := \z_1 := \ma \z = \ma(\l,\ph)$, then
\begin{equation}
	D \ma_\z = \left[ 
	\begin{array}{cc} 
		\ds -\frac{\sin \ph} {\sin \ph_1} - \frac{k \tau}  
		{\sin \ph_1} & 
		\ds \frac{\tau} {\sin \ph_1} \ \ \\ \\
		\ds k + k_1 \frac{\sin \ph} {\sin \ph_1} + 
		\frac{ k k_1 \tau} {\sin \ph_1} \ \ &
		\ds -1 - \frac{k_1 \tau} {\sin \ph_1}
	\end{array}
	\right] , 
	\label{dmap}
\end{equation}
where $\tau = \tau(\z)$ is the traveling time (equivalently, the
distance in $\ta_4$) between the two collision points. Also, $k$
(resp.~$k_1$) is the curvature of $\partial \ta$ in $\z$
(resp.~$\z_1$). We adopt the convention that in our semi-dispersing
\bi\ the curvature is non-negative. Another geometric relation that
can be simply checked is
\begin{equation}
	\ph_1 = \ph \pm \beta,
	\label{ph-phprime}
\end{equation}
$\beta$ being the angle between the tangent lines to $\partial \ta_4$ at
$\z$ and $\z_1$ (see Fig.~\ref{figk}). The sign depends on whether these
two tangent lines meet to the right or to the left of the segment
joining $\z$ and $\z_1$.

\fig{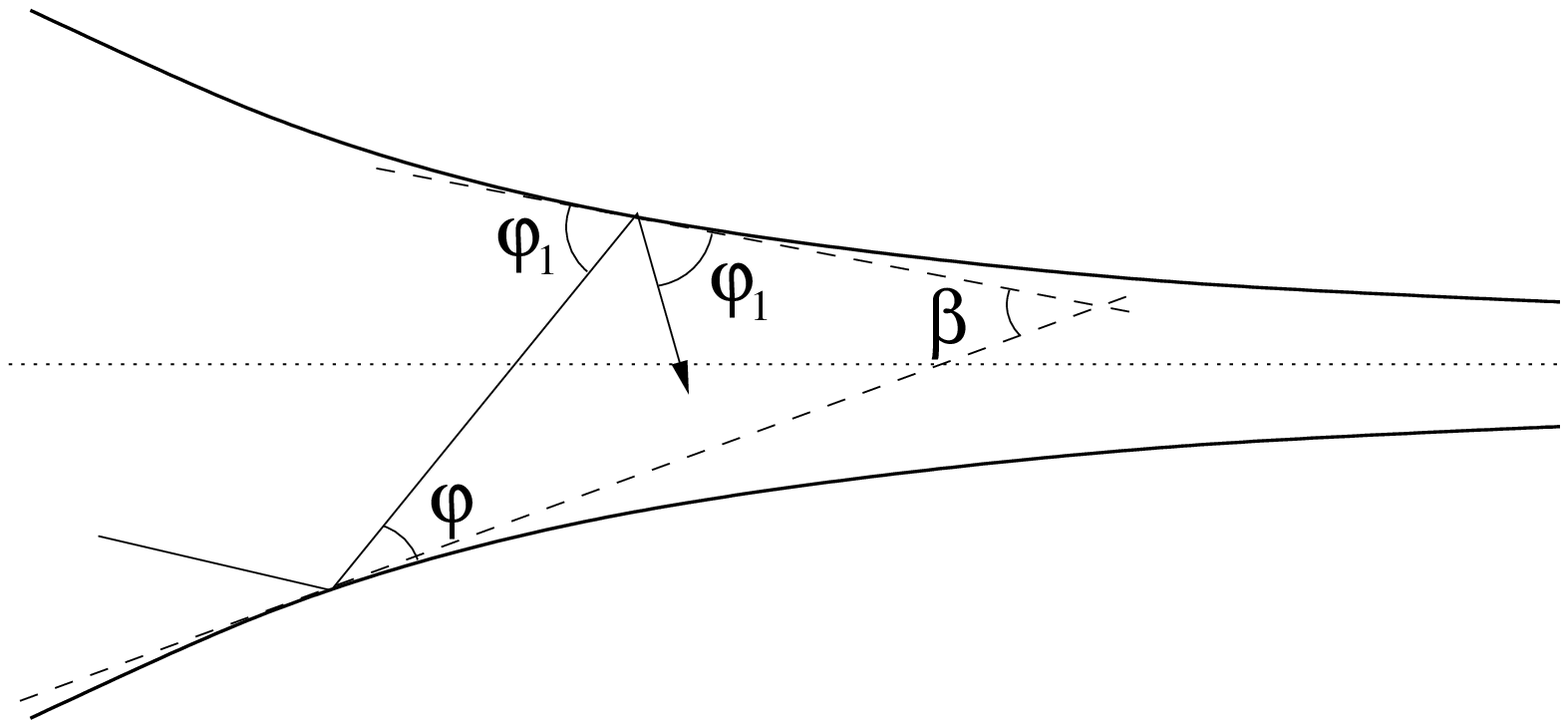} {3.5in} {The relation between $\ph$ and $\ph_1$.}

We conclude this preliminary section by presenting what is known about
the behavior of the \tr ies in the vicinity of the cusp(s) or the
corner. We return to the language of the \bi\ flow (as opposed to the
Poincar\'e map), and recall the convention that the material point
stops when it hits a vertex.

\begin{proposition}
	No \bi\ semi\o\ falls into the vertex $V$ or the ``vertex
	at infinity'' unless it shoots directly there. In other
	words, if $(x(t),y(t))$ is an \o\ in $\ta$, 
	\begin{eqnarray*}
		\lim_{t \to \pm\infty} (x(t),y(t)) = V &
		\Longleftrightarrow & (x(t),y(t)) = V, \ \forall 
		t \ge(\le) t_0; \\
		\lim_{t \to \pm\infty} x(t) = +\infty &
		\Longleftrightarrow & y(t) = 0, \ \forall t;
	\end{eqnarray*}
	the last condition specifying the \tr y that runs along the 
	$x$-semiaxis.
	\label{prop-eo}
\end{proposition}

\proof As far as $V$ is concerned, if we have a corner there, the
result comes from elementary considerations that are, anyway,
mentioned in \cite{bs}; one can even give an upper bound on the number
of rebounds that the \tr y performs before moving away from $V$. In the
case that $V$ has a cusp, the assertion is contained in \cite{v} and can be
easily deduced by the ``a gallon of water won't fit a pint-sized
cusp'' argument of \cite{k}.

For the non-compact cusp, the result was first derived by Leontovich
in 1962 \cite{l} (see also \cite{le}).
\qed

\sect{The cone field}
\label{sec-cone}

We use here the construction of stable and unstable directions for
dispersing \bi s, as originally developed by Sinai in \cite[\S 2]{s2}
(found in clean form, e.g., in \cite[\S 2]{g}), although we re-express
those results in the more modern language of invariant cones.

On $\ts \ps$, the tangent bundle of $\ps$, let us define
\begin{equation}
	\co^{+}(\z) := \co(\z) := \rset{ (d\l,d\ph) \in \ts_{\z} \ps } 
	{ d\l \cdot d\ph \le 0 }; 
	\label{cone}
\end{equation}
that is, the second and fourth quadrant of $\ts_{\z} \ps$, using the
natural basis $\{ \partial/\partial \l, \partial/\partial \ph
\}_{\z}$. This endows $\ts \ps$ with a continuous bundle of Lagrangian
cones, sometimes referred to as \emph{unstable cones}. The alternate
signs of $D\ma_\z$ in (\ref{dmap}) show that $D\ma_\z \co(\z) \subset
\Int \co(\ma \z)$, where $\Int \co$ denotes the interior of $\co$
\emph{together with 0}. We say that the above cones are \emph{strictly
invariant} under the action of $\ma$ \cite{w,lw}. 

\begin{remark}
	This definition of the cone field does not correspond to the one
	devised by Wojtkowski in \cite{w} for dispersing \bi s. Rather,
	it is the transliteration of the long-known properties of
	monotonic curves in $\ps$ under the action of $\ma^{\pm 1}$
	\cite[Cor.\ 2.2]{s2}. These cones could be rightfully called
	Sinai cones.
\end{remark}

Later on, we will also need a cone bundle for $\ma^{-1}$. Since the
time-reversed map can be treated in the same way as $\ma$, its cones
$\co^{-}(\z)$ will be defined analogously, as the first and third
quadrant of $\ts_{\z} \ps$.

\begin{remark}
	The fact that the two cone bundles $\co^{+}$ and $\co^{-}$ are
	(almost) complementary to each other is purely accidental. It
	is true that when the dynamical system is Hamiltonian (more
	precisely, when it preserves a symplectic form whose volume
	element is absolutely continuous w.r.t.\ the standard volume
	element, and viceversa), as in our case, defining $\co^{-}$ as
	the closure of the complement of $\co^{+}$ still leads to
	essentially all the sought results \cite{lw}. But, when
	possible, it is more convenient to define each cone bundle
	separately, based on the map that makes it invariant.
	\label{rk5}
\end{remark}

Denote $\si^{-}_n := \bigcup_{i=0}^{n-1} \ma^i \si^{-}$.  For $(\l_{-n},
\ph_{-n}) := \z_{-n} := \ma^{-n} \z = \ma^{-n} (\l, \ph)$, we define
the $n$-th nested cone as
\begin{equation}
	\co_n(\z) := D\ma^n_{\z_{-n}} \co(\z_{-n}) = \rset{
	(d\l,d\ph) \in \ts_{\z} \ps } { a_n \le \frac1 {\sin
	\ph} \, \frac{d\ph}{d\l} \le b_n }, 
	\label{n-cone}
\end{equation}
It can be computed from (\ref{dmap}) and (\ref{cone}) that
\begin{equation}
	a_n = -\frac{k}{\sin \ph} + \frac1 {\ds -\tau_{-1} + 
	\frac1 {\ds -\frac{2k_{-1}}{\sin \ph_{-1}} + \frac1 {\ds
	\ddots { \atop \ds \frac1 {-\tau_{-n}} } } } }
	\label{an}
\end{equation}
and 
\begin{equation}
	b_n = -\frac{k}{\sin \ph} + \frac1 {\ds -\tau_{-1} + 
	\frac1 {\ds -\frac{2k_{-1}}{\sin \ph_{-1}} + \frac1 {\ds
	\ddots { \atop \ds \frac1 {\ds -\tau_{-n} + \frac1 
	{\ds -\frac{2k_{-n}}{\sin \ph_{-n}} } } } } } }.
	\label{bn}
\end{equation}
Here $\tau_{-i}$ and $k_{-i}$ are respectively $\tau(\z_{-n})$ and
$k(\z_{-n})$, for $i = 1,2, ... , n$ (see also \cite[\S
2]{g}). Hence $a_n$ and $b_n$ are the left and right approximants of
the continued fraction defined by (\ref{an})-(\ref{bn}). The next
result shows that this continued fraction converges.

\begin{proposition}
	For all $\z \not\in \si^{-}_\infty := \bigcup_{i=0}^{\infty} 
	\ma^i \si^{-}$, 
	\begin{displaymath} 
		\lim_{n \to \infty} a_n = \lim_{n \to \infty} b_n =: 
		\chi^u(\z).
	\end{displaymath} 
	Hence, as $n \to \infty$, $\co_n(\z)$ converges, in the sense
	of decreasing sets, to a subspace $E^u(\z) \subset
	\ts_{\z} \ps$. We call it the unstable subspace. This is a 
	line of slope $\chi^u(\z) \sin \ph$ w.r.t.~the basis $\{
	\partial/\partial \l, \partial/\partial \ph \}_{\z}$.
	\label{prop-cont-frac}
\end{proposition}

\proof Since the terms of the continued fraction (\ref{an})-(\ref{bn})
are negative, classical results \cite[\S 2]{s2} show that a necessary
and sufficient condition for convergence is
\begin{equation}
	\sum_{i=-1}^{-\infty} \left( \tau_i + \frac{2k_i}{\sin \ph_i}
	\right) = +\infty.
\end{equation}
Assuming the contrary, $\tau_i$ would tend to zero. This would imply
that the backward \o\ of $\z$ zig-zags into $V$ or into the cusp at
infinity, in contradiction with Proposition \ref{prop-eo}.
\qed

It will be very important in the remainder to estimate the expansion
of a vector in $(d\l, d\ph) \in \co(\z)$. From (\ref{dmap}), using the
notation previously introduced:
\begin{eqnarray}
	(d\l_1)^2 &\ge& \left( \frac{\sin \ph} {\sin \ph_1} \right)^2
	\left( 1 + \frac{k \tau} {\sin \ph} \right)^2 d\l^2;
	\label{expan-dl} \\ 
	(d\ph_1)^2 &\ge& \left( 1 + \frac{k_1 \tau} {\sin \ph_1}
	\right)^2 d\ph^2.  
	\label{expan-dph}
\end{eqnarray}  
Sometimes the pseudo-metric $(d\l \sin\ph)^2$ is called
\emph{Z-metric} \cite{kss}; it has the property that it is strictly
increasing for vectors in the unstable cone.

\sect{Neighborhoods of the singularity lines}
\label{sec-sing}

The problem with the direction field defined above is that it is hard
to integrate, due to its scanty continuity properties. Morally its
integral lines are the unstable curves, i.e., curves with the property
that the distance between two points on the curve decreases when
$\ma^{-1}$ is iterated many times (see Definition \ref{def-lsum} for
the precise details).

The techniques we deploy here to construct these objects stem from the
original idea of Sinai, which relies upon \me\ estimates of the
tubular neighborhoods of the singularity set. However, a big
complication arises: since the $\si^{i\pm}$ are curves of infinite
length, and are nicely embedded in $\ps$, every $\eps$-neighborhood
has infinite \me. At least if one uses the ordinary Riemannian
distance.  On the other hand, in all of the arguments, it suffices to
\me\ distances along the unstable direction. To this purpose let us
define the \emph{unstable distance} $d^u(\z,\w)$ between two points
$\z$ and $\w$ as the infimum length of differentiable curves $t
\mapsto \gamma(t)$ joining $\z$ with $\w$, and such that
$d\gamma/dt(t) \in \co_1(\gamma(t))$ (check definition
(\ref{n-cone})). Every such curve is henceforth referred to as an
\emph{unstable curve}. Having introduced $d^u(\z,\w)$, the unstable
distance between a point and a set is defined in the usual way.

In order to state the central result of this section, let us introduce
some notation. Denote $\si^{0+} := \R_{0}^{+} \times \{ \pi \}$ and
$\si^{0-} := \R_{0}^{+} \times \{ 0 \}$. Also, for $A \subset \ps$ and
$\eps>0$,
\begin{equation}
	A_{[\eps]} := \rset{\z\in\ps} {\du (\z,A) \le \eps}.
	\label{a-eps}
\end{equation}

\begin{theorem}
	Let $\mathrm{Leb}$ denote the Lebesgue \me\ on $\ps$, and assume
	\emph{(A1)-(A4)}. For $i=0,1,2$, $\mu( \si^{i\pm}_{[\eps]} ) \le
	\mathrm{Leb} ( \si^{i\pm}_{[\eps]} ) \ll \eps$, as $\eps \to
	0^{+}$.  
	\label{thm-udist}
\end{theorem}

\proof We prove the statement for $\si^{2+}$. Later we will see how to
adapt the proof to neighborhoods of the other curves.

Let $\gamma$ be an unstable curve, shorter than $\eps$, having one
endpoint (say $\gamma_0$) on $\si^{2}$. Suppose for the moment that
the curve lies beneath $\si^{2}$.  $\gamma$ has negative slope, thus
can be reparameterized to be the graph of a \fn\ $\ph:
[\l_0,\l_0+\delta] \longrightarrow (0,\pi)$, with $\ph(\l_0) = \ph_0$
and $(\l_0,\ph_0) = \gamma_0$. (The reader will forgive the abuse of
notation regarding $\ph$.) It follows that $\delta \le \eps$. By
definition of unstable curve, the tangent vector to $\gamma$ at $\z$
belongs to $\co_{1}(\z)$. Hence, applying (\ref{an}) with $n=1$ gives
\begin{equation}
	\frac{d\ph}{d\l} \ge -k(\l) - \frac{\sin \ph} {\tau_{-1}
	(\l,\ph)}; \qquad \ph(\l_0) =\ph_0. 
	\label{udist-20}
\end{equation}
We consider this differential inequality in $U := [\l_0, \l_0 + \eps_0]
\times [\ph_0 - b, \ph_0]$, $\eps_0$ to be fixed later, and $b$ an
unimportant sufficiently large number. The situation is shown in
Fig.~\ref{figg}.

\fig{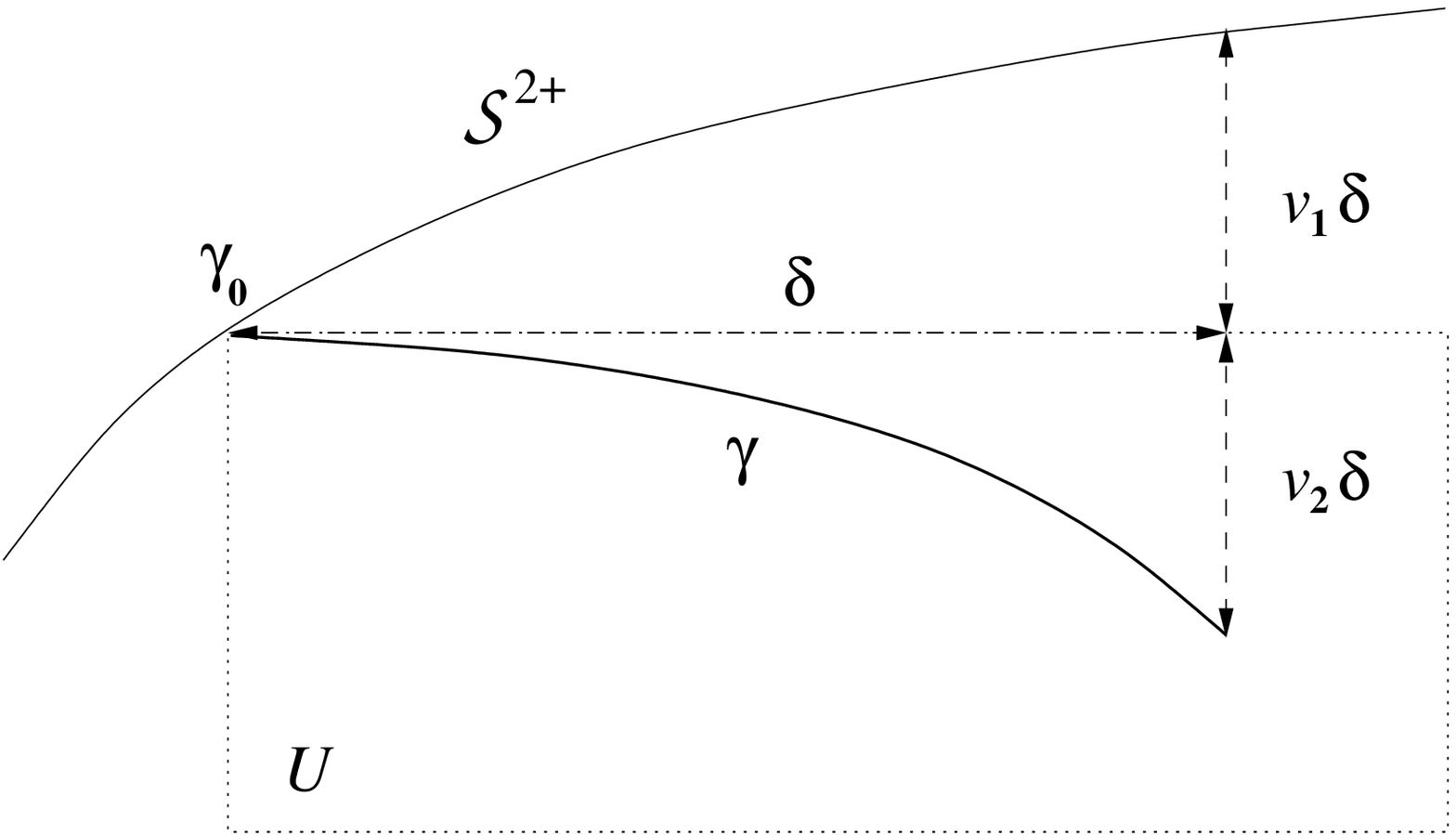} {3.5in} {The unstable curve $\gamma$.}

\fig{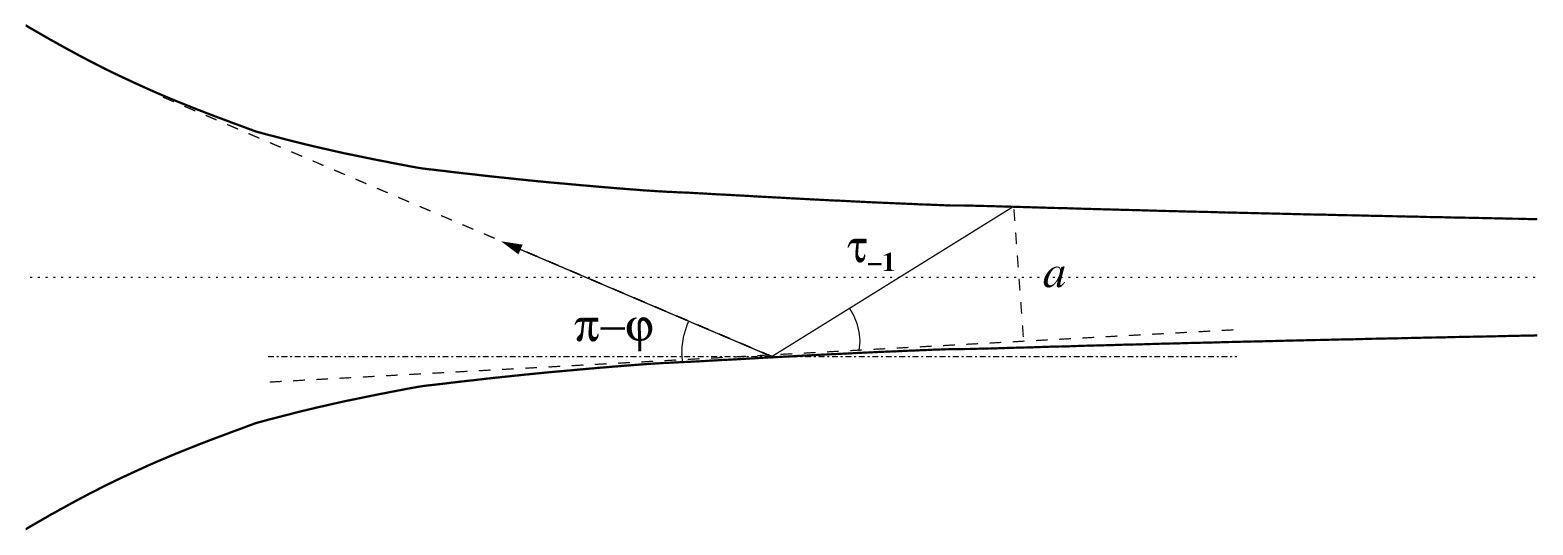} {4.5in} {The definition of $a$, in the case $(\l, \ph) \in
\si^{2}$.}

Define $a(\l,\ph) := \tau_1(\l,\ph) \sin\ph$. This is the length of the
``almost vertical'' segment depicted in Fig.~\ref{figf}. Inequality
(\ref{udist-20}) becomes
\begin{equation}
	\frac{d\ph}{d\l} \ge -k(\l) -\frac{\sin^2 \ph} {a(\l,\ph)}.  
	\label{udist-30}
\end{equation}

We want to replace the r.h.s.\ of the above with a simpler bound, in
order to solve the differential inequality.  First of all, we substitute
$k(\l)$ with $k_M := k(\bar{r}) := \max_{[\l_0, \l_0+b]} k$. Then we
notice that $a(\l,\ph)$ is decreasing in $\ph$, hence its minimum on $U$
is achieved at some point $(\hat{\l}, \ph_0) \in \partial U$: we denote
it by $a_m$ and use it to turn (\ref{udist-30}) into
\begin{equation}
	\frac{d\ph}{d\l} \ge -k_M -\frac{\sin^2 \ph} {a_m} \ge -k_M
	-const \, \frac{(\pi - \ph)^2} {a_m}   
	\label{udist-40}
\end{equation}
in $U$. Introducing $\tilde{\ph}(\l) := \pi - \ph(\l + \l_0)$ and
absorbing $const$ into $a_m$ (with no damage for the proof) we finally
get the differential inequality
\begin{equation}
	\frac{d\tilde{\ph}}{d\l} \le k_M +\frac{\tilde{\ph}^2} {a_m};
	\qquad \tilde{\ph}(0) = \tilde{\ph}_0,
	\label{udist-50}
\end{equation}
which is solved by
\begin{equation}
	\tilde{\ph}(\l) \le \tan \left( \sqrt{ \frac{k_M}{a_m} } \l +
	\arctan \left( \frac{\tilde{\ph}_0} {\sqrt{k_M a_m}} \right)
	\right)  \sqrt{k_M a_m}.
	\label{udist-60}
\end{equation}

What we are interested in, here, is the maximum vertical distance
between $\si^{2}$ and $\gamma$, when $\gamma$ is defined in the domain
$[\l_0, \l_0 + \delta]$. As illustrated in Fig.~\ref{figg}, this is the
sum of two quantities, which we conveniently denote $v_1 \delta$ and
$v_2 \delta$. The $v_i$'s ($i=1,2$) are actually \fn s of $\gamma_0$,
the initial condition of $\gamma$, and of $\delta$. For reasons that
will be clear later, we need to show that the dependence on $\delta$ can
be eliminated by two upper bounds that are integrable, as \fn s of
$\gamma_0 = (\l_0, \ph_0) \in \si^{2}$. More precisely, this means as
\fn s of $x$, where $x = x(\l_0)$ is the $x$-coordinate of the point of
$\partial \ta$ otherwise parameterized by $\l$. In other words, $x(\l)$
is the inverse of
\begin{equation}
	\l(x) := \int_0^x \sqrt{1 + (\f'(t))^2} \, dt.
	\label{udist-70}
\end{equation}
In the reminder, we will liberally switch from $\l_0$ to $x$.

Now, as far as $v_1$ is concerned, we notice that $\si^{2}$ is the
graph of an increasing \fn\ $g: \R^{+} \longrightarrow (0,\pi)$. $g$ has
at most one point of non-regularity, and thus is eventually convex
down. Therefore $v_1(x, \delta)\, \delta \le g'(x)\, \delta$, and $g'$
is obviously integrable. As regards $v_2$, $\forall \delta \le \eps_0$,
\begin{eqnarray}
	v_2(x, \delta) &\le& \max_{[\l_0, \l_0+\delta]} \left|
	\frac{d\ph}{d\l} \right| = \frac{d\tilde{\ph}}{d\l}
	(\eps_0) \le  \label{udist-80}  \\
	&\le& \left( 1 + \tan \left( \sqrt{ \frac{k_M}{a_m} } 
	\eps_0 + \arctan \left( \frac{\tilde{\ph}_0} {\sqrt{k_M
	a_m}} \right) \right)^2 \right) k_M.  \nonumber
\end{eqnarray}
The last inequality was obtained by plugging (\ref{udist-60}) into
(\ref{udist-50}).  We study the asymptotics of the quantities
contained above. First of all, if $\bar{x} := x(\bar{\l})$,
\begin{equation}
	k_M = k(\bar{\l}) = \frac{\f''(\bar{x})} {\left( 1 +
	(\f'(\bar{x}))^2 \right)^{3/2}} \sim \f''(\bar{x}).
	\label{udist-85}
\end{equation}
Now integrate (A4) to get $\forall \bar{x} \ge x$, $\f''(\bar{x}) \le
\f''(x) e^{c (\bar{x} - x)}$, for some positive $c$.  Since $\bar{\l} -
\l_0 \le \eps_0$, then $\bar{x} - x < \eps_0$. Therefore $\f''(\bar{x})
\sim \f''(x)$, i.e.,
\begin{equation}
	k_M \sim \f''(x)
	\label{udist-90}
\end{equation}
Moving on, it is clear from Fig.~\ref{figf} that
\begin{equation}
	\tilde{\ph}_0 = \pi-\ph_0 = \arctan |\f'(x)| + \arctan
	|\f'(x_t)|.  
	\label{udist-95}
\end{equation}
Therefore
\begin{equation}
	\tilde{\ph}_0 \sim |\f'(x)| + |\f'(x_t)| \sim |\f'(x)|,
	\label{udist-100}
\end{equation}
the last relation coming from (A2). 

The case of $a_m = a(\hat{\l}, \ph_0)$ is a little more involved. We
have already established that $a$ decreases when $\ph$ increases, hence
\begin{equation}
	a_m > \tau_{-1} (\hat{\l}, \hat{\ph}) \sin \hat{\ph},
	\label{udist-105}
\end{equation}
with $\hat{\ph}$ such that $(\hat{\l}, \hat{\ph}) \in \si^{2}$.  Define
$\hat{x} := x(\hat{\l})$ and $\hat{x}_t := x_t(\hat{x})$.
\begin{equation}
	\sin \hat{\ph} \sim |\f'( \hat{x} )| + |\f'( \hat{x}_t )| \sim
	|\f'( \hat{x} )| + 2 |\f'( \hat{x}_t )| \sim \arctan |\f'( 
	\hat{x} )| + 2 \arctan |\f'( \hat{x}_t ), 
	\label{udist-110}
\end{equation}
where the first relation is the analogue of (\ref{udist-100}) and the
second comes from (A2).  Looking back at Fig.~\ref{figf}, we see that
the rightmost term in (\ref{udist-110}) is the slope of the segment of
\tr y from $(\hat{\l}, \hat{\ph}) \in \si^{2}$ to its backward image,
a point that we denote by $(\l_{-1}, \ph_{-1})$. The length of said
segment is $\tau_{-1} (\hat{\l}, \hat{\ph})$. Therefore, from
(\ref{udist-105}), if we call $x_{-1} := x(\l_{-1})$,
\begin{equation}
	a_m \gg \f(\hat{x}) + \f(x_{-1}) > \f(\hat{x}). 
	\label{udist-120}
\end{equation}
In analogy with (\ref{udist-85})-(\ref{udist-90}), we integrate the
second inequality of (\ref{a9}) to obtain an exponential estimate for
$\f$, this time from below. $\forall \hat{x} \ge x$, $\f(\hat{x}) \ge
\f(x) e^{-c (\hat{x} - x)}$.  Therefore (\ref{udist-120}) implies that
$a_m \gg \f(x)$. On the other hand, considering the case when $\ph$ is
approximately $\pi/2$, it is evident that $a_m < 2\f(x)$. Hence,
\begin{equation}
	a_m \sim \f(x).
	\label{udist-130}
\end{equation}

Armed with (\ref{udist-90}), (\ref{udist-100}) and (\ref{udist-130}), we
can now consider (\ref{udist-80}). The argument of the arctan is bounded
by (A3), and so the arctan is less than some $\lambda < \pi/2$.  Also,
from (\ref{a9}), $\f''(x) \ll \f(x)$, thus $k_M/a_m \ll 1$. Let us fix
$\eps_0$ so small that $\sqrt{ k_M/a_m } \, \eps_0 < \pi/2 -
\lambda$. In this way we obtain that $v_2(x,\delta) \ll k_M \sim
\f''(x)$, which is integrable.

We are finally ready to use the integrability of $v_1$ and $v_2$. Denote
by $\si'_{[\eps]}$ the portion of $\si^{2}_{[\eps]}$ that lies beneath
$\si^{2}$. A given point in $\si'_{[\eps]}$ can be connected to a point
of $\si^{2}$ through a curve $\gamma$ with support of length $\delta <
\eps$. Therefore the the above construction (see once again
Fig.~\ref{figg}) implies
\begin{equation}
	\mathrm{Leb}(\si'_{[\eps]}) < \int_0^\infty \eps [ v_1(x - \eps,
	\eps) + v_2(x - \eps, \eps) ] \, dx \le const \cdot \eps.
	\label{udist-140}
\end{equation}
For the sake of rigor, let us remark that for $x \in [-\eps_0,0)$ we
have defined $v_i(x,\eps)$ in an arbitrary suitable way.  

\skippar

We now must prove the same result for the part of $\si^{2}_{[\eps]}$
that stands above $\si^{2}$. This will be even easier. In fact,
differential inequality (\ref{udist-20}) continues to hold. Let us
consider it in $U := [\l_0 - \eps_0, \l_0] \times [\ph_0, \ph_0 +
b]$. We simplify it by introducing the maximum of $k$ there, $k_M :=
k(\bar{\l})$ and the minimum of $\tau_{-1}$, $\tau_m := \tau_{-1}
(\hat{\l}, \ph_0)$, for some $\bar{\l}, \hat{\l} \in [\l_0 - \eps_0,
\l_0]$. This is so because the latter \fn\ is increasing in
$\ph$. Performing again the change of coordinate $\tilde{\ph}(\l) := \pi
- \ph(\l + \l_0)$, the inequality is turned into
\begin{equation}
	\frac{d\tilde{\ph}}{d\l} \le k_M +\frac{\tilde{\ph}} {\tau_m};
	\qquad \tilde{\ph}(0) = \tilde{\ph}_0.
	\label{udist-150}
\end{equation}
We do not even need to solve it. In fact, since $d\tilde{\ph}/d\l > 0$
(because the tangent to $\gamma$ always belongs to the unstable cone),
it is evident that
\begin{equation}
	\max_{[-\delta, 0]} \frac{d\tilde{\ph}} {d\l} =
	\frac{d\tilde{\ph}} {d\l} (0) = k_M +\frac{\tilde{\ph}_0}
	{\tau_m} =: v_2(x,\delta),
	\label{udist-160}
\end{equation}
which is the analogue of (\ref{udist-80}). Concerning the asymptotics of
these quantities, we see that $k_M \sim \f''(x)$, using another
exponential estimate. Also, $\tilde{\ph}_0 \sim |\f'(x)|$, by
(\ref{udist-100}). As to $\tau_m$, one has that $\tau_m > \tau_{-1}
(\hat{\l}, \hat{\ph})$, with $(\hat{\l}, \hat{\ph}) \in \si^{2}$. Then
$a' := \tau_{-1} (\hat{\l}, \hat{\ph}) \sin \hat{\ph}$ is completely
similar to $a_m$ in the previous case. We can use
(\ref{udist-110})-(\ref{udist-130}) to show that $\tilde{\ph_0} /
{\tau_m} \sim \tilde{\ph}_0^2 / a' \ll k_M$. Therefore (\ref{udist-160})
implies that $v_2(x,\delta) \ll \f''(x)$, which is of course integrable.

The integrability of $v_1$ is straightforward, since the convexity of
$g$ implies $v_1(x, \delta)\, \delta \le g'(x-\eps_0)\, \delta$. 

\skippar

This completes the proof of the result for the set $F^{2+}_{[\eps]}$.
Proving the corresponding statement for all the other neighborhoods is
now just a corollary of the above. Take for instance
$F^{1+}_{[\eps]}$.  The r.h.s.\ of (\ref{udist-20}) is decreasing in
$\ph$ (in other words the cones shrink as we approach the upper
boundary vertically). Hence the solutions will have smaller slope (in
absolute value) and $v_2$ (and $v_1$ too, for that matter) will be
smaller, making (\ref{udist-140}) still hold true.

The case of $F^{2-}_{[\eps]}$ is also ``overestimated'' by the above
computations. In fact, passing from a neighborhood of $\si^{2+}$ to
its symmetrical $\si^{2-}$, we observe that $\sin (\pi-\ph) = \sin
\ph$ and $\tau_{-1} (\l, \pi-\ph) = \tau (\l, \ph) > \tau_{-1} (\l,
\ph)$, for $\ph > \pi/2$ and $\l$ large. Once again, the r.h.s.\ of
(\ref{udist-20}) becomes smaller and the previous estimate of $v_2$
largely suffices. Furthermore, the maximum vertical distance between
$\gamma$ and $\si^{2-}$, in this case, is $|v_2 - v_1|$, which makes
the bound even more redundant. The other cases are now clear.  
\qed

\begin{corollary}
	There exists a \me\ $\pi$ defined on the singularity set,
	of finite mass, such that for every \emph{closed} $A \subseteq 
	\si^{+} \cup \si^{-}$, 
	\begin{displaymath}
		\mathrm{Leb} ( A_{[\eps]} ) \le \pi(A) \, \eps,
	\end{displaymath}
	for sufficiently small $\eps>0$. This \me\ is absolutely 
	continuous w.r.t.\ the one-dimensional Lebesgue \me\ on the 
	singularity set. 
	\label{cor-udist}
\end{corollary}

\proof This assertion is what Euclid would have called a
\emph{porism}, in the sense that it is derived from the proof of the
previous theorem, rather than from its statement. At any rate, the
argument is obvious. For a closed arc $A$ in $\si^{i\pm}$, one defines
$\eps \pi(A)$ by simply restricting the integral in (\ref{udist-140})
to the appropriate domain.  
\qed

\sect{Local stable and unstable manifolds}
\label{sec-lsum}

The \ex\ and \ac\ of the invariant manifolds for a \hyp\ system is
given by \emph{Pesin's theory}. It was first developed by Pesin for
smooth systems \cite{p} and has been later adapted and generalized in
many ways (see, e.g., \cite{ps} and its references).

In the case of \bi s, the dynamics is complicated by the presence of
the singularities, which generate a whole new mechanism for two nearby
points to have far-away forward (or backward) images. Extending
Pesin's theory to non-smooth \hyp\ systems was the goal of Katok and
Strelcyn's 1986 major work \cite{ks}. Among their results, the one
that concerns us here is the \ex\ and \ac\ of local stable and
unstable manifolds (LSUMs) at almost every point, for a very general
class of systems; a fine-tuned theorem that requires only reasonable
conditions on the non-regularity set.  This is the main,
irreplaceable, ingredient in the proof of local \erg ity using Hopf's
idea.

The problem is that this theorem, like most in the literature, assumes
the invariant \me\ to be finite, which is not the case of $\ps$!
Practically speaking, this means that we have to work out our own
results for the \hyp\ structure of $(\ps, \ma, \mu)$. This section is
devoted to the \ex\ of LSUMs and the following section to their \ac.

\skippar

We start by giving a definition of LSUM which is sufficient for our
purposes. In many instances, one can postulate (and then prove) much
more about these objects \cite[\S 6]{kh}.

\begin{definition}
	Given a point $\z \in \ps \setminus \si^{+(-)}$, we define a
	{\em local (un)stable manifold $\wsu$ for $\ma$ at $\z$} to be
	a $C^1$ topological disk containing $\z$, and such
	that:
	\begin{itemize} 
		\item[(a)] The tangent space to $\wsu$, at every
		point, is included in the (un)stable cone of every
		order, i.e., $\forall \w \in \wsu$, $\ts_\w \wsu 
		\subset \bigcap_n \co_n(\w)$, and it has the maximal 
		dimension there (1 in our case);

		\item[(b)] $\forall \w \in \wsu$, $|\ma^n \w
		- \ma^n \z| \to 0$, as $n \to +\infty(-\infty)$;

		\item[(c)] If $\wsu_0$ is another such manifold, then
		so is $\wsu \cap \wsu_0$. 
	\end{itemize}
	If the convergence in (b) is exponential, we say that $\wsu$
	is \emph{exponentially} (un)stable.
	\label{def-lsum}
\end{definition}

The above conditions ensure that, if $\wu$ is a LUM at $\z$, then a
subdisk of $\ma^{-1} \wu$ is a LUM at $\ma^{-1} \z$. In fact, by
assumption, $\ma^{-1}$ is smooth in a neighborhood of $\z$ and so at
least for a sufficiently small neighborhood $U$ of $\ma^{-1} \z$,
$\ma^{-1} \wu \cap U$ is a smooth topological disk of the right
dimension. Furthermore, for each $\w$ in this disk and each $n$,
$\ts_\w \ma^{-1} \wu \subset \co_n(\w)$, since $\ts_{\ma \w} \wu
\subset \co_{n+1} (\ma \w)$. In the same way, \emph{(b)} and
\emph{(c)} are easily seen to hold for $\ma^{-1} \wu \cap U$.

Property \emph{(c)} guarantees ``uniqueness'', in a certain sense. For
this reason, we allow the (customary) abuse of notation and call an
(un)stable manifold at $\z$, \emph{the} (un)stable manifold at $\z$,
denoting it by $\wsu(\z)$.

\skippar

We move on to the theorem of \ex\ of LSUMs. As we have mentioned later,
the technical problem is that $\mu(\ps) = \infty$. On the other hand,
the singularity lines, responsible for ``cutting'' the invariant
manifolds, are contained in a finite-\me\ set. Say we choose $\ps_0$ as
in Fig.~\ref{figh}. So, since the system is recurrent, one might think
of working with $\ma_0$, the map induced by $\ma$ on $\ps_0$, prove the
\ex\ of LSUMs for this subsystem and then ``push them forward'' to the
whole of $\ps$ by using $\ma$, since there are no singularities there.

But this is not so simple, since $\ma_0$ has a much bigger
discontinuity set than $\ma$. In fact, the set $\ps_0 \cap
\ma^{-n} \partial \ps_0$ represent the borderline between points that
take $n$ $\ma$-iterations to come back to $\ps_0$, and points that
take $n+1$ iterations. Therefore a new discontinuity is induced there.

We will use a strategy, however, that relies on the idea above, and
the additional fact that these new discontinuities do not really have
to do with the ``physics'' of the system. The new boundary was put
there arbitrarily and, somehow, can be deformed whenever is
convenient. For this reason, $\partial \ps_0$ might deserve the name
of \emph{fuzzy boundary}.

These concepts will be made rigorous in the next theorem. Recall the
notation (\ref{a-eps}).

\begin{theorem}
	Let $\ps$ be a Riemannian manifold, embedded in $\R^N$, and
	$(\ps,\ma,\mu)$ an invertible, recurrent, dynamical system on
	it.  Denote the discontinuity set of $\ma$ by $\si$. Assume that
	for some $\alpha$, the following holds:
	\begin{itemize}
		\item[(a)] $\mu \left( (\si \cup \partial\ps )_{[\eps]}
		\right) \ll \eps^\alpha$, for $\eps \to 0^{+}$.

		\item[(b)] These exists a continuous, invariant cone
		bundle $\co$, such that $\forall \z \in \ps$, $\bigcap_n
		\co_n(\z) = E^u(\z)$, a subspace of $\ts_\z
		\ps$. ($\co_n(\z)$ is defined as in
		\emph{(\ref{n-cone})}.)

		\item[(c)] There exists an increasing norm $\| \,
		\cdot \, \|$ for cone vectors, that is, $\forall \z
		\in \ps \setminus \si$, $\exists \kappa(\z)>1$ such
		that $\forall v \in \co(\z)$, $\| D\ma_\z v 
		\|_{\ma \z} \ge \kappa(\z) \, \| v \|_\z$.

		\item[(d)] Let us denote by $H$ the set where the
		\emph{expansion factor} $\kappa$ is not bounded away
		by 1, i.e., $H := \rset{\z} {\exists \z_n \to \z,
		\kappa (\z_n) \to 1}$. Then $\mu \left( H_{[\eps]} 
		\right) \ll \eps^\alpha$, when $\eps \to 0^{+}$.

		\item[(e)] Denoted by $| \, \cdot \, |$ the Riemannian
		norm on $\ts \ps$, and taken two \fn s $0 < p \le q$ 
		such that $\forall z \in \ps \setminus \si$,
		\begin{displaymath}
			p(\z) \, \|\, \cdot \,\|_\z \le |\, \cdot 
			\,|_\z \le q(\z) \, \|\, \cdot \,\|_\z, 
		\end{displaymath}
		then $p$ is locally bounded below, and $q(\z) \ll 
		[\du(\z, \si \cup \partial\ps)]^{-\beta}$.
	\end{itemize}
	Then, for $\mu$-a.e.\ $\z$, the local unstable manifold
	$\wu(\z)$ exists.

	\skippar

	Furthermore, let us take a $\ps_0 \subseteq \ps$, $\mu(\ps_0) <
	\infty$, such that $(\si \cup \partial\ps)_{[\eps_0]}
	\subseteq \ps_0$, for some $\eps_0 > 0$. Then the $\wu(\z)$ are
	exponentially expanding w.r.t.\ the return times to
	$\ps_0$. This means that, given a $\z \in \ps_0$ for which
	$\wu(\z)$ exists, and denoted by $\{ -n_k \}_{k\in\N}$ the
	sequence of its return times in the past, then $\exists C,
	\lambda >0$ such that
	\begin{displaymath} 
		\forall \w \in \wsu, \quad |\ma^{-n_k} \w - 
		\ma^{-n_k} \z| \le C \, e^{-\lambda k}, \quad 
		\mbox{as } k \to \infty.
	\end{displaymath} 
	\label{thm-ex}
\end{theorem}

\begin{remark}
	The statement of the theorem looks rather cumbersome because we
	have tried to present it in a certain generality. Since, as we
	have recalled, the literature is not very generous in terms of
	results for infinite-\me\ dynamical systems, the hope is that
	this theorem finds application beyond the scope of the present
	paper.
\end{remark}

At any rate, in concrete examples one may expect some of the
conditions to be trivially satisfied. For instance, in our case,
verifying \emph{(a)-(d)} will be immediate, as we will see
later. However, it turns out that hypothesis \emph{(e)} does not hold!
Indeed we must substitute it with a different set of
requirements. Nevertheless, for the sake of clarity---and in the
spirit of the above paragraph---we have decided to state Theorem
\ref{thm-ex} in the given form. This contains all the relevant
ideas, and most technical points. Theorem \ref{thm-ex2} will work out
the necessary adaptations for use on our \bi.

\skippar

\proofof{Theorem \ref{thm-ex}} Without loss of generality, assume that
also $H_{[\eps_0]} \subseteq \ps_0$. As anticipated, we denote by $\ma_0$
the return map onto $\ps_0$. This is well defined due to the recurrence.
Let us define
\begin{eqnarray}
	A_k &:=& \rset{\z \in \ps_0} { \du(\ma_0^{-k} \z, H \cup \si
	\cup \partial \ps) < \frac1 {(k+1)^{2/\alpha}} } = \nonumber \\ 
	&=& \ma_0^k \left( \left( H \cup \si \cup \partial \ps
	\right)_{[(k+1)^{-2/\alpha}]} \right) ,
	\label{ex-10}
\end{eqnarray}
having used notation (\ref{a-eps}). Hypotheses \emph{(a), (d)} guarantee
that $\mu(A_k) \ll (k+1)^{-2}$. Therefore, denoting
\begin{equation}
	A := \{ \, \{A_k\} \mbox{ infinitely often } \} :=
	\bigcap_{m\in\N} \bigcup_{k\ge m} A_m,
	\label{ex-20}
\end{equation}
we have that $\mu(A)=0$ by the Borel-Cantelli Lemma. Via an easy
argument, then, $\forall \z \in \ps_0 \setminus A$, $\exists C_1(\z) >
0$ such that, $\forall k \in \N$,
\begin{equation}
	\du(\ma_0^{-k} \z, H \cup \si \cup \partial \ps) \ge \frac{2
	C_1} {(k+1)^{2/\alpha}}.  
	\label{ex-30}
\end{equation}

By hypothesis, $H$ is a null-\me\ set. On its complement we define
\begin{equation}
	\psi(\z) := \inf_{ \w \in \bu(\z, \du(\z,H)/2) } \log 
	\kappa(\w) > 0,
	\label{ex-40}
\end{equation}
It is crucial to notice here that the ball in the above definition is
a $\du$-ball of $\ps$ and not $\ps_0$! In other words, we seek the
infimum of $\log \kappa$ in a neighborhood of $\z$ that can exceed
$\ps_0$.

Applying Lemma \ref{lemma-s-arg} in the Appendix to the dynamical system
$(\ps_0, \ma_0, \mu)$, gives that there is a $B \subset \ps_0$,
$\mu(B)=0$, such that, if $\z \in \ps_0 \setminus B$,
\begin{equation}
	\lim_{m\to\infty} \frac1m \sum_{k=1}^{m} \psi( \ma_0^{-k} 
	\z ) > 0. 
	\label{ex-50}
\end{equation}

\skippar

Now let us fix a $\z \in \ps_0 \setminus (A \cup B)$, and let $\{ -n_k
\}$ be its sequence of past return times, as in the statement of the
theorem. Since $\psi$ is positive, (\ref{ex-50}) implies that $\exists
\lambda > 0$ such that, $\forall m \in \Z^{+}$,
\begin{equation}
	\sum_{k=1}^{m} \psi( \ma^{-n_k} \z ) \ge \lambda m.
	\label{ex-60}
\end{equation}
A point $\w$ in a neighborhood of $\ma^{-n_m} \z$ is ``good'' if,
for all $j=0, .... , m$,
\begin{equation}
	\ma^{n_m - n_j} \w \in \bu \left( \ma^{-n_j} \z, \frac{C_1} 
	{(j+1)^{2/\alpha}} \right).
	\label{ex-65}
\end{equation}
Thus, using (\ref{ex-30}),
\begin{equation}
	\ma^{n_m - n_j} \w \in \bu \left( \ma^{-n_j} \z, \frac{ \du( 
	\ma^{-n_j} \z, H \cup \si \cup \partial \ps) } {2} \right).
	\label{ex-70}
\end{equation}
In particular, if $\w$ is good, the lower bound (\ref{ex-40}) holds
generous for $\w$ and its first $m$ forward images. The dynamics we are
talking about here is $\{ \ma^{n_m - n_j} \}$, the one generated by the
return times of $\z$. These are not necessarily the same return times as
$\ma^{n_m} \w$: in fact, the balls that appear in (\ref{ex-70}) may very
well exceed $\ps_0$. This implements the aforementioned idea of
$\partial \ps_0$ as a fuzzy boundary.

For $\w$ good and $v \in \co(\w)$,
\begin{equation}
	\| D\ma^{n_m} v \|_{\ma^{n_m} \w} \ge \exp \left\{
	\sum_{j=0}^{m-1} \psi( \ma^{n_m - n_j} \w ) \right\} \| v
	\|_\w \ge e^{\lambda m} \, \| v \|_\w, 
	\label{ex-80}
\end{equation}
because of (\ref{ex-60}).  We naturally call good an unstable curve
$\gamma$ which contains $\ma^{-n_m} \z$ and is made up of good
points. If we denote by $\len_\|$ the length of a curve in the $\|
\,\cdot\, \|$ metric, then (\ref{ex-80}) implies that a good curve
$\gamma$ verifies
\begin{equation}
	\len_\| (\ma^{n_m} \gamma) \ge e^{\lambda m} \, \len_\|
	(\gamma). 
	\label{ex-90}
\end{equation}

A sufficiently short (in the sense of $\len_\|$) $\gamma$ is
obviously good. We claim that $\gamma$ can be elongated in such a
way that it remains good and also
\begin{eqnarray}
	\len_\| (\ma^{n_m - n_j} \gamma) &\le& C_2 \, 
	e^{-\lambda j}, \qquad j=1, ... , m; \nonumber \\
	\len_\| (\ma^{n_m} \gamma) &=& C_2, \label{ex-100}
\end{eqnarray}
for some $C_2 = C_2(\z)$ to be determined as follows: By (\ref{ex-70})
and (\ref{ex-30}),
\begin{equation}
	\du(\ma^{n_m - n_j} \w, \si \cup \partial \ps) \ge 
	\frac{C_1} {(j+1)^{2/\alpha}}.  
	\label{ex-110}
\end{equation}
Therefore, by hypothesis \emph{(e)}, there is a $C_3$ such that
\begin{equation}
	q (\ma^{n_m - n_j} \w) \le C_3 \, (j+1)^{2\beta/\alpha}.
	\label{ex-120}
\end{equation}
We use the above to define $C_2$ as any number that verifies, for all $k
\in \N$, 
\begin{equation}
	C_2 \, C_3 \, (k+1)^{2\beta/\alpha}\, e^{-\lambda k} \le 
	\frac{C_1} {(k+1)^{2/\alpha}}.
	\label{ex-130}
\end{equation}

We proceed to prove (\ref{ex-100}). Assume first that $\gamma$ stays
good as the elongation is performed. Violating (\ref{ex-100}) amounts
to finding a $j$, $1 \le j \le m$, such that $\len_\| (\ma^{n_m - n_j}
\gamma) = C_2 \, e^{-\lambda j}$ and $\len_\| (\ma^{n_m} \gamma) <
C_2$, but this would contradict (\ref{ex-90}) with $j$ replacing $m$
and $\ma^{n_m - n_j} \gamma$ (also a good curve) replacing
$\gamma_j$.

It remains to show that $\gamma$ remains good until it reaches
situation (\ref{ex-100}), i.e., as long as $\len_\| (\ma^{n_m - n_j}
\gamma) \le C_2 \, e^{-\lambda j}$, $j=0, ... , m$. From
(\ref{ex-120}), 
\begin{equation}
	\len (\ma^{n_m - n_j} \gamma) \le C_3 \, 
	(j+1)^{2\beta/\alpha} \, \len_\| (\ma^{n_m - n_j} \gamma),
	\label{ex-140}
\end{equation}
recalling that $\len$ is the Riemannian length. Therefore, with
the help of (\ref{ex-130}),
\begin{equation}
	\len (\ma^{n_m - n_j} \gamma) \le C_2 \, C_3 \,
	(j+1)^{2\beta/\alpha} \, e^{-\lambda j} \le \frac{C_1}
	{(j+1)^{2/\alpha}}.
	\label{ex-150}
\end{equation}
Hence, by definition of $\du$, all points of $\ma^{n_m - n_j} \gamma$
are as close to $\ma^{-n_j} \z$ as (\ref{ex-65}) prescribes. This
completes the proof of the claim.

\begin{remark}
	Notice that, by (\ref{ex-70}), none of the $\ma^{n_m - n_j}
	\gamma$, for $j=0, ... , m$, can be cut by $\si$. This is all
	the more true for the other iterates $\ma^{n_m - n} \gamma$, $n
	= 0, ... , n_m$, $n \ne n_j$. In fact, for $n_j < n < n_{j+1}$,
	$\ma^{-n} \z \not\in \ps_0$ and, by construction, $\ma^{-n} \z$
	is even further away from $\si \cup \partial\ps_0$ than
	$\ma^{-n_j} \z$. Therefore, by \emph{(e)}, (\ref{ex-120}),
	(\ref{ex-140}) hold with $n$ in the place of $n_j$. Add that
	$\len_\| (\ma^{n_m - n} \gamma) \le \len_\| (\ma^{n_m - n_j}
	\gamma)$, and (\ref{ex-150}) holds too with $n$ replacing $n_j$.
	\label{rk10}
\end{remark}

We have now worked out the most technical part of this proof: we have
managed to reconduct to a situation in which certain curves are
exponentially contracting (in the Riemannian metric) up to some time in
the past, as seen in (\ref{ex-150}). The only peculiarity is that this
rate of contraction is attained w.r.t.\ the return times of a point $\z$
to $\ps_0$. Furthermore, these curves do not see the singularity
lines. Pesin's key idea is precisely that disks made up of such curves
are natural approximations for the unstable manifolds. So, apart from
the fact that we use a different time scale, the reasoning will now
become very standard in the context of Pesin's theory and thus the
exposition will be a little less detailed.

Given $\z \in \ps \setminus (A \cup B)$ as above, let $\Delta_m$ be a
smooth topological disk, centered in $\ma^{-n_m} \z$, good, lying
``inside'' the cone bundle (i.e., $\forall \w \in \Delta_m$, $\ts_\w
\Delta_m \subset \co(\w)$), and with the maximal dimension there. Call
this dimension $\nu \in \Z^{+}$.

Although $\Delta_m$ needs to be small to be good, we can choose it large
enough that $\ma^{n_m} \Delta_m$ is ``macroscopic''.  More precisely, we
require $\Delta_m$ to contain a topological disk of radius $C_2
e^{-\lambda m}$ in the $\| \,\cdot\, \|$ metric. This means that
\begin{equation}
	\inf \rset{ \len_\| (\gamma) } { \gamma \mbox{ smooth curve} 
	\, \subset \Delta_m \mbox{ linking } \ma^{-n_m} \z \mbox{ to }
	\partial \Delta_m } \ge C_2 \, e^{-\lambda m}.  
	\label{ex-160}
\end{equation}
By the first inequality of \emph{(e)}, $\len_\| (\, \cdot \,) \le C_4
\, \len (\, \cdot \,)$ in a neighborhood of $\z$. This guarantees that
there is a topological disk of Riemannian radius $C_2/C_4$ inside
$\ma^{n_m} \Delta_m$, that we can name $\mathcal{B}_m$. In fact,
assume the contrary. Then there exists a curve $\eta \subset \ma^{n_m}
\Delta_m$, starting at $\z$ and reaching $\partial \ma^{n_m}
\Delta_m$, such that $\len (\eta) < C_2/C_4$, which gives $\len_\|
(\eta) < C_2$. Therefore $\ma^{-n_m} \eta$ is good; hence, by
(\ref{ex-90}), $\len_\| (\ma^{-n_m} \eta) < C_2 e^{-\lambda m}$. But
$\ma^{-n_m} \eta$ links $\ma^{-n_m} \z$ to some $\partial \Delta_m$,
and this contradicts (\ref{ex-160}).

\skippar

In compliance with the plan we have anticipated, let us define:
\begin{equation}
	\wu(\z) := \lim_{m \to \infty} \mathcal{B}_m.
	\label{ex-170}
\end{equation}
The limit here is intended in a certain $C^1$ Hausdorff distance. More
precisely, if $\mathcal{B}$ and $\mathcal{B}'$ are two $C^1$ (closed)
disks, then their distance is defined as
\begin{equation}
	dist (\mathcal{B}, \mathcal{B}') := \max_{\z \in \mathcal{B}}
	dist ( \z, \mathcal{B}' ) + \max_{\w \in \mathcal{B}'} dist (
	\w, \mathcal{B} ), 
	\label{ex-180}
\end{equation}
where, with abuse of notation, we have denoted by the same symbol the
distance between two sets and the distance between a point and a
set. The latter is
\begin{equation}
	dist ( \z, \mathcal{B}' ) := \min_{\w \in \mathcal{B}'} \{ |\z -
	\w| + d_{G_l} ( \ts_\z \mathcal{B}, \ts_\w \mathcal{B}') \}.
	\label{ex-190}
\end{equation}
The norm and the distance on r.h.s.\ above are those inherited by the
embedding in $\R^{N}$. More specifically, $d_{G_\nu}$ is any distance in
$G_\nu (\R^N)$, the space of $\nu$-dimensional planes in $\R^N$. The
distance defined by (\ref{ex-180}) is complete. We will use Lemma
\ref{lemma-tecn1} in the Appendix to prove that limit (\ref{ex-170})
exists.

For all $m \ge 0$, denote by $K_m$ the compact neighborhood of $\z$
made up of points $\z'$ such that $\ma^{-n_m} \z'$ is good. In other
words, $K_m$ is the intersection, for $j = 0, \cdots, m$ of the
forward images of the balls in the r.h.s.\ of (\ref{ex-65}). Of course
$K_{m+1} \subseteq K_m$. In particular, points of $K_m$ stay away from
$\si_{n_m}^{-} = \bigcup_{i=0}^{n_m-1} \si^{-}$, so $\ma^{-n_m}$ is a
diffeomorphism between $K_m$ and its image. This implies that
$\co_{n_m}$ varies continuously on $K_m$. Moreover, if $g_m(\z')$
denotes the size of $\co_{n_m}(\z')$, i.e.,
\begin{equation}
	g_m(\z') := \max \rset{ d_{G_\nu} (X - Y) } { X,Y\, \nu
	\mbox{-dim.\ subspace of } \co_{n_m}(\z') }, 
	\label{ex-200}
\end{equation}
then one has $g_m(\z') \searrow 0$, as $m \to +\infty$, by condition
\emph{(b)}. At this point Lemma \ref{lemma-tecn1} tells us that this
convergence occur somehow ``uniformly'', although on shrinking sets. But
this is sufficient to see that, for $j>m$ large enough,
$dist(\mathcal{B}_m, \mathcal{B}_j) \le \eps$. In fact, both $\ma^{n_m}
\Delta_m$ and $\ma^{n_j} \Delta_j$ are contained in $K_m$ and their
tangent spaces are uniformly close over the two disks; furthermore,
since the disks have at least $\z$ in common, their points are also
close.

The completeness of $dist$ proves that $\mathcal{B}_m$ has a
limit. Moreover, one can see that the limit does not depend on the
choice of $\Delta_m$. In fact the above argument works as well if we
replace $\Delta_j$ with some other good $\Delta_j'$, so that also
$dist(\mathcal{B}_m, \mathcal{B}_j') \le \eps$, for the same $m$ and
$j$. This gives $\wu(\z) = \lim_m \mathcal{B}_m = \lim_j
\mathcal{B}_j'$.

For almost every $\z \in \ps \setminus \ps_0$, one obviously defines
$\wu(\z)$ as $\ma^{-n} \wu(\ma^n \z)$, if $n$ is the smallest positive
integer s.t.\ $\ma^n \z \in \ps_0$.  

\skippar

It remains to show that $\wu(\z)$ verifies the axioms of Definition
\ref{def-lsum}.  \emph{(a)} is just obvious by construction. \emph{(c)}
is more or less as direct: in fact, fixed $\z \in \ps_0$ for simplicity,
and taken given another $\wu_0(\z)$, one can construct $\Delta_j'$
simply by taking $\ma^{-n_j} \wu_0(\z)$, and possibly by extending it in
an arbitrary way, should it be smaller than the size prescribed by
(\ref{ex-160}).  By the above argument, re-applying $\ma^{n_j}$ and
taking the limit gives again $\wu(\z)$. Lastly, estimating the middle
term of (\ref{ex-150}) with some $C_5 \, e^{\lambda' j}$, for a certain
$\lambda' < \lambda$, shows that, $\forall \z' \in \wu(\z)$,
$|\ma^{-n_j}(\z') - \ma^{-n_j}(\z)| \le C_5 \, e^{\lambda' j}$, since
the distance in $\R^N$ is certainly less than or equal to the unstable
distance on $\ps$. This proves the last statement of Theorem
\ref{thm-ex}. The fact that $|\ma^{-n}(\z') - \ma^{-n}(\z)|$ becomes
small even for $n \ne n_j$ has been explained in Remark \ref{rk10}. This
verifies Definition \ref{def-lsum}, \emph{(b)}, whence the theorem.
\qed

Let us check that hypotheses \emph{(a)-(d)} of Theorem \ref{thm-ex} hold
for our non-compact \bi: \emph{(a)} is true by Theorem
\ref{thm-udist}. \emph{(b)} holds by the results in Section
\ref{sec-cone}, in particular Proposition \ref{prop-cont-frac}. The
increasing norm in \emph{(c)} is
\begin{equation}
	\| (d\l, d\ph) \|_{(\l, \ph)}^2 := \sin^2 \ph \, d\l^2 + d\ph^2,
	\label{ex-a3}
\end{equation}
see (\ref{expan-dl})-(\ref{expan-dph}). Also, $H = \emptyset$, and so
there is nothing to prove in \emph{(d)}.

From (\ref{ex-a3}), $p \equiv 1$, and $q(\l, \ph) = 1/\sin\ph$.
Therefore, as one can easily see, $\emph{(e)}$ fails to hold in our
case. The next theorem circumvents this problem. Actually, the
arguments will even resemble more closely those used for finite-\me\
dynamical systems. Except that, for infinite \me, the formulation is
heavier than Theorem \ref{thm-ex}.

\begin{theorem}
	The assertions of Theorem \ref{thm-ex} also hold if the
	estimate on $q$ in (e) is replaced by the following:
	\begin{itemize}
		\item[(f)] There exists a $\beta>0$ such that 
		$\int_{\ps_0} q^\beta \, d\mu < \infty$.

		\item[(g)] There exist $\eps_0, C'>0$ such that
		\begin{displaymath} 
			\sup_{ \w \in \bu(\z, \du(\z, \si \cup 
			\partial\ps )/2) } q(\w) \, \le C' \, q(\z), 
		\end{displaymath}
		uniformly in $\z$, every time $\du(\z, \si \cup 
		\partial\ps) \le \eps_0$.

		\item[(h)] Fixed a $\z \in \ps_0$, with $\{ -n_j \}$
		the sequence of its past return times to $\ps_0$,
		then, for $n_j < n < n_{j+1}$, $q(\ma^{-n} \z) \le C''
		q(\ma^{-n_j} \z)$, with $C''$ not depending on $\z$.
	\end{itemize}
	\label{thm-ex2}
\end{theorem}

\proofof{Theorem \ref{thm-ex2}} The main fact that we lose, if we give
up $\emph{(e)}$ in Theorem \ref{thm-ex}, is (\ref{ex-120}). That is, we
do not know whether, for $\w$ good, the ratio between the two metrics
grows polynomially along the backword \o\ of $\ma^{n_m} \w$. When this
happens, the growth is eventually tamed by the exponential contraction
in the increasing norm.

We have to reconstruct this situation: By \emph{(f)} and a suitable 
Chebychev-type inequality,
\begin{equation}
	\mu \left( \rset{\z \in \ps_0} {q(\z) > k^{2/\beta} } \right) 
	\ll k^{-2}.
	\label{ex2-10}
\end{equation}
In a way totally analogous to (\ref{ex-10})-(\ref{ex-30}), one concludes
that, for $\z$ outside a null-\me\ set, $\exists C_6 = C_6(\z)$ such that
\begin{equation}
	q(\ma_0^{-k} \z) = q(\ma^{-n_k} \z) \le C_6\, k^{2/\beta}.
	\label{ex2-13}
\end{equation}
Fixed any such $\z$, fast-forward to (\ref{ex-65})-(\ref{ex-70}): if
$\w$ is good and $m \ge j \ge j_0$, for some $j_0 = j_0(\eps_0)$, then
\begin{equation}
	\du(\ma^{n_m - n_j} \w, \ma^{-n_j} \z) \le \min \left\{ \eps_0,
	\frac{ \du( \ma^{-n_j} \z, \si \cup \partial \ps) } {2}
	\right\}.  
	\label{ex2-17}
\end{equation}
Hence
\begin{equation}
	q(\ma^{n_m - n_j} \w) \le C_5 \, j^{2/\beta}.
	\label{ex2-20}
\end{equation}
In fact, for $j \ge j_0$, the above comes from (\ref{ex2-13}) and
\emph{(g)}, with $C_5 = C'\, C_6$; the finitely many remaining values
of $j$ can be included by adjusting $C_5$. (\ref{ex2-20}) replaces
(\ref{ex-120}), which is what we wanted to do.

The remaining hypothesis has to do with Remark \ref{rk10}. Without
\emph{(h)} it might happen that, for $n_j < n < n_{j+1}$, $\ma^{n_m -
  n} \gamma$ is so long, in the Riemannian length, that it can reach
$\si$. Instead, by \emph{(h)} and (\ref{ex2-20}), (\ref{ex-120}) holds
as well with $n$ in the place of $n_j$, and so do
(\ref{ex-140})-(\ref{ex-150}).  
\qed

\fig{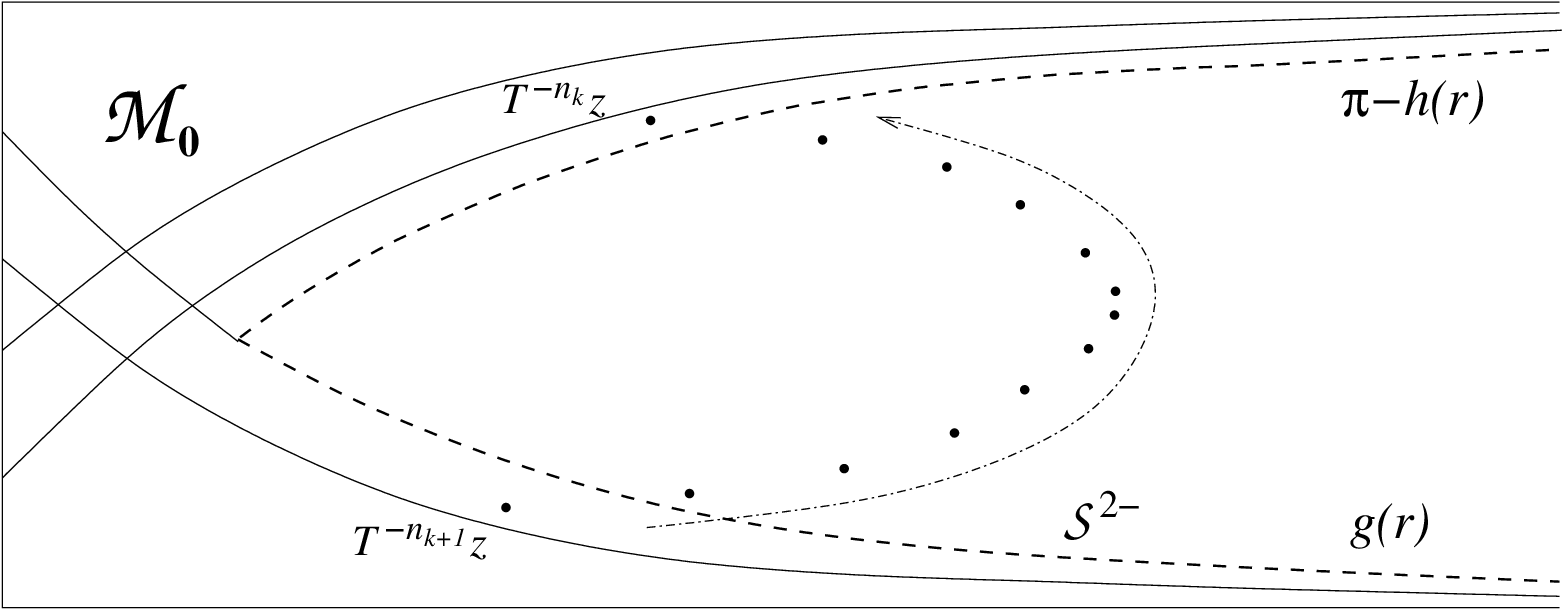} {5in} {A suitable choice of $\ps_0$.}

In order to use the above result, we have to define a suitable
$\ps_0$. Let us chose it like in Fig.~\ref{figh}. More precisely, we
define it in an arbitrary way in a compact region of $\ps$ and, for
$\l$ (or $x$) large, we ask that $\partial \ps_0 \cap \Int \ps$ be
composed of two curves. The lower curve is given by $\si^{2-}$. With
the usual correspondence $\l \longleftrightarrow x$---see
(\ref{udist-70})---this is the graph of the \fn
\begin{equation}
	\l \mapsto g(\l) := \arctan |\f'(x)| + \arctan |\f'(x_t)|  
	\sim |\f'(x)|;
	\label{ex-a5}
\end{equation}
see (\ref{udist-95})-(\ref{udist-100}). (Notice that we have already
encountered this \fn\ in the proof of Theorem \ref{thm-udist},
although what we denoted $g$ there was the graph of $\si^{2+}$, which
is $\pi - g$ here.)  As concerns the upper curve, this must lie below
a certain neighborhood $\si^{2+}_{[\eps_0]}$. From (\ref{udist-80}),
(\ref{udist-90}) and (\ref{udist-140}), we know that the thickness of
$\si^{2+}_{[\eps_0]}$ is of the order of $\f''(x)$. By (\ref{a9}) this
is asymptotically bounded by $|\f'(x)|$. Hence the sought curve can be
chosen as the graph of a \fn\ $\l \mapsto \pi - h(\l)$, with $h(\l)
\sim g(\l) \sim |\f'(x)|$.  Therefore \emph{(f)} holds with $\beta =
1$ since, by definition
\begin{equation}
	\int_{\ps_0} q(\l, \ph) \, d\mu(\l, \ph) = \int_{\ps_0} d\l d\ph
	= \mathrm{Leb}(\ps_0) < \infty.
	\label{ex-a10}
\end{equation}

\fig{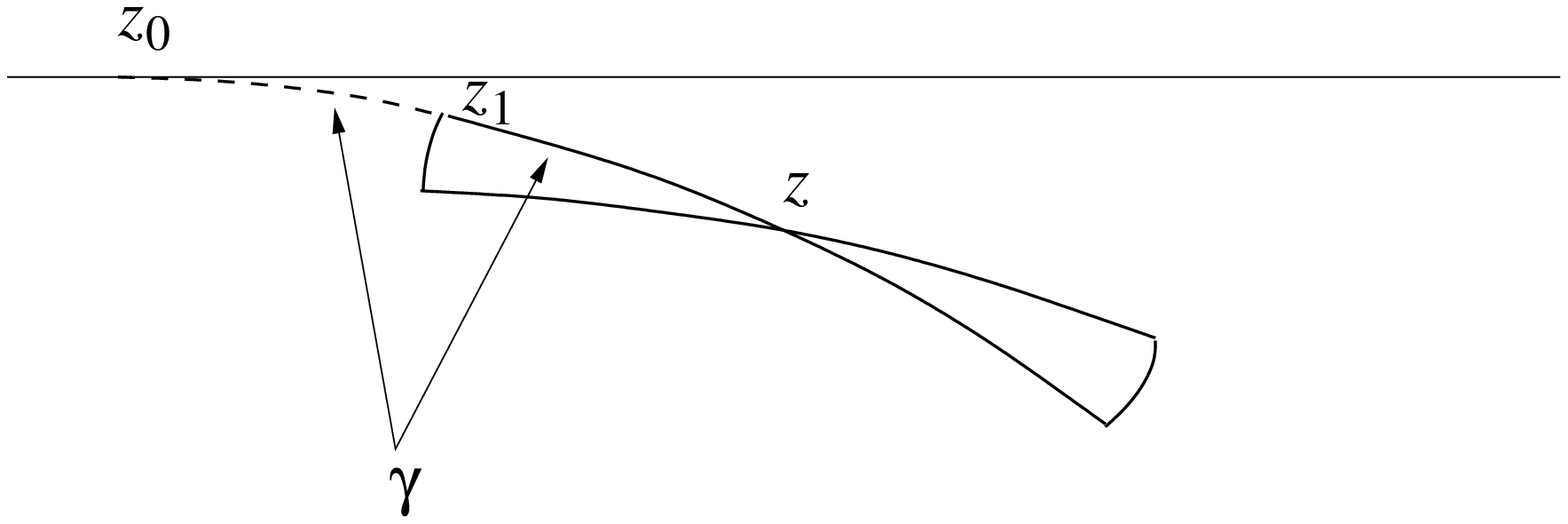} {4in} {The ``bowtie'' $\bu(\z, \du(\z, \partial\ps )/2)$.}

Verifying \emph{(g)} will be a trifle more boring. First of all, we can
forget about $\si$, since $q$ only diverges at $\partial\ps$.
Fig.~\ref{figi} shows that $\bu(\z, \du(\z, \partial\ps )/2)$ looks like
a ``bowtie''. If $\z = (\l, \ph)$ is sufficiently close to $\partial\ps$
(in the Riemannian sense) the bowtie is fairly horizontal and the
maximum of $q(\w)$ is achieved at the indicated point $\z_1 = (\l_1,
\ph_1)$. This point cuts the curve $\gamma$ in two parts of equal
length. $\gamma$ is exactly the type of unstable curve that we have
studied in the proof of Theorem \ref{thm-udist}. It is the graph of a
\fn\ $\l \mapsto \ph(\l)$ that satisfies
\begin{equation}
	\frac{d\ph}{d\l} = -k(\l) - \frac{\sin \ph} {\tau_{-1}
	(\l,\ph)}; \qquad \ph(\l_0) =\pi; 
	\label{ex-a15}
\end{equation} 
see (\ref{udist-20}). We need to prove that
\begin{equation}
	\frac{q(\z_1)} {q(\z)} = \frac{\sin\ph} {\sin\ph_1} = 
	\frac{\sin \ph(\l)} {\sin \ph(\l_1)} \le C'.  
	\label{ex-a20}
\end{equation}
Defining $\tilde{\ph} (\l) := \pi - \ph(\l + \l_0)$, as in Section
\ref{sec-lsum}, and noting that $\sin\ph = \sin \tilde{\ph} \sim
\tilde{\ph}$, for $\ph$ close to $\pi$, a sufficient condition for
(\ref{ex-a20}) is
\begin{equation}
	\frac{\tilde{\ph}(2 \bar{\l} )} {\tilde{\ph}( \bar{\l} )} 
	\le C_7, 
	\label{ex-a30}
\end{equation}
having called $\bar{\l} := \l_1 - \l_0$.  The above holds since
$\tilde{\ph}$ is a convex increasing \fn\ and $2 (\l_1 - \l_0) > \l -
\l_0$. A suitable version of the Lagrange Mean Value Theorem ensures
that $\exists \hat{\l} \in (0,\bar{\l})$ such that
\begin{equation}
	\frac{\tilde{\ph}(2\, \bar{\l}_1 )} {\tilde{\ph}( \bar{\l}_1 )}
	= \frac{ \ds 2 \frac{d\tilde{\ph}} {d\l} (2\hat{\l}) } { \ds
	\frac{d\tilde{\ph}} {d\l} (\hat{\l}) }.  
	\label{ex-a40}
\end{equation}
The denominator is bounded below by $k( \l_0 + \hat{\l} )$, via
(\ref{ex-a15}). As to the numerator, that is the r.h.s.\ of a
differential equation that we know how to estimate, from the proof
of Theorem \ref{thm-udist}---at least for $\gamma$ shorter than some
$\eps_0$. Indeed, fix $\eps_0$ as in Theorem \ref{thm-udist}.  If we
regard $\gamma$ as having an initial point in $( \hat{\l}, \tilde{\ph}
(\hat{\l}) )$, then (\ref{udist-80})-(\ref{udist-130}) say that
$d\tilde{\ph} / d\l (2\hat{\l}) \le C_8 k_M$, $C_8$ not depending on
$\z_0$. But $k_M = \max_{[\l_0 + \hat{\l}, \l_0 + \eps_0]} k$ was shown
to be asymptotically of the same order as $k( \l_0 + \hat{\l} )$.  Hence
(\ref{ex-a40}) is bounded above.

\fig{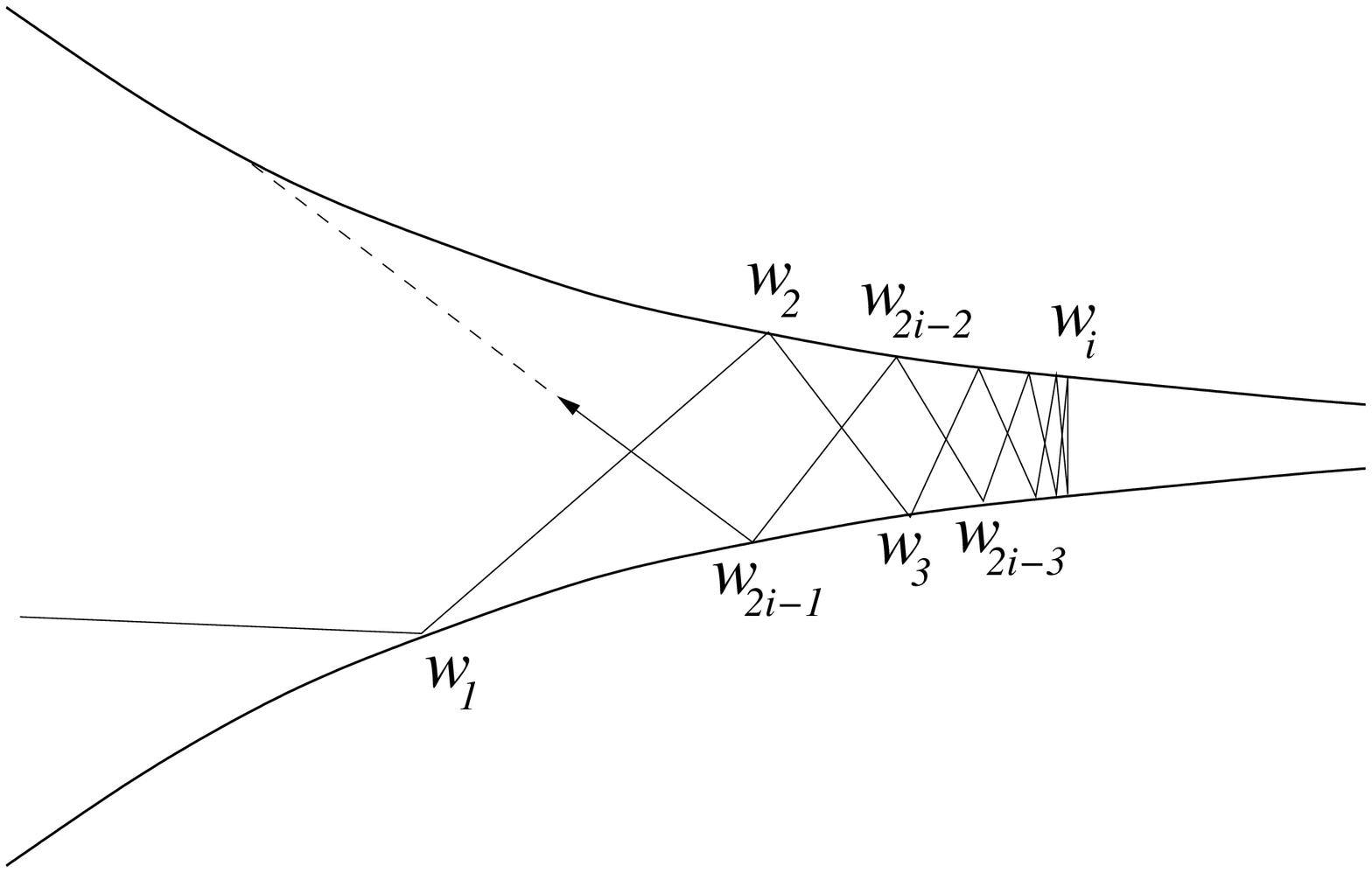} {4in} {An example of a \bi\ \tr y moving towards the cusp
and coming back. (This trajectory correspond exactly to the orbit of
Fig.~\ref{figh}, provided $\w_1 := \ma^{-n_{k+1}} \z$ and $\w_{2i-1}
:= \ma^{-n_k} \z$.) This illustration shows how one can think of the
two halves of the \tr y (the part moving right and the part moving
left) as the boundary of a dispersing beam of \tr ies originated at
$\w_i$.  Denote $(\l_j, \ph_j) := \w_j$. Then $j \mapsto \ph_j$ is
increasing, with $\ph_i$ very close to $\pi/2$.  Also, for $j<i$, the
arc $(\l_{j+2}, \l_{2i-j-2})$ lies entirely to the right of the arc
$(\l_j, \l_{2i-j})$. Hence, at most one other forward point $\w_l$
(namely $\w_{j+1}$), is such that of $\l_j \le \l_l \le \l_{2i-j}$.}

We now show that \emph{(h)} holds. Denote $\z_{-n} := (\l_{-n},
\ph_{-n}) := \ma^{-n} \z$. When $\z_{-n_k}$ belongs to any compact
subset of $\ps_0$, the portion $\{ \z_{-n} \}_{n = n_k+1}^{n_{k+1}-1}$
of its \o\ lies in a compact subset of $\ps$ (a consequence of
Proposition \ref{prop-eo}) where $q$ is bounded. Thus, we only need to
prove \emph{(h)} for $\z_{-n_k}$ lying far on the right of $\ps_0$;
either above the graph of $\pi - h$, or below $\si^{2-}$.

In the former case, the backward \o\ of $\z_{-n_k}$ is depicted in
Fig.~\ref{figh} and corresponds to a \bi\ \tr y going (back in time)
towards the cusp and coming back, as shown in Fig.~\ref{figj}. The
caption of that figure explains that there can be at most one value of
$n$ (precisely $n = n_{k+1} - 1$) for which $\l_{-n} > \l_{-n_k}$ and
$1/q(\z_{-n}) = \sin \ph_{-n} \le \sin \ph_{-n_k} = 1/q(\z_{-n_k})$.
For this $n$, it is not hard to get convinced that $\sin \ph_{-n} \sim
\sin \ph_{-n_k}$. As for the other values of $n$ for which $\sin
\ph_{-n} < \sin \ph_{-n_k}$ (to avoid confusion we point out that
there are none, in Fig.~\ref{figh}), $\z_{-n}$ must lie to the left of
$\z_{-n_k}$, and above $\si^{2-}$ by construction. Hence, from the
asymptotics $g(\l) \gg h(\l)$,
\begin{equation}
	\sin \ph_{-n} \sim \ph_{-n} \ge g(\l_{-n}) \ge g(\l_{-n_k}) \gg
	h(\l_{-n_k}) \ge \ph_{-n_k} \sim \sin \ph_{-n_k}.
\end{equation}
On the other hand, there is nothing to prove for the case in which
$\z_{-n_k}$ lies below $\si^{2-}$. In fact, the past \bi\ \tr y of that
point crosses the $y$-axis in $\ta_4$: therefore $\z_{-n_k-1}$ lies
above $\si^{2+}$, hence in $\ps_0$.

\sect{Absolute continuity}
\label{sec-ac}

The purpose of this section is to establish the \ac\ of the LSUMs
w.r.t.\ the invariant \me\ $\mu$. Later on we will specify precisely
what this means.

We need to introduce yet another cross-section for the \bi\ flow: the
cross-section induced by countably many \emph{transparent walls} $G_n
:= \{X_n\} \times [0, \f(X_n)]$, as depicted in Fig.~\ref{figl}.  We
choose $X_n$ ($n>1$) such that $\f(X_n) = n^{-3}$, and consider only
line elements based in one of the $G_n$'s. The phase space is $\ps_1
:= \bigsqcup_{n\ge 1} \left( \ps^{r,n} \sqcup \ps^{l,n} \right)$, with
$\sqcup$ denoting the disjoint union. $\ps^{l,n}$ is defined as $(0,
n^{-3}) \times (0,\pi)$ and its points indicated by $(\l, \ph)$; the
position variable $\l$ is the $y$-coordinate of the point in $G_n$,
and the direction variable $\ph$ is the counterclockwise angle ($\le
\pi$) between the velocity vector and the $y$-direction. Thus, line
elements in $\ps^{l,n}$ point left, whence the notation. $\ps^{r,n}$
is formally defined in the same way, but $\l$ equals $n^{-3}$ minus
the $y$-coordinate of the point, and $\ph$ is the counterclockwise
angle between the unit vector and the negative $y$-direction. Line
elements in $\ps^{r,n}$ point right (Fig.~\ref{figl}).

\fig{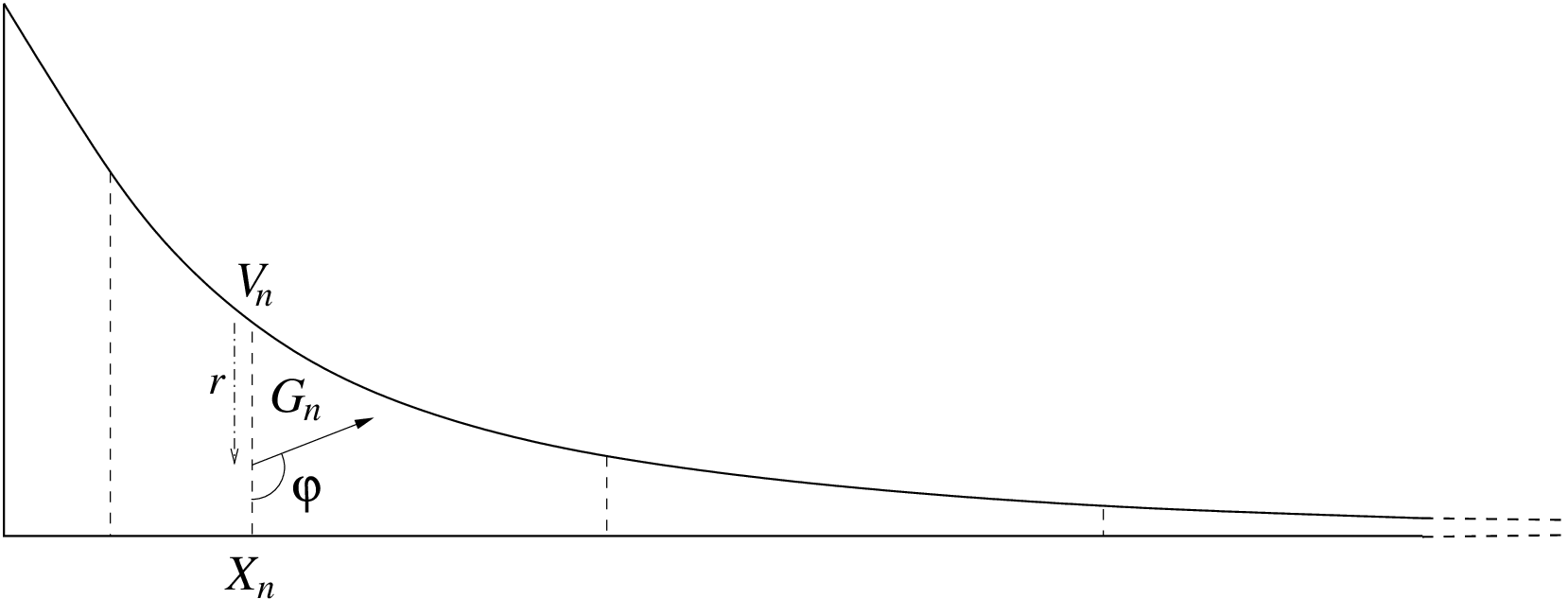} {5in} {The definition of $G_n$, $V_n$, and $\ps^{r,n}$.}

The Poincar\'e map, which we name $\ma_1$, is defined on all points of
$\ps_1$ that would not result in a tangency or the hitting of a vertex
$V_n$ (see Fig.~\ref{figl} for the definition of $V_n$). We call the
set of these excluded points $\sir$ and, for $i=l,r$, denote
$\sir^{i,n} := \sir \cap \ps^{i,n}$. A sketch of these two sets is
given in Fig.~\ref{figm}, and an explanation follows momentarily,
after an important remark.

\fig{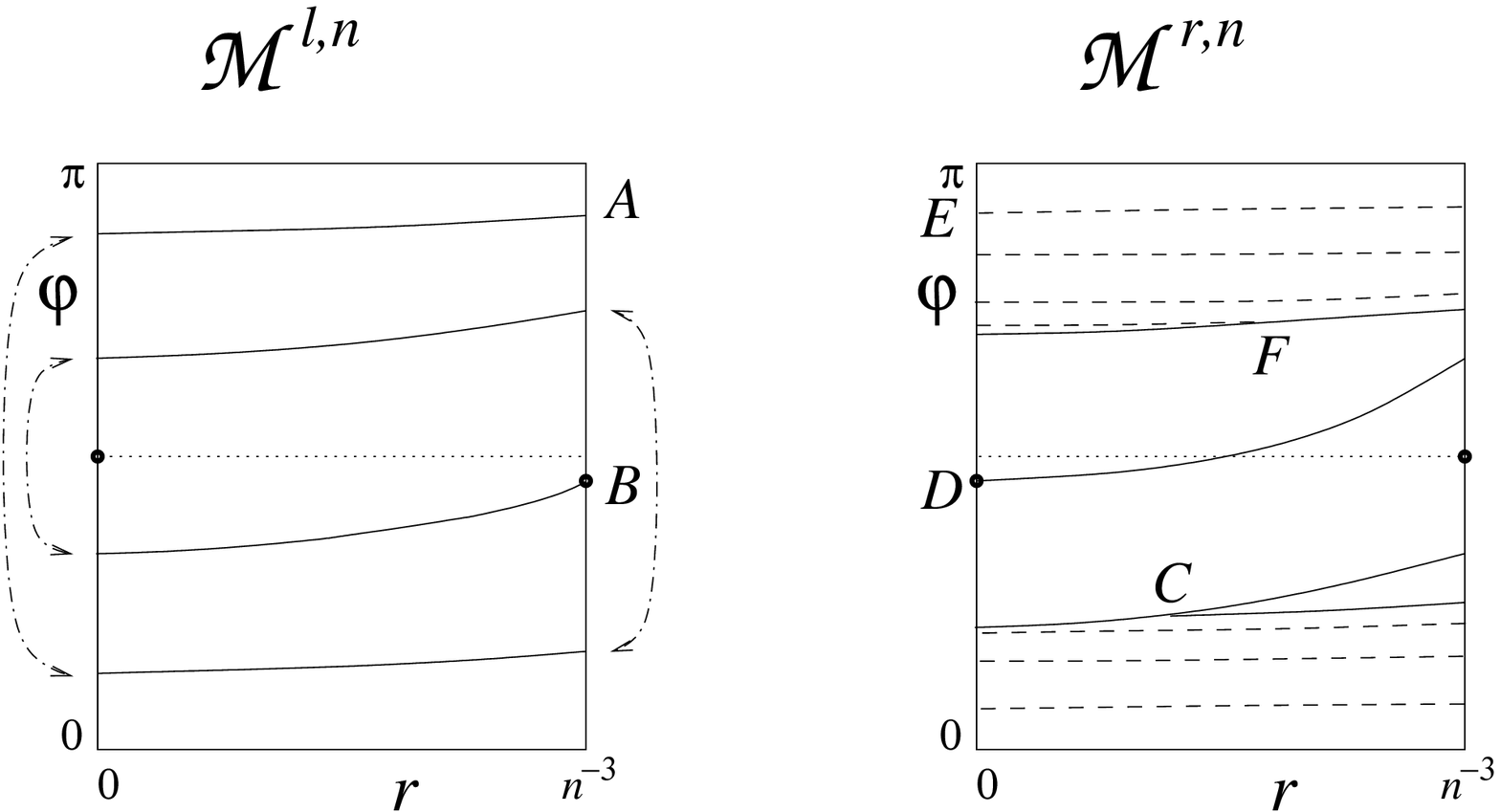} {5in} {The singularity lines $\sir^{l,n} \in \ps^{l,n}$ and
$\sir^{r,n} \in \ps^{r,n}$. The bullets represent the fixed points
for the identifications in the vertical segments of $\partial
\ps^{i,n}$, $i=l,r$.  These identifications are made explicit, for
some points, in the left picture. Notice that the distance from $B$
(or $D$) to the line $\ph = \pi/2$ is $\beta_n$.}

\begin{remark}
	For a point $\z \in \ps_1 \setminus \sir$ whose \bi\ \tr y 
	hits $\uw$ $k$ times before crossing the next transparent wall
	(say $G_m$), $D (\ma_1)_\z$ equals the product of the 
	differentials $D\ma_{\z_i}$ corresponding to the rebounds on 
	$\uw$ (at suitable $\z_i$, $i = 0, ... , k-1$), times 
	$-D\ma_{\z_k}$, where $D\ma_{\z_k}$ would correspond to a 
	\emph{rebound} on $G_m$---as opposed to a crossing. This is 
	so because the transparent wall can be regarded as a ``double 
	bouncer'', i.e., causing two instantaneuous collisions, one 
	from and one towards $G_m$. The transfer matrix between them 
	is of the form (\ref{dmap}) with $\ph_1 = \pi - \ph$, and 
	$k = k_1 = \tau = 0$; that is, minus the identity.
	\label{rk20}
\end{remark}

It is not hard to figure out that $\sir^{i,n}$ is a collection of
curves, each being the graph of an increasing \fn\ of $\l$.  Denote by
$\beta_n := \arctan (| \f'(X_n) |)$ the angle between the horizontal
direction and the tangent line to $\uw$ in $V_n$. In Fig.~\ref{figm},
in the left picture, $\sir^{l,n}$ can be regarded as a continuous
curve, once we identify $(0,\ph) \leftrightarrow (0,\pi - \ph)$ and
$(n^{-3},\ph) \leftrightarrow (n^{-3},\pi - 2\beta_n - \ph)$; which
are the proper identifications for line elements based in $(X_n,0)$
and $V_n$, respectively. This continuous curve, from point $A$ to
point $B$, emcompasses all initial conditions in $\ps^{l,n}$ that end
up in $V_{n-1}$, or hit $\uw$ tangencially.

As concerns $\ps^{r,n}$, the suitable identifications are $(0,\ph)
\leftrightarrow (0,\pi - 2\beta_n - \ph)$ and $(n^{-3},\ph)
\leftrightarrow (n^{-3},\pi - \ph)$. There are two possible sources of
singularity in this case: Initial conditions that end up in $V_{n+1}$
or hit $\uw$ tangentially (they correspond to the solid curve running
from $C$ to $D$); and initial conditions that move off to the right,
come back and hit $V_n$ (dashed curve from $E$ to $F$). One can
recognize that the self-intersection in $C$ corresponds to the \bi\ 
\tr y of Fig.~\ref{fign}. Furthermore, $F$ corresponds to a \tr y that
hits $V_{n+1}$ almost vertically and then continues its motion to the
left to hit $V_n$.

\skippar

What was discussed above should convince one that the number of lines in
$\sir^{i,n}$ is related to the maximum number of rebounds against $\uw$,
for points in $G_n$. Let us call this latter integer $M_n$.  For the
rest of this section, we use the symbols $\ll, \gg, \sim$ for the
asymptotics $n \to +\infty$.

\begin{lemma}
	Assuming \emph{(A5)}, $M_n \ll n^{\xi_0}$, with $\xi_0 :=
	6\xi - 1$.  
	\label{lemma-mn}
\end{lemma}

\proof For a material point starting in $G_n$, the maximum number of
$\uw$-rebounds before crossing one of the walls $G_{n-1}, G_n$, or
$G_{n+1}$, is evidently given by the \tr y shown in Fig.~\ref{fign}.

\fig{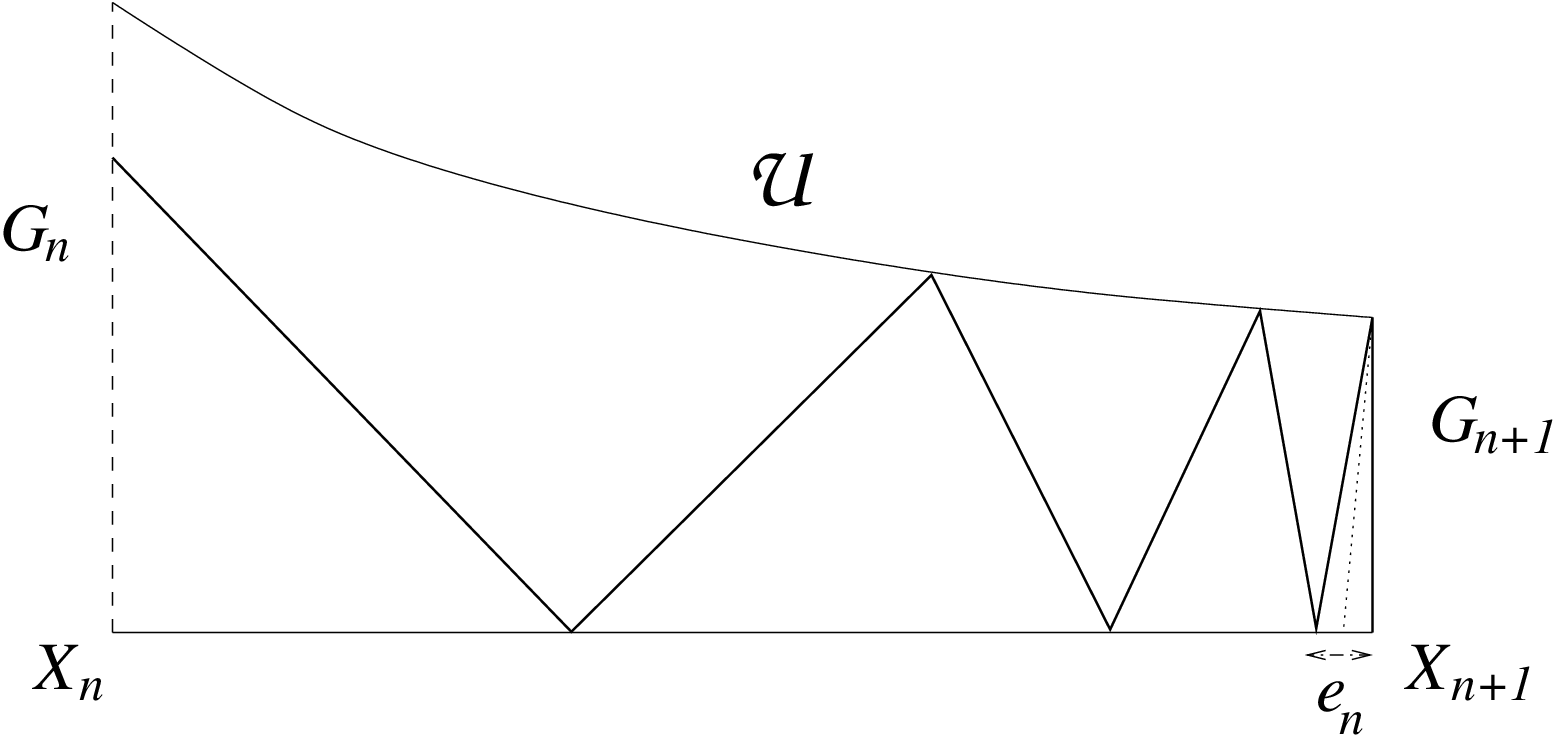} {3.5in} {Among the \tr ies that start in $G_n$, this one
achieves the maximum number of rebounds against $\uw$, before
intersecting another transparent wall. The depicted polyline is run over
twice, its velocity being reversed at the point $(X_{n+1},0)$.}

Define $e_n$ as in the picture, that is, 
\begin{equation}
	e_n := \f(X_{n+1}) \tan( 2 \beta_{n+1} ) \sim \f(X_{n+1}) \,
	|\f'(X_{n+1})| = (n+1)^{-3} \, |\f'(X_{n+1})|.  
	\label{mn-10}
\end{equation}
One sees that
\begin{equation}
	M_n \ll \frac{X_{n+1} - X_n} {e_n},
	\label{mn-20}
\end{equation}
the implicit constant being at most two, or so (the \tr y is run over
twice). Now, $\f'(X_{n+1}) = \f'( \f^{-1} ((n+1)^{-3}) ) = 1/
(\f^{-1})' ((n+1)^{-3})$. For the sake of the notation, let us name
$g := \f^{-1}$ and $t_n := n^{-3}$. Then (\ref{mn-10}) becomes
\begin{equation}
	e_n \sim \frac{ t_{n+1} } { |g'(t_{n+1})| }.
	\label{mn-30}
\end{equation}
Also,
\begin{equation}
	X_{n+1} - X_n = g(t_{n+1}) - g(t_n) \le |g'(t_{n+1})| \, 
	(t_n - t_{n+1}),
	\label{mn-40}
\end{equation}
having used the monotonicity of $\f'$ (hence $g'$). Now, (\ref{mn-30})
and (\ref{mn-40}) turn (\ref{mn-20}) into
\begin{equation}
	M_n \ll |g'(t_{n+1})|^2 \, \frac{ t_n - t_{n+1} } { t_{n+1} } 
	\sim |g'(t_{n+1})|^2 \, \frac1n.
	\label{mn-50}
\end{equation}
But from (A5)
\begin{equation}
	|g'(t_{n+1})| = \frac1 { | \f'( \f^{-1} (t_{n+1}) ) | } \ll 
	\frac1 { [ \f ( \f^{-1} (t_{n+1}) ) ]^\xi } \sim t_n^{-\xi} =
	n^{3\xi}, 
	\label{mn-60}
\end{equation}
which, together with (\ref{mn-50}), finishes the proof of the lemma.
\qed

Coming back to the number of singularity lines in $\ps^{i,n}$, we can
now be more precise and see that $\sir^{l,n}$ has about $M_{n-1}/2$
lines, whereas $\sir^{r,n}$ has $M_n + M_{n+1}/2$ (in Fig.~\ref{figm},
the dashed lines plus the solid lines).

Furthermore, we are in the position to show that the stable and
unstable manifolds of $\ma$ in $\ps$ can be carried over to $\ps_1$.
We do this for the LUMs only, the other case being of course just the
same.  Let us consider $\ma_2$, the Poincar\'e map corresponding to
the cross-section $\ps_2 := \ps \sqcup \ps_1$. For $\z \in \ps_1$, set
$k$ to be the smallest positive integer for which $\ma_2^{-k} \in
\ps$; then define $\tilde{W}^u (\z) := \ma_2^k \, \wu(\ma_2^{-k} \z)$.
Although $\tilde{W}^u (\z)$ is contracting in the past, in general it
will not be a $C^1$ curve, due to the singularities $\sir$. We must
therefore ``prune'' it in such a way that all conditions of Definition
\ref{def-lsum} remain valid.

To this end, notice that, by Lemma \ref{lemma-tecn2} of the Appendix
$\mu \left( \sir_{[\eps]} \right)$ goes to zero like a power in $\eps$
(the notation $\sir_{[\eps]}$ is also introduced there).  Furthermore,
setting $q(\l,\ph) = 1/\sin\ph$ as in Section \ref{sec-lsum}, it is
evident that $\int_{\ps_1} q\, d\mu < \infty$. By the standard
arguments that we are familiar with, by now, these two facts imply
respectively that the backward images of a.e.~$\z \in \ps_1$, via
$\ma_1$, approach $\sir$ only polynomially, and the deformation
constant between the Riemannian distance and the $\| \,\cdot\,
\|$-distance also grows polynomially. Since $\ma_1^{-k} \, \tilde{W}^u
(\z)$ is contracting exponentially w.r.t.\ $\| \,\cdot\, \|$ (by Lemma
\ref{lemma-s-arg}, because $\mu(\ps_1) < \infty$), it follows that it
can fail to be smooth only for a finite number of $k$'s. Hence it can
be pruned in such a way that Definition \ref{def-lsum}, \emph{(b)},
holds, without the risk of reducing its length to zero. (The reader
that finds this argument too sketchy can check that Theorem
\ref{thm-ex2} applies to $(\ps_2, \ma_2, \mu)$.)

\begin{remark}
	At this point, it might be worthy to discuss why we needed to
	introduce the new Poincar\'e section $\ps_1$. Its main asset
	is simply that $\mu(\ps_1) < \infty$, which guarantees
	exponential contraction. One might reply that we had
	exponential contraction already, w.r.t.\ the return times to
	$\ps_0$. However, using that fuzzy cross-section, it is not
	clear how many $\ma$-iterations ($\uw$-rebounds) can occur
	between two returns to $\ps_0$; whereas $\ps_1$ was
	specifically designed to ensure that the number of
	$\ma$-iterations between two returns grows at most
	polynomially, as we go back in time (we will check this
	below). So, why did we not use the simpler dynamical system
	$(\ps_1, \ma_1, \mu)$ from the beginning, and avoid the
	cumbersome machinery of the previous sections?  The answer is
	that for the proof of the tail bound (see Lemma
	\ref{lemma-tail}), we need that the $\eps$-neighborhoods of
	the singularity set have \me\ of order $\eps$ or better:
	$\si_{[\eps]}$ satisfies this; for $\sir_{[\eps]}$ that is not
	clear.  
	\label{rk30}
\end{remark}

In the rest of the section we will prove that the stable and unstable
foliations verify the forthcoming definition. The term \emph{foliation}
is used in a rather sloppy way here. What we mean, of course, is the type
of object we have been dealing with, so far: a collection of short
leaves for a.e.\ point. \emph{Measurable foliation} would be a more
precise name, but we will not delve into these questions.

\begin{definition}
	A $\nu$-dimensional foliation $\mathcal{W}$ in $\R^N$ is said to
	be \emph{absolutely continuous} with respect to the \me\ $\mu$
	if the following happens: Given any cylinder $C$ endowed with an
	axis $\Theta$ (a cylinder is a set that, in some orthogonal
	frame, looks like $A \times \R^{N-\nu}$, with $A$ a Borel set in
	$\R^\nu$; and an axis is any $\{ x \} \times \R^{N-\nu}$, with $x
	\in A$), and any union $L$ of leaves $W(\z) \in \mathcal{W}$,
	transversal to $C$ (i.e., to all axes of $C$) and exceeding it
	(i.e., $\partial W(\z) \in \R^N \setminus C$), then we have
	\begin{displaymath}
		\mu(L \cap C) = 0 \qquad \Longrightarrow
		\qquad \mathrm{Leb}_{\Theta} (L \cap \Theta) = 0;
	\end{displaymath}
	where $\mathrm{Leb}_{\Theta}$ is the $(N - \nu)$-dimensional 
	Lebesgue \me\ on $\Theta$. 
	\label{def-ac}
\end{definition}

More precisely, the above states that the transversal \me\ on
$\mathcal{W}$ defined by $\mathrm{Leb}$ on $\Theta$ is absolutely
continuous with respect to $\mu$.

\skippar

In our case, the situation can be simplified. First of all, we can use
the 2-dimensional Lebesgue \me\ instead of $\mu$, since they are
\emph{equivalent} (absolutely continuous w.r.t.\ each other). Second,
consider the foliation $\mathcal{V}$ given by parallel straight lines
strictly contained in the constant cone field $\co^{-}$ (this is no
loss of generality). With a slight abuse of notation, we denote by
$\len$ not just the length given by the ordinary (Riemannian) distance
$d$, but also the 1-dimensional Lebesgue \me\ on any smooth
curve---e.g., on $\Theta \in \mathcal{V}$.  Also, it is obvious that
$\len$ on a straight line orthogonal to $\mathcal{V}$ is the
appropriate transverse \me\ for $\mathcal{V}$.

Now pick a $\Theta \in \mathcal{V}$ and a segment $I \subset \Theta$
such that, for $\len$-almost every $\z \in I$, $\wu(\z)$ exists (this
can be achieved for $\len$-almost every $\Theta$, by Fubini's
Theorem). Call $\wu_r(\z)$ the right part (say) of $\wu(\z)$, w.r.t.\ 
$\Theta$, and $L$ the union of all the $\wu_r(\z)$ based in $I$. If we
can show that $\exists d_0 > 0$ such that, for $\len$-a.a.\ $\Theta'
\in \mathcal{V}$ to the right of $\Theta$, with $d(\Theta', \Theta)
\le d_0$, $\len (\Theta' \cap L) > 0$, then we have proved that the
unstable foliation verifies Definition \ref{def-ac}. (Simply by using
Fubini and integrating $\Theta$ over $\mathcal{V}$.)  Needless to say,
the case of the stable foliation is completely analogous.

\begin{theorem}
	Assuming \emph{(A1)-(A5)}, the stable and unstable foliations
	in $\ps_1$ are absolutely continuous w.r.t.\ $\mu$.
	\label{thm-ac}
\end{theorem}

\proof Except for its final part (where we use Lemma \ref{lemma-mn} and
the \emph{ad hoc} construction of $\ps_1$) this proof is very
standard. We present it completely, however, because it is hard to
derive it as a rigorous corollary of any of the theorems available in
the literature. Archetypal results include \cite[\S 4]{g}, \cite[Part
II]{ks}, and \cite[\S 4]{ps}.

Set $A \subset \ps_1$ to be a full-\me\ set of points satisfying some
properties which will be unveiled during the course of the proof. To
begin with, every $\z \in A$ has a LUM.

Using the notation introduced above, take a $\Theta \in \mathcal{V}$
and a $I \in \Theta$ such that, for $\len$-a.e.\ $\z \in I$, $\z \in
A$. Fix any such $\z$ and name it $\z_0$. Denoting by $L_\eps$ the
union of all $\wu_r(\z)$, with $\z \in I$ and $\len (\wu_r(\z)) \ge
\eps$, then $\exists \eps_0 >0$ such that $\z_0 \in I \cap L_{\eps_0}$
and $\len( I \cap L_{\eps_0}) > 0$. Fix $d_0 := const \, \eps_0$; for
a suitable choice of $const$ we are assured that any $\wu_r(\z)$
longer than $\eps_0$ intersects any $\Theta'$ to the right of $\Theta$,
with $d(\Theta', \Theta) \le d_0$ (remember that, by the construction
of $\mathcal{V}$, the angle between the LUMs and the direction of
$\mathcal{V}$ is bounded from below). Thus, fixed such a $\Theta'$, we
define the \emph{holonomy map} $h$ on $I \cap L_\eps$ so that $h(\z)$
equals the unique point (by transversality) in $\wu_r(\z) \cap
\Theta'$. To simplify the notation, set $\w_0 := h(\z_0)$. See
Fig.~\ref{figo}.

\fig{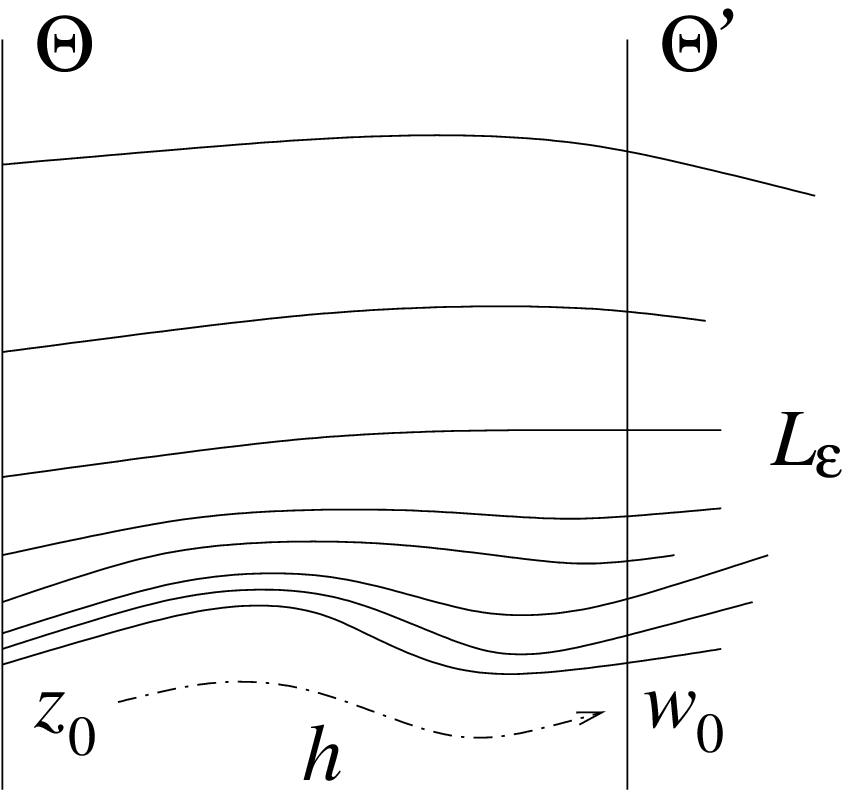} {2.5in} {The construction of the holonomy map $h$.}

Without loss of generality (i.e., possibly modifying $A$ by a null-\me\
set), $\z_0$ is a density point of $I \cap L_\eps$, via the Lebesgue
Density Theorem. Thus $\z_0$ is also an accumulation point of $I \cap
L_\eps$, and it makes sense to speak of the Jacobian of the map $h$
there, that is,
\begin{equation}
	Jh_{\z_0} := \lim_{\z \to \z_0} \frac{ |h(\z) - h(\z_0)| } 
	{ |\z - \z_0| }.
	\label{ac-10}
\end{equation}
The proof of Theorem \ref{thm-ac} consists precisely in showing that
the above limit exists and is positive.

The usual trick one employs is to pull back $h$ to the holonomy map $h_n$
between $\ma_1^{-n} \Theta$ and $\ma_1^{-n} \Theta'$, as sketched in
Fig.~\ref{figp}. Set $\z_{-n} := \ma_1^{-n} \z_0$ and $\w_{-n} :=
\ma_1^{-n} \w_0$. Then, by definition (\ref{ac-10}), $\forall n \in \N$,
\begin{equation}
	Jh_{\z_0} = J (\ma_1^{-n})_{\z_0} \, J(h_n)_{\z_{-n}} \, J
	(\ma_1^{n})_{\w_{-n}} = J(h_n)_{\z_{-n}} \, \prod_{k=0}^{n-1} 
	\frac{ J(\ma_1^{-1})_{\z_{-k}} } { J(\ma_1^{-1})_{\w_{-k}} }.  
	\label{ac-20}
\end{equation}
Introducing $u_{-k}$, a unit tangent vector to $\ma_1^{-k} \Theta$ at
$\z_{-k}$, and $v_{-k}$, a unit tangent vector to $\ma_1^{-k} \Theta'$ at
$\w_{-k}$, one sees that the quantities
\begin{equation}
	J(\ma_1^{-1})_{\z_{-k}} = \left| D(\ma_1^{-1})_{\z_{-k}} u_{-k}
	\right|, \qquad J(\ma_1^{-1})_{\w_{-k}} = \left|
	D(\ma_1^{-1})_{\z_{-k}} v_{-k} \right|
	\label{ac-30}
\end{equation}
are actual derivatives, since, by definition of $A$, $\wu(\z_0)$
(which contains $\z_0$ and $\w_0$) belongs in $\ps_1 \setminus
\sir^{-}_\infty$. Therefore, $\forall k$, a certain neighborhood of
$\wu(\z_{-k})$ is contained in a connected component of $\ps_1
\setminus \sir^{-}$, where $\ma_1^{-1}$ is smooth.

At this point, one would like to show that the product in (\ref{ac-20})
converges to a finite number, and $\lim_{n \to +\infty} J(h_n)_{\z_{-n}}
= 1$, since $\ma_1^{-n} \Theta$ and $\ma_1^{-n} \Theta'$ get closer and
closer to each other as $n$ grows. The argument, however, is a bit more
complicated. Indeed we will prove those two facts not for the Jacobians,
but for suitable approximations; precisely, quantities like the
following:
\begin{equation}
	R(h_n)(\z_{-n}, \z_{-n}') := \frac{ \len( \mathrm{arc} 
	(\w_{-n},\w_{-n}') ) } { \len( \mathrm{arc} (\z_{-n}, 
	\z_{-n}') ) },
	\label{ac-40}
\end{equation}
where $\w_{-n}' := h_n (\z_{-n}')$ and $\mathrm{arc}(\z,\z')$ denotes
the arc segment an the appropriate backward image of $\Theta$ or
$\Theta'$ (in this case, $\ma_1^{-n} \Theta$ and $\ma_1^{-n}
\Theta'$).

Recall the definition of ``goodness'' (\ref{ex-65}), used in the proof
of Theorem \ref{thm-ex}, and keep in mind that, for $(\ps_1, \ma_1,
\mu)$, we can use the ordinary distance, as opposed to the unstable
distance---because $\mu \left( \sir_{ \{\eps\}} \right) \ll
\eps^\alpha$, for some $\alpha>0$. Now, fix a $\z_0' \in I \cap
L_\eps$, which we think of as close to $\z_0$, and consider the
curvilinear quadrilateral $\qu$ specified by $\z_0$, $\w_0$, $\w_0' :=
h(\z_0')$, as illustrated in Fig.~\ref{figp}.  If $\lambda$ denotes
the same constant as in (\ref{ex-60}), set $\lambda_1 < \lambda$ and
$C>0$ such that $B(\z_{-n}, \, C \, e^{-\lambda_1 n} )$ is good
$\forall n \ge 0$, and $\wu_r(\z_0)$ is strictly contained in $B(\z_0,
C)$.  (For this we might have to cut $\wu_r(\z_0)$, but this is no
loss of generality: one will simply take smaller $\eps_0$ and $d_0$.)
Furthermore, let $n$ be the minimum integer such that $\qu_n :=
\ma_1^{-n} \qu$ is not contained in $B( \z_{-n}, \, C \, e^{-\lambda_1
  n} )$. (This is possible since $\mu(\qu_n)$ is constant in $n$,
while $\mu \left( B( \z_{-n}, \, C \, e^{-\lambda_1 n} ) \right)$
vanishes---although maybe not monotonically, since $\mu \ne
\mathrm{Leb}$.)

\fig{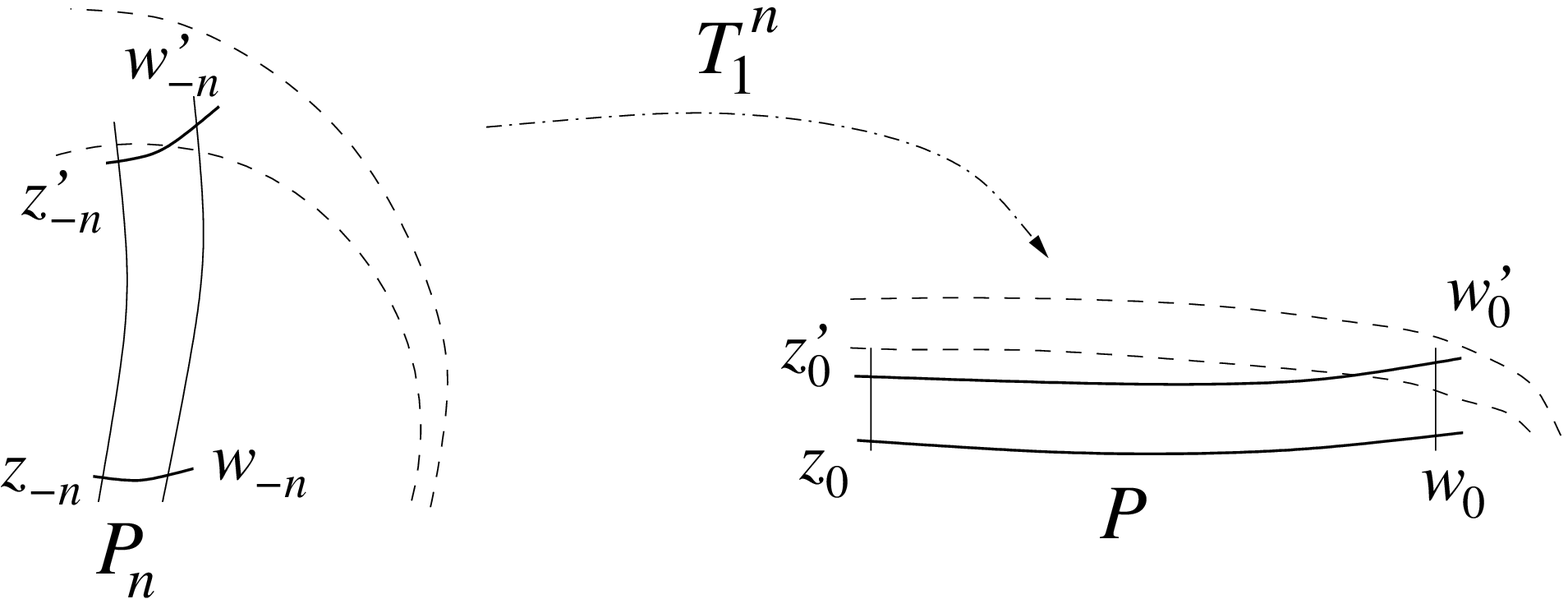} {4.5in} {The stretching of the curvilinear quadrilateral
$\qu$ backwards in time. In the left picture, the inner dashed curve
represents (part of) $\partial B( \z_{-n}, \, C_2 \, e^{-\lambda_1
n} )$, and the outer dashed curve represents $\ma_1^{-1} \partial
B( \z_{-n+1}, \, C_2 \, e^{-\lambda_1 (n-1)} )$. The dashed curves
n the right picture are the $\ma_1^n$-iterates of these curves.}

By construction, all points in $\qu_n$ are good, relative to $\z_0$.
Therefore, for all $\z \in I \cap L_\eps$, $\wu_r(\z)$ contracts at a
rate faster than $e^{-\lambda' n}$, for any $\lambda' \in (\lambda_1,
\lambda)$, see (\ref{ex-150}).  But the diameter of $\qu_n$ is larger
than $C \, e^{-\lambda_1 n}$. These two facts imply that, no matter
how deformed $\qu_n$ becomes, it will look more and more stretched
along the stable direction, as $n \to +\infty$. Moreover, its two long
opposite sides will have approximately the same length, as
Fig.~\ref{figp} tries to show. In fact not only is the distance
between $\z_{-n}$ and $\z_{-n}'$ about the same as the distance
between $\w_{-n}$ and $\w_{-n}'$, but also the two joining curves will
be rather ``parallel'', its tangent lines having to belong to the cone
field $\co_n$, which must be narrow for $n$ big---at least for almost
every $\z_0$.

This demonstrates that, as $\z_0' \to \z_0$ (hence $n \to +\infty$),
$R(h_n)(\z_{-n}, \z_{-n}') \to 1$.  We will have proved Theorem
\ref{thm-ac} when we are able to show that
\begin{equation}
	\lim_{n \to +\infty} \, \prod_{k=0}^{n-1} \frac{ R(\ma_1^{-1})
	(\z_{-k}, \z_{-k}') } { R(\ma_1^{-1}) (\w_{-k}, \w_{-k}') } 
	> 0.
	\label{ac-60}
\end{equation}
By the Lagrange Mean Value Theorem, $R(\ma_1^{-1}) (\z_{-k}, \z_{-k}')
= J(\ma_1^{-1})_{\bar{\z}_{-k}}$, for some $\bar{\z}_{-k} \in
\mathrm{arc} (\z_{-k}, \z_{-k}')$, and analogously for $R(\ma_1^{-1})
(\w_{-k}, \w_{-k}')$. Then, a sufficient condition for (\ref{ac-60})
is
\begin{equation}
	\lim_{n \to +\infty} \, \sum_{k=0}^{n-1} \left| \log
	J(\ma_1^{-1})_{\bar{\z}_{-k}} - \log
	J(\ma_1^{-1})_{\bar{\w}_{-k}} \right| < \infty.  
	\label{ac-70}
\end{equation}
Note that we do not use the notation $\sum_{k=0}^\infty$ because the
terms of the sum depend on $n$ too, through $\bar{\z}_{-k}$ and
$\bar{\w}_{-k}$.  We can apply the Mean Value Theorem to each term
above, to obtain
\begin{equation}
	| \bar{\w}_{-k} - \bar{\z}_{-k} | \, \left| \frac{\partial} 
	{\partial b_k} \log J(\ma_1^{-1})_{\tilde{\z}_{-k}} \right|,
	\label{ac-80}
\end{equation}
with $\tilde{\z}_{-k}$ lying on the segment between $\bar{\z}_{-k}$
and $\bar{\w}_{-k}$, and $b_k$ being a unit vector in the direction of
such segment.

Here is where the specific design of $\ps_1$ comes into play. For
every point $\z \in \ps_1$, denote by $M(\z)$ the number of
$\uw$-rebounds (i.e., $\ma$-iterations, or, to be more precise,
$\ma_2$-iterations) before the point first returns to $\ps_1$.  By
Lemma \ref{lemma-mn}, there exists a $C_1>0$ such that
\begin{equation}
	\rset{\z \in \ps_1} {M(\z) \ge C_1 \, k^{\xi_0}} \subseteq
	\bigsqcup_{n=k}^\infty \left( \ps^{r,n} \sqcup \ps^{l,n} 
	\right), 
	\label{ac-90}
\end{equation}
whose \me\ is of order $k^{-2}$. The associated series converges,
therefore we can apply the usual Borel-Cantelli argument to conclude
that, for all $\z$ in a full-\me\ subset of $\ps_1$ (which contains
$A$, without loss of generality), there exists a $C_2 = C_2(\z)$ for
which $M(\z_{-k}) = M(\ma_1^{-k} \z) \le C_2 \, k^{\xi_0}$.  

Now let us notice that for any $\w \in \qu_k$ (in particular for
$\tilde{\z}_{-k}$), the number of $\uw$-rebounds must be the same as
for $\z_{-k}$. In fact, if $M(\w) \ne M(\z_{-k})$ then $\w$ and
$\z_{-k}$ are separated by an $\sir$-singularity line, which is not
possible by the construction of $\qu_k$. Therefore $M( \tilde{\z}_{-k}
) \le C_2 \, k^{\xi_0}$. Decomposing $D (\ma_1^{-1})_{\tilde{z}_{-k}}$
into the product of $M(\tilde{z}_{-k})$ differentials of $\ma_2^{-1}$,
turns (\ref{ac-80}) into
\begin{equation}
	| \bar{\w}_{-k} - \bar{\z}_{-k} | \, \left|
	\sum_{i=0}^{M(\tilde{z}_{-k})-1} \frac{\partial} 
	{\partial b_k} \log J(\ma_2^{-1})_{\tilde{\z}_{-k,i}} \right| , 
	\label{ac-100}
\end{equation}
with $\tilde{\z}_{-k,i} = \ma_2^{-i} \tilde{\z}_{-k}$, for $i = 0,
... , M( \tilde{\z}_{-k} ) - 1$. The strategy is now rather clear:
Since $| \bar{\w}_{-k} - \bar{\z}_{-k} | \ll e^{-\lambda' k}$ and the
number of terms in the sum is a power-law in $k$, it suffices to check
that each such term is bounded by a power of $k$.

First of all, setting $b_{k,i} := D (\ma_2^{-i})_{\tilde{\z}_{-k}} b_k
/ | D (\ma_2^{-i})_{\tilde{\z}_{-k}} b_k |$, one observes that $|
\partial / \partial b_k \, \cdot \, | \le |\partial / \partial b_{k,i}
\, \cdot \, |$, by contraction. Hence, we are reduced to consider
directional derivatives of $\log J(\ma_2^{-1})$. By Remark \ref{rk20}
we can consider $D (\ma_2^{-1})$ as the differential of a regular \bi\
map. Therefore we can apply Lemma \ref{lemma-distor} of the Appendix
to $\tilde{\z}_{-k,i} =: (\tilde{\l}_{-k,i}, \tilde{\ph}_{-k,i})$ and
find a constant $C_3$ such that
\begin{equation}
	\left| \frac{\partial} {\partial b_{k,i}} \log
	J(\ma_2^{-1})_{\tilde{\z}_{-k,i}} \right| \le \frac{C_3} {
	\sin^2 \tilde{\ph}_{-k,i} \: \sin^4 \tilde{\ph}_{-k,i+1} },
	\label{ac-110}
\end{equation}
with the understanding that, when $i = M( \tilde{\z}_{-k} ) - 1$,
$\tilde{\ph}_{-k,i+1}$ really means the second coordinate of
$\ma_2^{-1} \tilde{\z}_{-k,i}$.  But this point belongs to $\qu_{k+1}$
and one can see that the forthcoming arguments are not invalidated.

\begin{remark}
	There are two issues to clarify in order to use Lemma
	\ref{lemma-distor}.  First, the lemma applies to \bi s with
	finite horizon. This problem is easily circumvented. Say, for
	instance, that we have a $\ma_2$-iteration that corresponds to
	a segment of \tr y going from $G_n$ to $G_{n-1}$ (this is the
	longest free path, for points in $\ps^{l,n}$). We can always
	divide it into $M_n$ segments of approximately the same
	length, by imagining as many transparent walls between
	$G_{n-1}$ and $G_n$. It is easy to see that this length is
	less than a quantity that depends only on the shape of the
	\bi.

	The second issue is to control $(\partial u_{-k,i} / \partial
	b_{k,i}) (\tilde{\z}_{-k,i})$ uniformly in $k$ and $i$. We
	only spend a few words on this, which is a standard argument
	in Pesin's theory. The fact is that the $\tilde{\z}_{-k,i}$'s
	belong to a sequence of good sets with respect to the same
	$\z_0$---more precisely, neighborhoods of $\{ \z_{-k,i}
	\}$. Within this sequence one enjoys some local form of
	uniform \hyp ity. Therefore, to the (backward) images of the
	(stable) direction field $\mathcal{V}$ one applies the line of
	reasoning of \cite[Thm.\ 6.2.8, Step 5]{kh}: the iterates of
	$\Theta, \Theta' \in \mathcal{V}$ (or rather, appropriately 
	short pieces thereof) approach uniformly the stable direction
	\emph{faster} than they get close to each other. See also
	\cite[3.10-3.14]{ps}.
\end{remark}

So, it remains to show that $q( \tilde{\z}_{-k,i} ) = 1 / \sin
\tilde{\ph}_{-k,i}$ grows like a power of $k$. This property is easily
checked for $q( \z_{-k,i} )$. Indeed, for the subsequence of $\{
\z_{-k,i} \}$ corresponding to the returns to $\ps$, we have already
proved it in Section \ref{sec-lsum}---see in particular Theorem
\ref{thm-ex2}, \emph{(f)} and \emph{(h)}. As regards the returns to
$\ps_1$, one uses the same arguments, given that also $\int_{\ps_1} q
\, d\mu < \infty$.

At this point, we cannot argue that the same property must hold for
$\tilde{\z}_{-k,i}$, being this point sufficiently close to
$\z_{-k,i}$ as to be good throughout the backward \o\ of $\z_0$.  In
other words, we cannot use Theorem \ref{thm-ex2}, \emph{(g)}, because
we are now dealing with $\ma_2$-iterations and, although the
Riemannian distance between $\tilde{\z}_{-k,i}$ and $\z_{-k,i}$
vanishes exponentially, the unstable distance might not.  However, one
can reason as follows: Since $\tilde{\z}_{-k,i} \in \qu_{k,i} :=
\ma_2^{-i} \qu_k$, whose sides shrink exponentially in $k$, then, at
least for large $k$, it must belong to the curvilinear triangle
$F_{k,i}$ defined like this: take the two sides of $\qu_{k,i}$ that
intersect in $\z_{-k,i}$ and prolong them arbitrarily, in the stable
and unstable direction, respectively, until their lengths are $d^s(
\z_{-k,i}, \partial \ps) / 2$ and $\du( \z_{-k,i}, \partial \ps) / 2$.
(These latter quantities are bigger than a negative power of $k$.)
Finally, connect the two resulting vertices by a segment. Since
$q(\z)$ is a \fn\ of one variable only, its maximum on $F_{k,i}$ is
always achieved by a point on the curvilinear sides. Then, a double
application of Theorem \ref{thm-ex2}, \emph{(g)}, for both unstable
and stable balls, shows that the value of $q$ along the two
curvilinear sides is comparable to $q( \z_{-k,i} )$. But this has the
right rate of growth, as we have recalled.  
\qed

We conclude this section by noting that, due to the a.e.\ smoothness of
$\ma_2$, the \ac\ that we have established for $\ps_1$ is immediately
proved for $\ps$ as well.

\sect{Local ergodicity theorem}
\label{sec-loc-erg}

The local \erg ity theorem we present in this section is a
generalization of the one formulated by Liverani and Wojtkowski in
\cite{lw}. We have chosen it because it utilizes invariant cones.
Although our goal is to extend it to a class of infinite-\me\ \sy s
that includes (some of) our \bi s, much of the proof is of local
nature, and is unaffected by the finiteness, or lack thereof, of the
invariant \me.  For this reason we will not give that part of the
proof, referring the reader to \cite[\S\S 8-12]{lw}. The only lemma
that needs modification is the so-called \emph{tail bound} \cite[\S
13]{lw}, which we restate and prove at the end of this section.

Theorem \ref{thm-loc-erg}, like all similar results, requires several
technical conditions. In order to state them we need a definition.

\begin{definition}
	A compact subset $A$ of $\R^N$ is called \emph{regular} if 
	it is a finite union of pieces $\Sigma_i$ of codimension-one 
	submanifolds, such that:
	\begin{itemize}
		\item[(a)] $\Sigma_i$ is the closure of its interior 
		(in the topology of the submanifold);

		\item[(b)] The pieces overlap at most on their
		boundaries, i.e., $\Sigma_i \cap \Sigma_j \subset 
		\partial \Sigma_i \cap \partial \Sigma_j$;

		\item[(c)] $\partial \Sigma_i$ is a finite union of 
		compact subsets of codimension-two submanifolds.
	\end{itemize}
	$A$ is called \emph{locally regular} if, $\forall \z \in
	\R^N$, there is a neighborhood $U$ of $\z$ such that $A \cap
	U$ is regular.
\end{definition}

For the sake of the format, we list the conditions before giving the
statement of the theorem. It is the goal of Section \ref{sec-erg-ta}
to provide a class of \bi s that verify (C1)-(C8).

\begin{itemize}
	\item[(C1)] \textsc{Phase space.} The phase space $\ps$ is an
	open, connected subset of $\R^{2\nu}$, with $\partial \ps$
	regular. $\R^{2\nu}$ is endowed with a symplectic form which
	is assumed to be equivalent to the standard one; i.e., the 
	symplectic volume element $d\mu$ is assumed to be absolutely 
	continuous w.r.t.\ the standard volume element
	$d\mathrm{Leb}$, and viceversa.

	\item[(C2)] \textsc{Map.}  The map $\ma$ is invertible and
	recurrent.  $\ma$ is not defined on the singularity set
	$\si^{+}$; and $\ma^{-1}$ is not defined on $\si^{-}$.
	(Morally $\si^{\pm} = \ma^{\mp 1} \partial \ps$, in the sense
	that one can construct ill-behaved extensions of $\ma$ and
	$\ma^{-1}$ for which that holds.) On $\ps \setminus \si^{+}$,
	$\ma$ preserves the symplectic form mentioned in (C1).
\end{itemize}

\begin{remark}
	Some of the restrictions formulated above are not really
	essential.  First, the choice of a linear space was only made
	to simplify the estimation of the \me\ of the tubular
	neighborhoods of certain regular sets (see below). A
	$2\nu$-dimensional symplectic manifold, embedded in $\R^N$,
	connected, and with bounded geometry, would have worked as
	well.  Second, requiring the dynamical \sy\ to be Hamiltonian
	(that is, a symplectomorphism where defined) has the effect
	that certain properties are symmetric for time reversal (e.g.,
	(C3) or \cite[\S 8]{lw}). Assuming those properties to hold
	for both directions of time would yield the same result. See
	also Remark \ref{rk5}.
\end{remark}

\begin{itemize}
	\item[(C3)] \textsc{Cone bundle.} There exists a cone bundle
	$\co^{+} = \co$, continuous where the map is defined, and
	\emph{eventually strictly invariant} for the map. This means
	that, for almost all $\z$, there is an $n = n(\z)$ such that
	$D\ma_\z^n \co(\z) \subset \Int \co( \ma^n \z)$.  (C1)
	guarantees that the same happens for $\co^{-}$, the cone
	bundle comprised by the closures of the cones $\R^{2\nu}
	\setminus \co^{+}$.
\end{itemize}

\begin{itemize}
	\item[(C4)] \textsc{Local regularity of singularity sets.} 
	Denoting, as we have done in the past, $\si_n^{\pm} :=
	\bigcup_{i=0}^{n-1} \ma^{\mp 1} \si^{\pm}$, we suppose that
	$\si_n^{+}$ and $\si_n^{-}$ are locally regular for all
	$n$.
      
	\item[(C5)] \textsc{Measure of tubular neighborhoods.} On
	$\si^{-}$ there is a finite \me\ $\pi_{-}$, such that, for 
	every closed subset $A$ of $\si^{-}$ (in the topology of 
	$\si^{-}$),
	\begin{displaymath}
		\mu( A_{[\eps]} ) \le \pi_{-}(A) \, \eps,
	\end{displaymath}
	for small $\eps$. Here, as introduced in (\ref{a-eps}),
	$A_{[\eps]}$ denotes the tubular neighborhood of $A$, of
	radius $\eps$, in the unstable distance. $\pi_{-}$ must be
	absolutely continuous w.r.t.\ $\mu_{\si^{-}}$, the \me\ 
	induced on $\si^{-}$ by $\mu$ and the ordinary distance.
	An analogous condition stands for $\si^{+}$ and the stable 
	distance.

	\item[(C6)] \textsc{Proper alignment of singularity sets.} For
	a codimension-one subspace of a linear symplectic space, the
	characteristic line is defined as the skew-orthogonal
	complement. (In our case, in two dimensions, the subspace and
	the characteristic line are the same thing.)  We assume that
	the tangent space to $\si^{+}$ at any point $\z \in \si^{+}$
	has a characteristic line contained in $\co^{-}(\z)$. The
	reversed condition holds for $\si^{-}$.

	\item[(C7)] \textsc{Non-contraction property around the
	singularity sets.} For every $\z_0 \in \ps$, there exist a
	neighborhood $U_0$ of $\z_0$, an $\eps_0 > 0$, and a $K > 0$, 
	such that, every time $\w \in \si_{[\eps_0]}^{-}$ and $\ma^n 
	\w \in U_0$, with $n > 0$, then
	\begin{displaymath} 
		|D\ma_\w^n v| \ge K |v|,
	\end{displaymath} 
	for $v \in \co_1(\w)$. Here $| \,\cdot\, |$ is the modulus of
	a vector in $\R^{2\nu}$ (i.e., the appropriate Riemannian
	norm). Also, the definition of $\co_1(\w)$ is the same as
	in (\ref{n-cone}). The analogous condition stands for the
	time-reversed dynamical \sy.
\end{itemize}

At this point, there is one last condition to formulate, the so-called
\emph{Sinai--Chernov Ansatz}. This can be given in two versions: a
very general one which, however, can be cumbersome to check in
some instances; and a more specific one for \sy s (like
semi-dispersing \bi s) that possess an increasing norm for unstable
vectors. For the sake of completeness we give both, although in the
remainder we will only work with the second version, more natural in
our framework.

For the first case we need to use the notion of \emph{expansion
coefficient} $\sigma_\co(L)$ for a linear symplectic map $L$ w.r.t.\ 
an invariant cone $\co$. We skip its rather lengthy definition which,
however, can be found in \cite[\S\S 4-6]{lw}. (After all, we never use
this object in the present work.)

\begin{itemize}
	\item[(C8)$_1$] \textsc{Sinai--Chernov Ansatz for the 
	expansion coefficient.} For $\pi_{-}$-a.e.\ $\z \in 
	\si^{-}$,
	\begin{displaymath}
		\lim_{n \to \infty} \sigma_\co ( D\ma_\z^n ) =
		+ \infty.
	\end{displaymath}
	Once again, the analogous condition stands for the
	time-reversed dynamical \sy.
\end{itemize}

Before presenting the alternative version of the Sinai--Chernov
Ansatz, we need to better specify what we need by increasing norm,
mainly to simplify definitions and proofs.

\begin{definition}
	The dynamical \sy\ defined above is said to have an
	\emph{increasing norm} for unstable vectors, $\| \,\cdot\,
	\|$, if this norm satisfies Theorem \ref{thm-ex}, (c) and (d),
	the latter with $H = \emptyset$, and is locally equivalent to
	the Riemannian norm $| \,\cdot\, |$.
	\label{def-inc-norm}
\end{definition}

\begin{itemize}
	\item[(C8)$_2$] \textsc{Sinai--Chernov Ansatz for the 
	increasing norm.} For $\pi_{-}$-a.e.\ $\z \in 
	\si^{-}$,
	\begin{displaymath}
		\lim_{n \to +\infty} \: \inf_{0 \ne v \in \co_1(\z)}
		\frac{ \|D\ma_\z^n v\|} {\|v\|} = + \infty.
	\end{displaymath}
	Of course, the analogous condition holds for $\si^{+}$ and 
	$\ma^{-1}$.
\end{itemize}

We are now in position to state the local \erg ity theorem.

\begin{theorem}
	Consider a dynamical \sy\ $(\ps, \ma, \mu)$ endowed with a
	\hyp\ structure (LSUMs at a.e.\ point, absolutely continuous
	w.r.t.\ $\mu$). Suppose, furthermore, that this \sy\ satisfies
	\emph{(C1)-(C8)$_i$} ($i=1$ or $2$). Then, for any $\z_0$ that
	possesses a semi\o\ (i.e., $\z_0 \in \ps \setminus
	\si_\infty^{+}$, or $\z_0 \in \ps \setminus \si_\infty^{-}$),
	there is a neighborhood $U$ of $\z_0$ belonging to one \erg\
	component of $\ma$.  \label{thm-loc-erg}
\end{theorem}

\begin{remark}
	 The statement can be strengthened to include points that only
	 possess a finite \o. Using the expansion coefficient,
	 Liverani and Wojtkowski \cite[\S 7]{w}, prove it for $\z$
	 such that $\sigma ( D\ma_\z^n ) > 3$, for some $n$, positive
	 or negative.
\end{remark}

For the purpose of stating and proving the tail bound lemma that,
together with the results in \cite[\S\S 8-12]{lw}, will grant Theorem
\ref{thm-loc-erg}, we need to recall just one fact from the arguments
that we omit.  Since it is crucial that the local stable and unstable
manifolds be as large as possible, one decides to prolong $\wsu(\z)$,
as defined in Section \ref{sec-lsum}, as much as it is compatible with
the requirements of Definition \ref{def-lsum} (see \cite[Thm.\
9.7]{lw}).  This implies that the boundary of the new, say, unstable
manifolds are made up of points of $\partial \ps$ or $\ma^i \si^{-}$,
for some $i\ge 0$ (no smoothness is possible across the
singularity!). For simplicity, we keep denoting any such ``grown'' LUM
with the same symbol $\wu(\z)$, and say that it is \emph{cut} by
$\partial \ps$ or $\ma^i \si^{-}$. Furthermore, looking back at the
proof of Theorem \ref{thm-ex}, we define the \emph{radius} of
$\wu(\z)$ to be the inf of $\len(\gamma)$ over all smooth curves
$\gamma \subset \wu(\z)$ that join $\z$ with $\partial \wu(\z)$.

\begin{lemma}
	For every $\z_0 \in \ps$, there is a neighborhood $U$ of 
	$\z_0$, and a $\delta_0 > 0$ such that, $\forall \eta > 0$, 
	$\exists M$ that verifies
	\begin{displaymath}
		\mu \left( \rset{\z \in U} {\wu(\z) \mbox{ \emph{has
		radius}} < \delta \mbox{ \emph{because it is cut by}} 
		\bigcup_{i=M+1}^\infty \ma^i \si^{-} } \right) \le
		\eta \delta,
	\end{displaymath}
	for every $\delta \le \delta_0$.
	\label{lemma-tail}
\end{lemma}

\begin{remark}
	As anticipated, we prove this lemma only under condition
	(C8)$_2$. Using (C8)$_1$ would only amount to minor changes in
	the proof which, anyway, can be reconstructed with the aid of
	\cite[\S 13]{lw}.
\end{remark}

\proofof{Lemma \ref{lemma-tail}} For the sake of the notation, we will
henceforth drop the super/subscripts from $\si^{-}$ and $\pi_{-}$.
For a linear map $L: \ts\ps_\z \longrightarrow \ts\ps_\w$ leaving the
cone bundle $\co$ inviariant, let us denote
\begin{equation}
	\sigma_{*} (L) := \inf_{0 \ne v \in \co(\z)} \frac{ \|Lv\| }
	{|v|}.
	\label{tail-10}
\end{equation}
Since the increasing norm and the ordinary norm are locally
equivalent, it follows from (C8)$_2$ that $\lim_{n \to \infty}
\sigma_{*} ( D\ma_\z^n ) = + \infty$, for $\pi$-a.e.\ $\z \in \si$.

Fix $h>0$, to be thought of as a small parameter. Since $\pi$ is
finite, one can choose $E_1 \subset \si$ such that $\pi( E_1 ) \le h$
and $\si \setminus E_1$ is bounded.  From (C4), for any compact subset
$B$ of $\ps$, $\si \cap B$ is also compact. Hence, without loss of
generality, we can also assume that $\si \setminus E_1$ is compact and
$\pi( \overline{E_1} ) \le h$, where the bar denotes, here and in the
sequel, the closure in $\si$.  Now, for every (large) parameter $s>0$,
there is an $M = M(h,s)$ such that
\begin{equation}
	E_2 := \rset{\z \in \si} { \sigma_{*} ( D\ma_\z^M ) \le s+1 }
	\label{tail-20}
\end{equation}
has \me\ $\pi (E_2) \le h$. Notice that $E_2$ is closed. As the reader
has apprehended, we are throwing away points of $\si$ that somehow might
give us complications. For reasons that will be clear later, we want
the remaining set to be compact. Eliminating the points in $E_2$ might
have destroyed this property, which we would like to recover now,
through another purge. The map $\ma^M$ is discontinuous at $\si_M^{+}$
which, by proper alignment, intersects $\si = \si^{-}$ only in pieces
of codimension-two manifolds. Therefore the discontinuity set of
$\left. \ma^M \right|_\si$ is a set of $\mu_\si$-\me (hence $\pi$-\me)
zero. Let us remove a neighborhood $E_3$ of this set (in $\si$), such
that $\pi( \overline{E_3} ) \le h$.

The above choice guarantees the compactness of 
\begin{equation}
	\si^s := \si \setminus \bigcup_{j=1}^3 E_j = \rset{\z \in \si
	\setminus (E_1 \cup E_3) } { \sigma_{*} ( D\ma_\z^M ) \le s+1 }
	\label{tail-30}
\end{equation}
because of the continuity of $\sigma_{*} ( D\ma_\z^M )$ in the above
domain. Moreover, $\sigma_{*} ( D\ma_\z^M )$ is also continuous in a
certain neighborhood of $\si^s$ in $\ps$. This neighborhood
necessarily contains $\si_{[c]}^s$, for $c$ small enough.

At this point, take $U_0$ and $\eps_0$ as in (C7).  Since the
increasing norm and the Riemannian norm are locally equivalent,
$\exists C_1>0$ such that, $\forall \w \in U_0$,
\begin{equation}
	\| \,\cdot\, \|_\w \le C_1\, | \,\cdot\, |_\w.
	\label{tail-40}
\end{equation}
Now take $U$, a smaller neighborhood of $\z_0$, such that $U_{[C_1
\delta_0]} \subseteq U_0$. (Should this condition lead to trouble---in
the sense that $\bu(\z_0, C_1 \delta_0)$ already exceeds $U_0$---we
can always take a smaller $\delta_0$ with no damage for the proof.)

The main idea behind the tail bound is to split $Y(\delta,M)$, the set
that appears in the statement of the lemma, into pieces whose \me\ is
easy to estimate. We proceed as follows. For $\z \in Y(\delta,M)$,
denote $m(\z)$ the smallest $i \ge M+1$ such that $\wu(\z)$ is cut by
$\ma^i \si$. Also set
\begin{equation}
	k(\z) := \# \rset{i = 1, ... , m(\z)-M} {\ma^{-i}\z \in U}.
	\label{tail-50}
\end{equation}
Why the returns to $U$ is this stretch of \o\ are important, we will
see later.  Let us introduce $Y_m^k := \rset{\z \in Y(\delta,M)}
{m(\z)=m,\, k(\z)=k}$. We claim that
\begin{equation}
	\ma^{-m} Y_m^k \cap \ma^{-m'} Y_{m'}^k = \emptyset, 
	\mbox{ for } m \ne m'. 
	\label{tail-55}
\end{equation}
In fact, suppose not and assume, say, that $m < m'$. If $\w \in
\ma^{-m} Y_m^k \cap \ma^{-m'} Y_{m'}^k$ then, for $\z := \ma^m \w$ and
$\z' := \ma^{m'} \w$, we would have $k(\z') > k(z)$, which is absurd.
From the claim and the invariance of $\mu$ we obtain that, for a fixed
$k \in \N$,
\begin{equation}
	\mu \left( \bigcup_{m>M} Y_m^k \right) \le \sum_{m>M} \mu
	\left( Y_m^k \right) = \sum_{m>M} \mu \left( \ma^{-m} Y_m^k
	\right) = \mu \left( \bigcup_{m>M} \ma^{-m} Y_m^k \right).
	\label{tail-60}
\end{equation}

By Definition \ref{def-inc-norm} (noting in particular that $H =
\emptyset$), we see that there exist a $\rho < 1$ such that, $\forall
\w \in U_0$ and $v \in \co(\w)$,
\begin{equation}
	\| D\ma_\w^{-n} v \|_{\ma^{-n} \w} \le \rho^j \| v \|_\w, 
	\label{tail-70}
\end{equation}
if $j$ is the number of returns to $U_0$ of the piece of \o\ $\left\{
\ma^{-i} \w \right\}_{i=1}^n$. The above holds \emph{uniformly} in
$\w \in U_0$. In fact, assuming that $U_0$ stays away from $\partial
\ps$, the amount of contraction (for unstable vectors and relative to
the increasing norm) at every return is bounded below by $\inf_{\w \in
U_0} \kappa(\w)^{-1} =: \rho$ (see Theorem \ref{thm-ex}, \emph{(c)}).

Fix $\z \in Y_m^k$ ($k\in\N$, $m\ge M+1$); there is by definition a
smooth curve $\gamma$ that connects $\z$ to a point $\z' \in \partial
\wu(\z) \cap \ma^m \si$, and such that $\len(\gamma) < \delta$. We
observe that
\begin{equation}
	\len_\| (\ma^{-n} \gamma) < C_1 \rho^k \delta,
	\label{tail-80}
\end{equation}
$n$ being the largest $i\le m-M$ such that $\ma^{-i} \z \in U$.  In
fact, by the definition of $U$, if $\ma^{-i} \z$ returns $k$ times to
$U$, $\ma^{-i} \gamma$ returns $k$ times to $U_0$, and one can apply
(\ref{tail-40}) and (\ref{tail-70}) to the points of $\gamma$.  Now,
$\ma^{-m} \z' \in \si$ and we have two cases.

\skippar

\noindent
\textsc{Case 1:} $\ma^{-m} \z' \in \bigcup_{j=1}^3 E_j$. Here we first
use (\ref{tail-80}), then switch to the Riemannian length, and finally
employ (C7) to estimate the maximum (Riemannian) expansion during the
time $i=n+1, ... ,m$. The net result is
\begin{equation}
	\len (\ma^{-m} \gamma) < \frac{C_2}K \, \rho^k \, \delta,
	\label{tail-90}
\end{equation}
for some $C_2>0$. The above is an \emph{a priori} estimate that can be
proved correct (by contradiction, for instance) once we choose
$\delta_0 \le \eps_0 / C_2$, so that $C_2 \rho^k \delta_0 \le \eps_0$,
$\forall k \le 0$. Therefore, setting $C_3 := C_2/K$,
\begin{equation}
	\ma^{-m} \z \in \ma^{-m} \gamma \subset \left( \bigcup_{j=1}^3
	\overline{E_j} \right)_{[C_3 \rho^k \delta]}, 
	\label{tail-100}
\end{equation}
whose \me, by (C5) and the estimates above, does not exceed $3 h C_3
\rho^k \delta$.

\skippar

\noindent
\textsc{Case 2:} $\ma^{-m} \z' \in \si^s$. In this case it is not hard
to see that
\begin{equation}
	\len (\ma^{-m} \gamma) < \frac{C_1}s \, \rho^k \, \delta.
	\label{tail-110}
\end{equation}
Once again, this is an \emph{a priori} estimate and everything works
rigorously provided $C_1 \delta_0 \le c$, where $c$ was
introduced before. Therefore, in analogy with (\ref{tail-100}),
\begin{equation}
	\ma^{-m} \z \in \si^s_{[C_1 s^{-1} \rho^k \delta]}. 
	\label{tail-120}
\end{equation}
Using again (C5) and estimating $\pi(\si^s)$ from above by $\pi(\si)$,
we conclude that the \me\ of the above set is less than or equal to
$\pi(\si) C_1 s^{-1} \rho^k \delta$.

\skippar

The previous two estimates, together with
(\ref{tail-55})-(\ref{tail-60}), yield
\begin{equation}
	\mu \left( \bigcup_{m>M} \ma^{-m} Y_m^k \right) \le \left[ 3\,
	h \, C_3 + \frac{ \pi(\si) \, C_1} s \right] \rho^k \, \delta,
	\label{tail-130}
\end{equation}
whence
\begin{equation}
	\mu \left( Y(\delta,M) \right) \le \left[ 3\, h \, C_3 +
	\frac{ \pi(\si) \, C_1} s \right] \frac \delta {1 - \rho}.
	\label{tail-140}
\end{equation}
For a given $\eta>0$, the coefficent of $\delta$ above can be made
smaller than $\eta$, if $h$ and $s$ are chosen suitably small and
large, respectively. These, in turn, determine how big $M$ must be for
the statement of Lemma \ref{lemma-tail} to hold true.  
\qed

The most valuable consequence of Theorem \ref{thm-loc-erg} is, of
course, the \emph{global} \erg ity of some of our \sy s.

\begin{proposition}
	If the \bi\ map $\ma$ introduced in Section \ref{sec-prelim}
	is locally \erg, in the sense of Theorem \ref{thm-loc-erg},
	then it is \erg.
	\label{prop-erg}
\end{proposition}

\proof We do this in two steps. First, we show that only a countable
number of points in $\ps = (0,+\infty) \times (0,\pi)$ can fail to
verify Theorem \ref{thm-loc-erg}, that is, to be in the interior of an
\erg\ component. Second, we observe that removing these points leaves
$\ps$ connected, nevertheless. Ergo, there is only one \erg\
component.

As concerns the first assertion, we evaluate the cardinality of
$\si_\infty^{+} \cap \si_\infty^{-}$: the sets $\si_\infty^{+}$ and
$\si_\infty^{-}$ are countable unions of smooth curves, respectively
increasing and decreasing (in fact, $\si_\infty^{-}$ is even a mirror
image of $\si_\infty^{+}$).  Therefore there can be at most one point
of intersection for each pair of increasing--decreasing curves. This
means, at most countably many points.

The proof of the second assertion is the contents of Lemma
\ref{lemma-conn} in the Appendix.  
\qed

\sect{Ergodic tables}
\label{sec-erg-ta}

In view of Proposition \ref{prop-erg}, we devote the last part of this
work to checking that the \fn s
$$
	\f(x) = C x^{-p}, \quad C,p>0,
	\eqno{\mathrm{(E1)}}
$$
give rise to \sy s that verify Theorem \ref{thm-loc-erg}---thus
yielding examples of \erg\ \bi s with a non-compact cusp. Moreover,
some of these tables have an infinite area, which is especially nice,
for the reasons we have outlined in the introduction. 

To start with, we will fix $C:=1$, since that constant never plays a
role in our computations. Conditions (C1) through (C6) from Section
\ref{sec-loc-erg} are rather easy to establish. (For (C4) one might
notice that $\si^\pm$ is composed of three smooth curves, two of which
are unbounded. Thus, $\si_n^\pm$ will comprise a finite number of
smooth curves, some of which failing to be compact only because of
their unboundedness. Hence, \emph{local} regularity is guaranteed.)

As concerns (C8)$_2$, we derive from (\ref{ex-a3}) and
(\ref{expan-dl})-(\ref{expan-dph}) that
\begin{equation}
	\inf_{0 \ne v \in \co(\z)} \frac{ \|D\ma_\z v\|} {\|v\|} 
	\ge \min \left\{ \left( 1 + \frac{k \tau} {\sin\ph} \right), 
	\left( 1 + \frac{k_1 \tau} {\sin\ph_1} \right) \right\},
	\label{erg-5}
\end{equation}
with the usual notation $\z = (\l, \ph)$, etc. It is implicitly
written in (A4) that $\f'' > 0$, so the curvature of $\uw$ is always
positive; and continuous, by the assumptions on the differentiability
of $\f$. Therefore $k \tau$ and $k_1 \tau$ are bounded below when the
point is sufficiently far away from $V$ and from the cusp at
infinity. But Proposition \ref{prop-eo} ensures that, for \emph{every}
\o, this happens infinitely many times.

\skippar

It remains to verify the non-contraction property (C7). This is quite
often the hardest property to check for a \bi\ \sy\ (e.g., \cite[\S
14]{lw}). As a matter of fact, our formulation of (C7) is so
cumbersome precisely because we are only able to prove non-contraction
in its weakest useful form. Only in this part of the paper do we need
more properties of $\f$ than just (A1)-(A5).

\fig{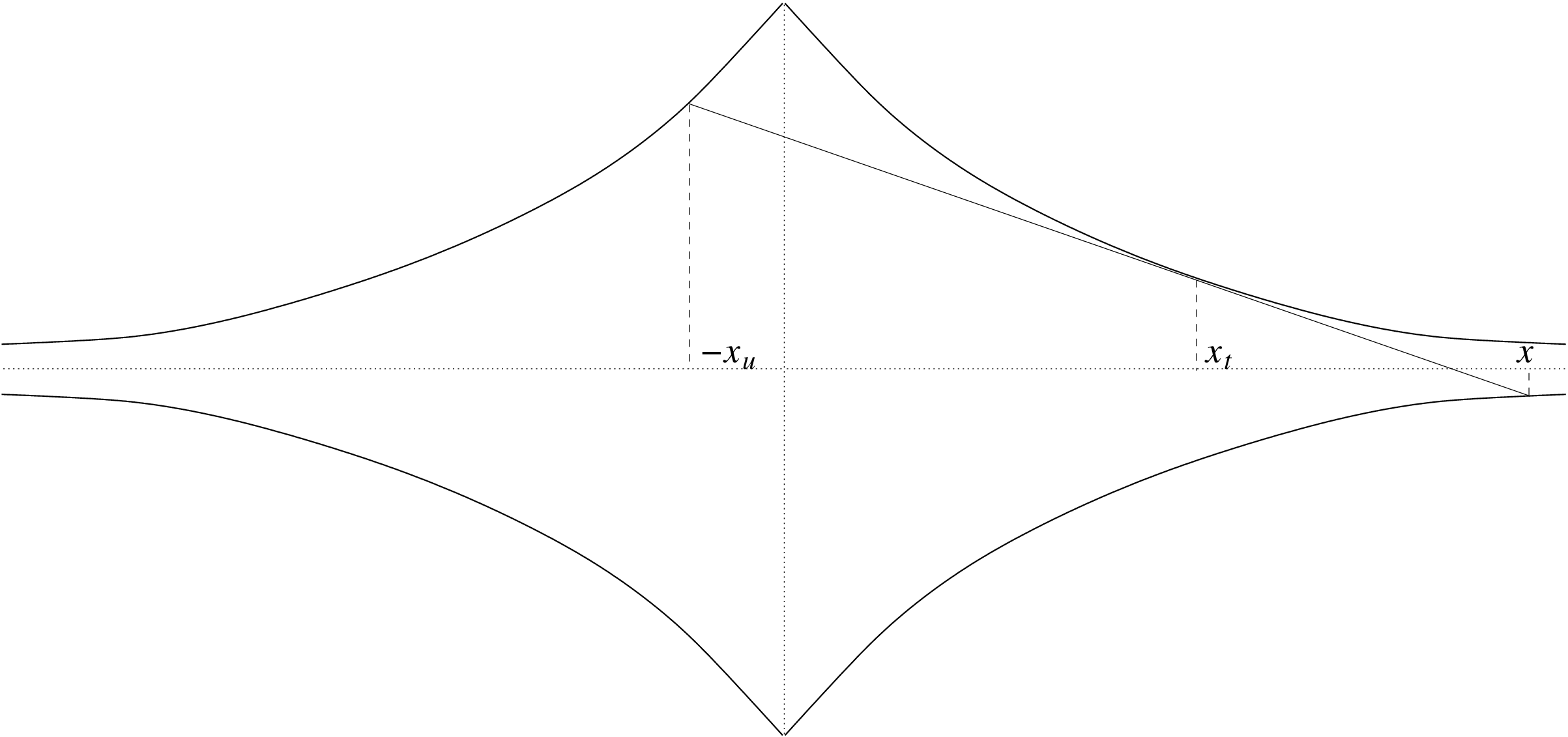} {5in} {The definition of $x_u$.}

Let us start by illustrating a feature of $\f$ that is very similar to
(A2). Looking at Fig.~\ref{fige} and recalling the definition of
$x_t = x_t(x)$, consider the tangent line to $\partial
\ta_4$ at $(x_t, \f(x_t))$. Then denote by $-x_u = -x_u(x) <0$ the
abscissa of the point at which this line intersects $\partial \ta_4$ in
the second quadrant. We verify that
\begin{equation}
	|\f'(x_u)| \ll |\f'(x)|.
	\label{a7}
\end{equation}
In fact, in analogy with (\ref{cond-xt}), $x_u$ is uniquely determined
by
\begin{equation}
	\frac{\f(x) + \f(x_u)} {x + x_u} = -\f'(x_t).
	\label{cond-xu}
\end{equation}
Using the fact that, for our $\f$, $x_t/x = const$ (cf.\ Section
\ref{sec-prelim}), we can write an equation similar to
(\ref{verif-a2-1}) and conclude that $x_u/x = const$, too; whence
(\ref{a7}).

At this point, let us introduce $\l_T$, a large number to be
determined later.  Denoting $\ps_T := \rset{(\l, \ph) \in \ps} {\l <
\l_T} = (0,\l_t) \times (0,\pi)$, we single out the line elements
relative to $\ta_T$, a certain truncated \bi\ whose four-fold copy,
$\ta_{4,T}$, appears in Fig.~\ref{figq}. Given the shape of $\si^{-}$,
it is obvious that there is an $\eps_0$ so small that
$\si_{[\eps_0]}^{-} \cap \ps_T$ stays away from the line $\ph = 0$
(see Fig.~\ref{figs}). Therefore, for $\w \in \si_{[\eps_0]}^{-} \cap
\ps_T$ and $v$ an unstable vector based in $\w$, one has
\begin{equation}
	|D\ma_\w^n v| \ge \min \left\{ 1 , \frac{\sin\ph} {\sin\ph_n}
	\right\} |v| \ge C_1 |v|, 
	\label{erg-10}
\end{equation}
for some positive $C_1$ and all $n$. (We have changed notation since
(\ref{erg-5}): here and for the rest of the section $(\l, \ph) := \w$
and $(\l_n, \ph_n) := \w_n := \ma^n \w$.)

\fig{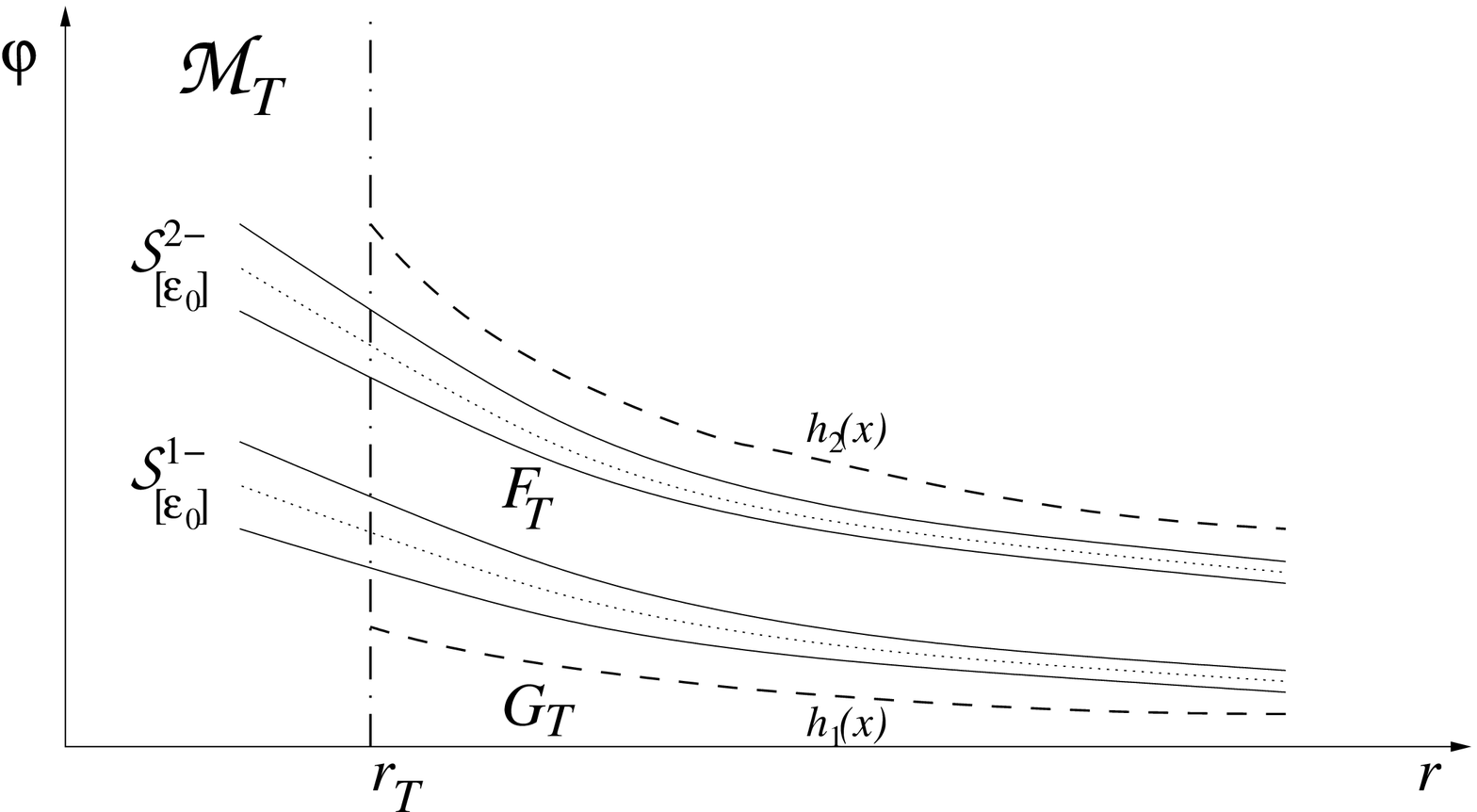} {5in} {The definition of $F_T$ and $G_T$.}

It remains to consider the case $\w \in \si_{[\eps_0]}^{-} \setminus
\ps_T$. We will see later that it pays off to be more general and take
$\w \in F_T$, a set that we introduce now with the aid of
Fig.~\ref{figs}. First of all, $F_T \subset \ps \setminus \ps_T$ and
$F_T \supset \si_{[\eps_0]}^{-} \setminus \ps_T$. The leftmost part of
$\partial F_T$ belongs to the segment $\l=\l_T$. We know from Section
\ref{sec-lsum} (see in particular Fig.~\ref{figh}, definition
(\ref{ex-a5}) and following paragraph) that $\si^{2-}$ and the upper
and lower boundaries of $\si_{[\eps_0]}^{2-}$ are the graphs of three
\fn s that behave asymptotically like $|\f'(x)|$. The same is true for
$\si^{1-}$. This implies that it is possible to take the lower and
upper boundaries of $F_T$ to be the graphs of two functions $h_j$
($j=1,2$) such that
\begin{equation}
	h_j(x) = K_j |\f'(x)|.
	\label{erg-12}
\end{equation}
On $K_j$ we will also impose some extra conditions later on. Finally,
let us call $G_T$ the region to the right of $\l=\l_T$ and below the
graph of $h_1$.

At this point, we assume that $\l_T$ is big enough, so that $\z_0 \in
\ps_T$. Then we take $U_0$ to be strictly contained in $\ps_T$.
Furthermore, without loss of generality, $\l_T$ is so large that every
$\w \in F_T$ points very much torwards the right, when regarded as a
unit vector based somewhere on $\uw$; e.g., we can make sure that the
\o\ of $\w$ makes a minimum number of rebounds to the right before
starting to move left.

Here we are going to use the arguments expounded at the end of Section
\ref{sec-lsum}, about a \tr y moving towards the cusp at infinity and
coming back---see in particular Fig.~\ref{figj} and its caption. Fix a
$\w$ as above and call $m$ the number of rebounds the corresponding
\tr y performs in the part of $\ta_4$ that lies to the right of the
truncated \bi; in other words, $\w_m$ is the last rebound before the
material point either crosses the $y$-axis or hits the dispersing part
of $\partial \ta_{4,T}$ (or both).  Fig.~\ref{figq} shows some
examples of $\w_m$, together with $\w'$, which is the velocity vector
right before the collision at $\l$ (more precisely, $\w'$ is a
translation, along the \bi\ \tr y, of $\w_{-1} := \ma^{-1} \w$).  As
explained in Section \ref{sec-lsum}, $\w_m$ and $-\w'$ can be thought
of as the (oriented) boundary of a dispersing beam of \o s originating
in a point further right into the cusp (more or less the point where
the \tr y starts moving left).  Since the beam is dispersing, its
focus (the intersection between the straight lines defined by $\w'$
and $\w_m$) lies outside $\ta_4$, as Fig.~\ref{figq} illustrates.

\fig{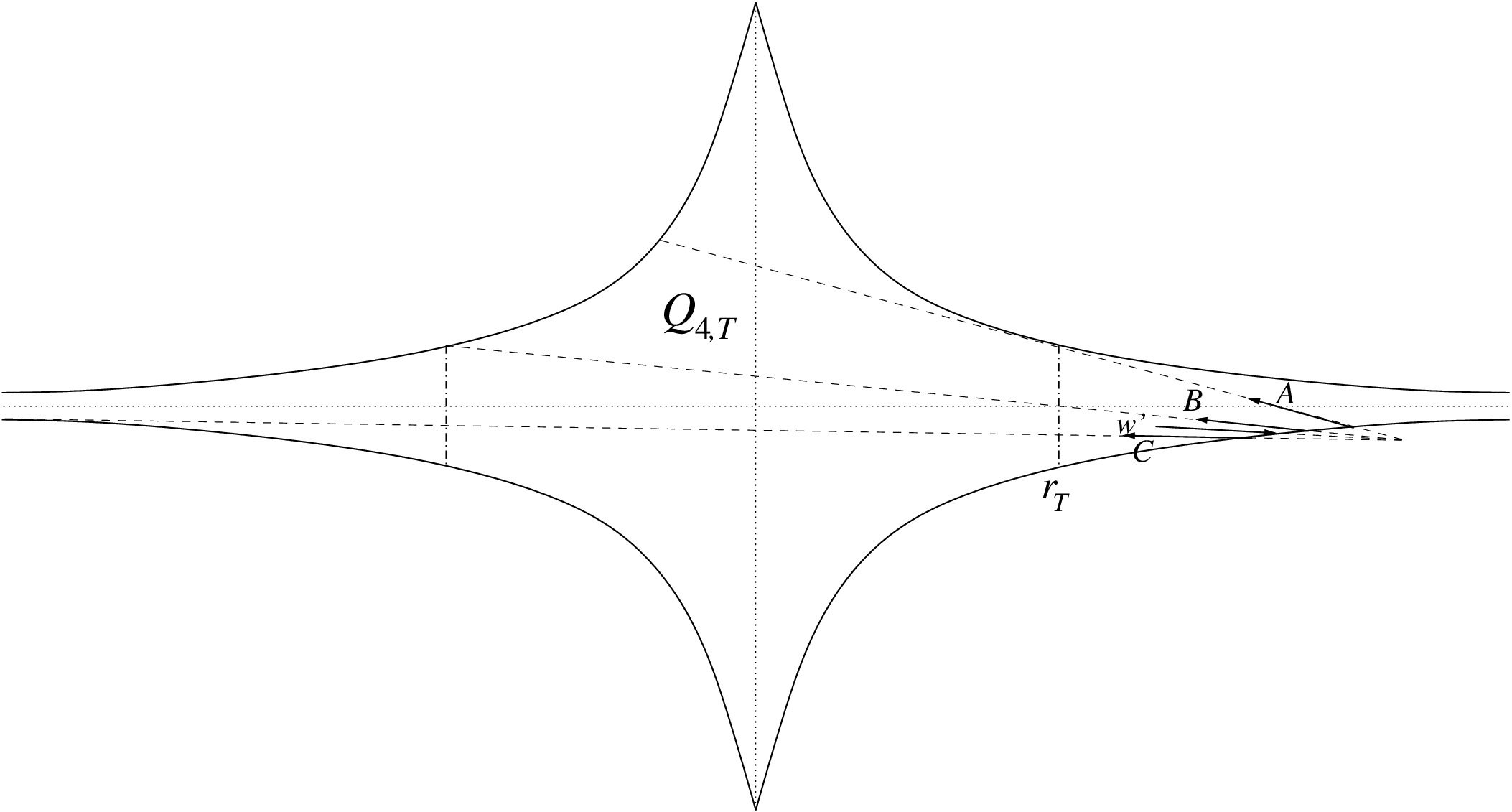} {5.5in} {Given $\w'$, the velocity vector of a \tr y
traveling towards the cusp, and necessarily coming back, we present
several possibilities for $\w_m$, the last rebound in the region to
the right of $\ta_{4,T}$.  These possibilities are chosen from the 
same dispersing beam, i.e., they have a common focus.}

\begin{remark}
	One might point out that in Fig.~\ref{figq} there is no need
	for $\l$ and $\l_m$ to lie on the same piece of $\partial
	\ta_4$, that is, on the same copy of $\uw$.  This is
	absolutely true, and in fact we have assumed nothing of the
	sort: $\w_m$ should be based in the first-quadrant or in the
	fourth-quadrant copy of $\uw$, depending on $k$ being odd or
	even. The choice of Fig.~\ref{figq} is purely illustrative and
	does not affect any of the forthcoming calculations.
\end{remark}

We make a brief digression in order to derive an inequality that will
be crucial in the remainder. Suppose we have a finite portion of an
\o\ (say $\{ \w_i \}_{i=0}^m$) and we want to estimate the amount of
horizontal expansion for unstable vectors. We can do better than
simply use (\ref{expan-dl}) recursively, for $i = 0, ..., m$. In fact,
a repeated application of Lemma \ref{lemma-2-boun} of the Appendix,
together with (\ref{expan-dl}) itself, proves in particular that
\begin{equation}
	(d\l_{m+1})^2 \ge \left( \frac{\sin \ph} {\sin \ph_{m+1}}
	\right)^2 \left( 1 + \frac{k} {\sin\ph} \left( \sum_{i=0}^m
	\tau_i \right) \right)^2 d\l^2.  
	\label{erg-15}
\end{equation}
In other words, what the above is saying is simply that the amount of
horizontal expansion for unstable vectors can only decrease if one
considers all rebounds (apart from the starting point $\w = \w_0$) to
take place against flat pieces of the boundary. And this is, after
all, obvious for semi-dispersing \bi s.

Therefore, considering the finite segment of \o\ $\{ \w_i \}_{i=0}^m$,
if we take $v \in \co_1(\w)$ and set $\bar{\tau} := \sum_{i=0}^m
\tau_i$, we obtain
\begin{equation}
	\frac{|D\ma_\w^{m+1} v|} {|v|} \ge C_2 \left( \frac{\sin\ph}
	{\sin\ph_{m+1}} + \frac{k \bar{\tau}} {\sin\ph_{m+1}}
	\right),  
	\label{erg-20}
\end{equation}
for some $C_2 \in (0,1)$.  In fact, let us observe that, if $v =:
(d\l, d\ph) \in \co_1(\w)$, then $|d\l| \ge C_2 |v|$, because
$\co_1(\w)$ becomes thinner and more horizontal as $\w$ stays in $F$
and moves to the right---cf.\ (\ref{udist-20}).  (That is, one can
actually select $C_2$ arbitrarily close to 1, provided $\l_T$ is big
enough.) The remaining part of estimate (\ref{erg-20}) is just
(\ref{erg-15}).

It is easy to see that there are now only three cases, concerning
$\w_{m+1}$: $\w_{m+1} \in F_T$, $\w_{m+1} \in G_T$, and $\w_{m+1} \in
\ps_T$.

\skippar

\noindent
\textsc{Case 1:} $\w_{m+1} \in F_T$, that is, that the \tr y of $\w_m$
crosses the truncated \bi\ and hits the second or third-quadrant
portions of $\partial \ta_4$, with an incidence angle not so close to
zero. An example is vector $B$ of Fig.~\ref{figq}. We claim that the
r.h.s.\ of (\ref{erg-20}) can be made bigger than 1 if $\l_T$ was
previously selected to be large enough. In formula, 
\begin{equation}
	|D\ma_\w^{m+1} v| > |v|.
	\label{erg-25}
\end{equation}
Since $\w \in F_T$, and due to ({\ref{erg-12}),
\begin{equation}
	\sin\ph \gg |\f'(x)|.
	\label{erg-30}
\end{equation}
To estimate $\sin\ph_{m+1}$ we consider the worst case. This occurs
when the focus of the beam lies very close to $\partial \ta_4$ (that
is, the beam is as dispersing as it can be), and $\w_m$ describes a
segment of \tr y tangent to $\uw$. This case is labeled by $A$ in
Fig.~\ref{figq}. (Actually, vector $A$ of Fig.~\ref{figq} even
overestimates the worst case, since its \tr y does not reach the
region to the left of $\ta_{4,T}$; however, there are in general
segments of \tr y that reach the left region and are tangent to
$\uw$.)  At the limit, when the focus of the beam lies on $\partial
\ta_4$, $\l = \l_m$.  In this case, with the aid of Fig.~\ref{fige},
\begin{equation}
	\sin\ph_{m+1} \sim |\f'(x_t)| + |\f'(x_u)| \ll |\f'(x)|,
	\label{erg-40}
\end{equation}
the last estimate coming from (A2) and (\ref{a7}). Using once again
the correspondence $\l \longleftrightarrow x$ defined by
({\ref{udist-70}), we have $k(\l) \sim \f''(x)$. Therefore, putting
everything together,
\begin{equation}
	\frac{\sin\ph} {\sin\ph_{m+1}} \left( 1 + \frac{k \bar{\tau}}
	{\sin\ph} \right) \ge C_3 + C_4 \, \frac{\f''(x)} {|\f'(x)|}
	\, \bar{\tau}, 
	\label{erg-50}
\end{equation}
where $\bar{\tau} = \bar{\tau}(\w) = \bar{\tau} (x,\ph)$. Since for
the $\f$'s we are considering, $(\f''/\f')(x) = const/x$, the claim in
(\ref{erg-25}) will be settled once we have proved the following
lemma---which we do at the end of the section.

\begin{lemma}
	For $\f$ as in \emph{(E1)} and $\l_T$ fixed,
	\begin{displaymath}
		\lim_{x \to +\infty} \min_{(\l(x),\ph) \in F_T} \frac
		{\bar{\tau}(x,\ph)} {x} = +\infty.
	\end{displaymath}
	\label{lemma-tbar}
\end{lemma}

\noindent
\textsc{Case 2:} $\w_{m+1} \in G_T$. All the estimates that we have
produced in Case 1 hold even more generously in this case, since $\sin
\ph_{m+1}$ is much smaller than the worst case (\ref{erg-40}).
However, for a reason that is going to be clear momentarily, we want
to avoid the case in which the angle of incidence of $\w_{m+1}$, on
the third-quadrant copy of $\uw$ is too close to zero, e.g., vector
$C$ of Fig.~\ref{figq}.

So we look at the next iteration of $\ma$. Since $\w_{m+1} \in G_T$,
and (using (\ref{ph-phprime}), for instance),
\begin{equation}
	\ph_{m+2} = \ph_{m+1} + \arctan |\f'(x_{m+1)}| + \arctan
	|\f'(x_{m+2})|.  
	\label{erg-55}
\end{equation}
it is true that $\w_{m+2} \in F_T$, at least if in (\ref{erg-12})
$K_1$ and $K_2$ were chosen small and large enough,
respectively. Therefore
\begin{equation}
	\sin\ph_{m+2} \ll |\f'(x_{m+1})| <  |\f'(x)|,
	\label{erg-57}
\end{equation}
as one sees that $x > x_{m+1}$. Now we use (\ref{erg-20}) with $m+2$
instead of $m+1$ and, exploiting also the other inequalities presented
above, we conclude that
\begin{equation}
	|D\ma_\w^{m+2} v| > |v|.
	\label{erg-58}
\end{equation}

\skippar

Before moving on to Case 3, let us explain why we had decided to
distinguish Case 2 from Case 1. The reason is because, this way, we
can set $\tilde{\w}$ to be $\w_{m+1}$ or $\w_{m+2}$, depending on Case
1 or 2; then $\tilde{\w} \in F_T$. So one can consider the previous
three cases on $\tilde{\w}$, too. In particular we see that, as long
as the \tr y keeps oscillating between the left and the right cusp and
fails to bounce within $\ta_{4,T}$,
\begin{equation}
	|D\ma_\w^{m_j+1} v| > |v|,
	\label{erg-60}
\end{equation}
where the $m_j$ are the analogues of $m$ or $m+1$, at the future
returns from the ``excursions'' in the cusp.

\skippar

\noindent
\textsc{Case 3:} $\w_{m+1} \in \ps_T$. Estimate (\ref{erg-25}) holds
in this case as well; after all, we have derived it by considering
situation $A$ of Fig.~\ref{figq}, which is the worst possibility even
when $\w_{m+1} \in \ps_T$.

As we will see later, we would like to have also
\begin{equation}
	\sin\ph_{m+1} \ge C_5, 
	\label{erg-70}
\end{equation}
for some $C_5 = C_5(\l_T)$.  This is in general not true, as
Fig.~\ref{figq} shows. We can have $\w_m \not\in \ps_T$ and
$\sin\ph_{m+1}$ arbitrarily close to zero (in fact, the segment of \tr
y originated by $\w_m$ can hit $\partial \ta_4$ indefinitely close to
tangentially). If this is the case, however, we can fix things at the
next rebound, recycling the ideas used before. It is simple to verify
that, chosen $C_5$ small enough, if $\sin\ph_{m+1} > C_5$, then
$\w_{m+1}$ is based in the first-quadrant part of $\partial \ta_{4,T}$
(in fact, hitting the second or third-quadrant boundary would imply
that $\ph_{m+1} > \arctan |\f'(x_T)| =: M$).  Furthermore, the next
rebound will occur in the second quadrant, with an angle of incidence
$\ph_{m+2}$ bigger than, e.g., $M$.

Finally we notice that, if we do decide to consider the next rebound,
then $|D\ma_\w^{m+2} v| > |v|$, as $\sin\ph_{m+2}$ is smaller than
$\sin\ph_{m+1}$ would have been, had $w_m$ been tangent. In either
case, therefore, we are fine.

\skippar

We are ready to verify (C7), at last.  If, for some $n>0$, $\w_n$ is to
belong in $U_0$ (hence in $\ps_T$), there must be a positive integer
$l$ such that $\w_{m_l +1} \in \ps_T$. Excluding the possibility that
we have to consider $\w_{m_l +2}$ (this would not change much, as we
have just seen), one can apply (\ref{erg-60}) and (\ref{erg-70}) for
$m=m_l$, to obtain
\begin{equation}
	\frac{|D\ma_\w^n v|} {|v|} = \frac{|D\ma_\w^{m_l +1} v|} {|v|}
	\frac{|D\ma_\w^n v|} {|D\ma_\w^{m_l +1} v|} > \min \left\{ 1 ,
	\frac{\sin\ph_{m_l+1}} {\sin\ph_n} \right\} \ge C_5,
	\label{erg-80}
\end{equation}
which settles the non-contraction property.

\skippar

It remains to give the proof that was held off earlier.

\skippar

\proofof{Lemma \ref{lemma-tbar}} First of all, it is clear that
$\bar{\tau}(x,\ph)$ is a decreasing \fn\ of $\ph$, at least for $\ph$
small. Then, by definition of $F_T$, 
\begin{equation}
	\min_{(\l(x),\ph) \in F_T} \bar{\tau}(x,\ph) = \bar{\tau}(x,
	h_2(x)) =: \bar{\tau}_m (x),  
\label{tbar-10}
\end{equation}
with the customary misuse of notation $h_2(x) = h_2(\l(x))$.  So we
are reduced to studying the \tr y of $\w := (\l(x), h_2(x))$, for large
values of $x$.  

Using the same notation as before, we call $x_n = x(\l_n)$ the
abscissa of the $n$-th collision point, whose line element is $\w_n =
(\l_n,\ph_n)$. By construction, $x_0 = x$. Setting $\alpha_n :=
\arctan |\f'(x_n)|$ and rephrasing (\ref{erg-55}) gives
\begin{equation}
	\ph_{n+1} = \ph_n + \alpha_n + \alpha_{n+1}.
	\label{tbar-20}
\end{equation}
With a bit of geometry (taking perhaps a look at Fig.~\ref{figf}), we
check that
\begin{eqnarray}
	\tau_n \sin( \ph_n + \alpha_n ) &=& \f(x_{n+1}) + \f(x_n). 
	\label{tbar-30} \\
	\tau_n \cos( \ph_n + \alpha_n ) &=& x_{n+1} - x_n;
	\label{tbar-40} 
\end{eqnarray}
All these quantities ultimately depend on $x$.

\begin{lemma}
	If $\f(x) = x^{-p}$, $p>0$, there exists an increasing
	sequence $\{ \xi_n \}$ such that, for fixed $n$,
	\begin{displaymath}
		\lim_{x \to +\infty} \frac{x_n(x)} {x} = \xi_n.
	\end{displaymath}
	Furthermore
	\begin{displaymath}
		\lim_{n \to +\infty} \xi_n = +\infty.
	\end{displaymath}
	\label{lemma-tecn3}
\end{lemma}

\proof As concerns the first assertion, we will prove it by
induction. For $n=0$ there is nothing to prove, as $x_0 = x$ (whence
$\xi_0 = 1$). So let us assume that the limit above exists for all $i
\le n$ and try to show that it exists for $n+1$, too.

It will be convenient in the sequel to introduce the symbol
$\simeq$. Its meaning refines that of $\sim$: by definition $f(x)
\simeq g(x)$ states that $f(x) / g(x) \to 1$ as $x \to +\infty$.

Let us take the ratio of (\ref{tbar-30})-(\ref{tbar-40}):
\begin{equation}
	\tan( \ph_n + \alpha_n ) = \frac{ \f(x_{n+1}) + \f(x_n) } 
	{ x_{n+1} - x_n }.
	\label{tecn3-10}
\end{equation}
As $n$ is fixed, $\ph_n(x)$ and $\alpha_n(x)$ tend to zero, when $x
\to +\infty$. Therefore
\begin{equation}
	\tan( \ph_n + \alpha_n ) \simeq \ph_n + \alpha_n \simeq h_2(x) 
	+ |\f'(x)| + 2 \sum_{i=1}^n |\f'(x_i)|,
	\label{tecn3-20}
\end{equation}
since, from (\ref{tbar-20}), $\ph_n = \ph_0 + \alpha_0 + 2
\sum_{i=1}^{n-1} \alpha_i + \alpha_n$, and $\ph_0 = h_2(x)$. Once we plug
$\f(x) = x^{-p}$ in (\ref{tecn3-10})-(\ref{tecn3-20}) we obtain
\begin{equation}
	p \left( K_2 x^{-p-1} + x^{-p-1} + 2 \sum_{i=1}^n x_i^{-p-1}
	\right) \simeq \frac{ x_{n+1}^{-p} + x_n^{-p} } { x_{n+1} -
	x_n }, 
	\label{tecn3-30}
\end{equation}
having used (\ref{erg-12}), too. Divide both sides by $x^{-p-1}$:
\begin{equation}
	p \left( K_2 + 1 + 2 \sum_{i=1}^n \left( \frac{x_i}x
	\right)^{-p-1} \right) \simeq \frac{ \ds \left(
	\frac{x_{n+1}}x \right)^{-p} + \left( \frac{x_n}x \right)^{-p}
	} { \ds \left( \frac{x_{n+1}}x \right) - \left( \frac{x_n}x
	\right) }.  
	\label{tecn3-33}
\end{equation}
Let us name $H(x)$ the above l.h.s., including the \fn\ $(1 + o(x))$
that is implicitly meant by the $\simeq$ symbol. The induction
hypothesis implies that that $H(x)$ has a limit, as $x \to
+\infty$. This limit, denoted $\tilde{H}$, is evidently
positive. After a little algebra (\ref{tecn3-33}) becomes
\begin{equation}
	\left( \frac{x_{n+1}}x \right)^{-p} = H(x) \left(
	\frac{x_{n+1}}x \right) - H(x) \left( \frac{x_n}x \right) -
	\left( \frac{x_n}x \right)^{-p}.
	\label{tecn3-37}
\end{equation}
Since all the terms except $(x_{n+1}/x)$ are known have a limit, and
since $\tilde{H}>0$, it easy to see that the relation above, regarded
as an equation in the variable $(x_{n+1}/x)$, tends to an equation
that has only one solution. This must be $\xi_{n+1} := \lim_{x \to
+\infty} (x_{n+1}(x)/x)$.

\skippar

As concerns the second assertion of the lemma, we proceed by
contradiction. Suppose that $\xi_n \nearrow \tilde{\xi} < +\infty$, as
$n \to +\infty$. Applying the first assertion to (\ref{tecn3-33}) one
obtains
\begin{equation}
	p \left( K_2 + 1 + 2 \sum_{i=1}^n \xi_i^{-p-1} \right) =
	\frac{ \xi_{n+1}^{-p} + \xi_n^{-p} } { \xi_{n+1} - \xi_n }.
	\label{tecn3-40}
\end{equation}
Then the l.h.s.\ of (\ref{tecn3-40}) grows asymptotically like
$n$. But the numerator of the r.h.s.\ converges, implying that
$\xi_{n+1} - \xi_n \sim n^{-1}$, which in turn contradicts the
convergence of $\{ \xi_n \}$.  
\qed

We are just a step away from the proof of Lemma \ref{lemma-tbar}. In
fact, for any $M>0$ we can fix $n$ such that $\xi_n > M+1$. Then
\begin{equation}
	\frac{\bar{\tau}_m (x)} x = \frac1x \sum_{i=0}^{m(x)}
	\tau_i(x) \ge \frac1x \sum_{i=0}^n \tau_i(x) > \frac{x_n(x) -
	x} x > M,
	\label{tbar-50}
\end{equation}
for $x$ large enough. This means precisely that $\bar{\tau}_m (x)/x
\to +\infty$, as $x \to +\infty$, implying Lemma \ref{lemma-tbar}.
\qed

\sect{Acknowledgments}

I am indebted to G.~Del Magno, N.~Chernov and M.~Lyubich for many
enlightening discussions on the subject. I also wish to thank
C.~Liverani, R.~Markarian, N.~Sim\'anyi, Ya.~G.~Sinai and D.~Sz\'asz
for their useful inputs.

During the preparation of this paper I was invited to visit the
Institute of Mathematics of the Technical University of Budapest, and
the Max Planck Institute for Mathematics in the Sciences, in Leipzig.
I would like to thank both institutions for their nice hospitality.

Partial financial support from Universit\`a di Bologna, through
S.~Graffi, is also gratefully acknowledged.

\appendix

\sect{Appendix: Scattered lemmas}

\begin{lemma}
	Let $(\ps, \ma, \mu)$ be a dynamical system with $\mu(\ps) <
	\infty$, and $\psi$ a positive \fn\ on $\ps$. Then its \erg\
	average $\psi_{*}$ is positive almost everywhere.
	\label{lemma-s-arg}
\end{lemma}

\proof First of all, $\psi_{*}$ exists almost everywhere by the first
Birkhoff Theorem. Set $A := \rset {\z\in\ps} {\psi_{*}(z)=0}$; this is
an invariant set. If we assume that $\mu(A)>0$, then we can apply the
second Birkhoff Theorem to $(A, \ma, \mu)$. We obtain
\begin{equation}
	0 = \int_A \psi_{*} \, d\mu = \int_A \psi \, d\mu >0,
\end{equation}
which is absurd.
\qed

\begin{lemma}
	Let $\{ g_n \}_{n\in\N}$ be a family of positive \fn s defined
	on a compact set $K_0$. Assume that $g_n$ is continuous on a
	compact $K_n$, with $K_n \supseteq K_{n+1}$. If, furthermore,
	$\forall x \in K_0$, $g_n(x) \searrow 0$, as $n \to \infty$,
	then, in the same limit, $\max_{K_n} g_n \searrow 0$.
	\label{lemma-tecn1}
\end{lemma}

\proof Let $x_n \in K_n$ be defined by $\max_{K_n} g_n =
g_n(x_n)$. One checks that $\max_{K_n} g_n = g_n(x_n) \ge g_n(x_{n+1})
\ge g_{n+1} (x_{n+1}) = \max_{K_{n+1}} g_{n+1}$; the first inequality
comes from the definition of $x_n$ and the fact that $x_{n+1} \in K_n$,
whereas the second is a result of the monotonicity of $\{ g_n \}$.

Passing perhaps to a subsequence, $x_m \longrightarrow \bar{x} \in
\bigcap_n K_n$, as $m \to \infty$. Suppose now that the assertion is
false, that is, $\exists \eps > 0$ such that $g_n(x_n) \ge \eps$, for
all $n$.

Let us consider the pointwise convergence $g_n (\bar{x}) \searrow
0$. There exist an $N$ for which $g_N(\bar{x}) \le \eps/3$. Since
$x_m$ eventually belongs to $K_N$, and $g_N$ is continuous there,
$\exists M=M(N)$ such that, $\forall m \ge M$, $g_N(x_m)
\le 2\eps/3$.

Now, if $N \ge M$, we reach a contradiction since, from above,
$g_N(x_N) \le 2\eps/3$. If $N < M$, we use again the monotonicity of
the family and see that $g_M(x_M) \le g_N(x_M) \le 2\eps/3$.
\qed

\begin{lemma}
	For some $\eta > 1$, let $\ps^n := [0,n^{-\eta}] \times [0,1]$
	and let $\sir^n \subset \ps^n$ be the union of $M_n$ graphs of
	monotonic \fn s over $[0,n^{-\eta}]$. Define $\sir :=
	\bigsqcup_{n\in\N} \sir^n$ and denote by $\sir_{ \{\eps\} }$
	the tubular neighborhood of $\sir$ of radius $\eps$ (with
	respect to the ordinary distance). If $M_n$ grows at most
	polynomially in $n$, then $\mathrm{Leb} \left( \sir_{ \{\eps\}
	} \right)$ decays polynomially in $\eps$, as $\eps \to 0^{+}$.
	\label{lemma-tecn2}
\end{lemma}

\proof Without loss of generality, one can assume that $\sir^n$ is
made up of vertical segments in $\ps^n$ (in fact, the graph of a
monotonic \fn\ in $\ps^n$ has length less than $1 + n^{-\eta}$). The
worst situation, in the sense of least amount of overlap among the
tubular neighborhoods, occurs when these segments are equispaced. So,
let us consider this case. Suppose that $M_n \le C_1 n^\rho$. Then the
spacing between the segments $\sir^n$ is at least $n^{-(\eta+\rho)} /
C_1$. For a given $\eps > 0$, let $k$ be the maximum $n$ such that
$n^{-(\eta+\rho)} / C_1 \ge 2\eps$. For $n > k$, the \me\ of $\sir^k$ is
estimated by the \me\ of the entire $\ps^k$. Then
\begin{equation}
	\mathrm{Leb} \left( \bigsqcup_{n = k+1}^\infty \sir^n _{
	\{\eps\} } \right) \le \sum_{n = k+1}^\infty n^{-\eta} \le C_2
	(k+1)^{-\eta + 1} \le C_3 \, \eps^{ \frac{\eta - 1} {\eta +
	\rho} }.
\end{equation}
For $n \le k$, we are guaranteed that the $\eps$-neighborhoods of the
segments do not overlap, therefore
\begin{equation}
	\mathrm{Leb} \left( \bigsqcup_{n = 1}^k \sir^n _{ \{\eps\} }
	\right) \le \eps \, C_1 \sum_{n = 1}^k n^\rho \le C_4 \, \eps \,
	k^{\rho+1} \le C_5 \, \eps^{1 - \frac{\rho + 1} {\eta + \rho}
	} = C_5 \, \eps^{ \frac{\eta - 1} {\eta + \rho} }.
\end{equation}
The last two estimates prove the statement.
\qed

\begin{lemma}
	Let $\ma$ be the \bi\ map associated to a table $\ta$ with
	finite horizon, i.e., $\tau(\z) \le \tau_M$, $\forall
	\z$. Assume that the curvature $k(\l)$ and its derivative are
	bounded above by $k_M$ and $k'_M$, respectively. Set $J \ma_z
	:= |D\ma_z v(\z)|$, the Jacobian of $T$ relative to the smooth
	direction field $v \subset \mathrm{S} \ps$ (here
	$\mathrm{S}\ps$ is the unit tangent bundle of $\ps$; hence
	$|v(\z)|=1$). Then, for $b \in \mathrm{S} \ps_z$,
	\begin{displaymath}
		\frac {\partial} {\partial b} \log J \ma_z \le \frac 
		C {\sin^2\ph \: \sin^4\ph_1},
	\end{displaymath}
	with the notation of Section \ref{sec-prelim}. The constant 
	$C$ depends only on $\tau_M$, $k_M$, $k'_M$ and $| (\partial v
	/ \partial b) (\z)|$. 
	\label{lemma-distor}
\end{lemma}

\proof Looking back at (\ref{dmap}), we set $F(\z) := \sin\ph_1 \,
D\ma_\z$, so that the matrix elements of $F$ contain no annoying
denominators. By hypothesis, then,
\begin{equation}
	\| F(\z) \| \le C_1 = C_1(\tau_M,k_M),
	\label{distor-10}
\end{equation}
with $\| \,\cdot\, \|$ denoting (only in this proof!)\ the norm of a
matrix as a linear operator. Also by (\ref{dmap}), we know that $\det
D\ma_\z = \sin\ph / \sin\ph_1$, whence $\det F = \sin\ph \,
\sin\ph_1$. Therefore $F$ is invertible and $F^{-1}$ only contains
denominators of the form $\sin\ph \, \sin\ph_1$. It follows that
\begin{equation}
	\| F^{-1} \| \le \frac{C_1} {\sin\ph \, \sin\ph_1},
	\label{distor-13}
\end{equation}
which, in turn, gives
\begin{equation}
	|Fv| \ge \frac{\sin\ph \, \sin\ph_1} {C_1},
	\label{distor-17}
\end{equation}
as $|v|=1$.  For the sake of format, let us indicate the partial
derivative w.r.t.\ $b$ with the symbol $\partial_b$.  Via elementary
calculus we obtain
\begin{equation}
	\partial_b \log J \ma_z = \frac{ D\ma_z v(\z) \cdot
	\partial_b \, D\ma_z v(\z) } { |D\ma_z v(\z)|^2 } = \sin\ph_1 
	\, \frac{ Fv \cdot \partial_b \left( {\ds \frac{Fv} 
	{\sin\ph_1} } \right) } {|Fv|^2}.
	\label{distor-20}
\end{equation}
We focus on the most troublesome term in (\ref{distor-20}):
\begin{equation}
	\partial_b \left( \frac{Fv} {\sin\ph_1} \right) = \frac{
	(\partial_b F) v } {\sin\ph_1} + \frac{ F (\partial_b v) } 
	{\sin\ph_1} + Fv \, \partial_b \left( \frac1 {\sin\ph_1} 
	\right) =: Y_1 + Y_2 + Y_3. 
	\label{distor-40}
\end{equation}
We start with $Y_2$ which can be easily bounded above by $C_2 /
\sin\ph_1$, where $C_2 = C_2 (C_1, |\partial_b v|)$. As for $Y_3$, we
observe that $\partial_b \ph_1$ is simply the second component of the
vector $D\ma_\z b$, whose norm, by (\ref{distor-10}) and the fact that
$|b|=1$, does not exceed $C_1 / \sin\ph_1$. Hence, working out the
other terms, we end up with $|Y_3| \le C_1^2 / \sin^3\ph_1$. In order
to estimate $Y_1$, we notice that $F(\z)$ is indeed a polynomial in
the variables
\begin{equation}
	\sin\ph, \ \sin\ph_1, \ k=k(\l), \ k_1=k(\l_1), \ \tau = 
	dist_\ta(\l,\l_1).
	\label{distor-50}
\end{equation}
Here $dist_\ta(\l,\l_1)$ is the distance on the table $\ta$ between
the points represented by the coordinates $\l$ and $\l_1$.
(Incidentally, let us observe that this \fn\ is smooth, for $\l \ne
\l_1$, with derivatives bounded by 1 in absolute value.)

A given matrix element of $\partial_b F$ is then a finite sum of
products of the variables in (\ref{distor-50}) times the derivative of
\emph{one} of those variables. Since the \fn s $\sin, k$ and
$dist_\ta$ have bounded derivatives, the only singularity will occur
when, by implicit differentiation, $\partial_b$ hits $\l_1$ or
$\ph_1$. This corresponds to either component of $D\ma_\z b$, and the
differentiation gives rise to a singularity of the type $1/
\sin\ph_1$. Therefore there is a $C_3 = C_3(\tau_M,k_M,k'_M)$ such
that $|Y_1| \le C_3 / \sin^2\ph_1$.

Taking the worst case among the above estimates, we conclude that
there exists a $C_4 = C_4(\tau_M, k_M, k'_M, |\partial_b v|)$ such
that
\begin{equation}
	\partial_b \left( \frac{Fv} {\sin\ph_1} \right) \le \frac{C_4}
	{\sin^3\ph_1}.
	\label{distor-60}
\end{equation}
Plugging (\ref{distor-10}), (\ref{distor-17}) and (\ref{distor-60})
into (\ref{distor-20}) gives the assertion of the lemma.  
\qed

\begin{lemma}
	If $A$ is a countable subset of $\ps := I \times J$, with
	$I,J$ two non-degenerate intervals, then $\ps \setminus A$ is
	path-connected. 
	\label{lemma-conn}
\end{lemma}

\proof Take $\z_1, \z_2$, two distinct elements of $\ps \setminus A$,
and let $B$ denote the segment in $\ps$ whose points are equidistant
from $\z_1$ and $\z_2$.  For $\w \in B$, call $\gamma_\w$ the polyline
connecting $\z_1$ to $\w$ and $\w$ to $\z_2$. $\{ \gamma_\w \}$ is an
uncountable family of pairwise disjoint paths. Since only countably
many of them can intersect $A$, it follows that there exist infinitely
many paths connecting $\z_1$ to $\z_2$.  
\qed

\begin{lemma}
	Consider the finite segment of \tr y $\{ \z_0=\z, \z_1=\ma\z,
	\z_2=\ma^2\z \}$ of a \bi\ map $\ma$. Then, at the first order
	in $d\z$, and up to a minus sign, $d\z_2$ is given by a single
	differential a \bi\ map, with dynamical parameters:

	\noindent 
	Free path: $\ds \hat{\tau} = \tau_0 + \tau_1 + 2 \frac {k_1}
	{\sin\ph_1} \, \tau_0 \tau_1$

	\noindent
	Curvature at initial point $\z_0$: $\ds \hat{k}_0 = k_0 + 2
	\sin\ph_0 \, \frac {k_1} {\sin\ph_1} \, \frac {\tau_1} 
	{\hat{\tau}}$

	\noindent
	Curvature at final point $\z_2$: $\ds \hat{k}_2 = k_2 + 2
	\sin\ph_2 \, \frac {k_1} {\sin\ph_1} \, \frac {\tau_0} 
	{\hat{\tau}}$

	In other words, $D\ma_{\z_1} D\ma_{\z_0} = -M$, where
	$M(\hat{\tau}, \hat{\ph}_0, \hat{k}_0, \hat{\ph}_2,
	\hat{k}_2)$ is again a differential of the type
	\emph{(\ref{dmap})}. If we take the same initial and final
	angles as for the actual segment of \tr y, $\hat{\ph}_0 =
	\ph_0, \hat{\ph}_2 = \ph_2$, then the other three parameters
	are fixed as above.
	\label{lemma-2-boun}
\end{lemma}

\proof This is just a verification, which is made easier if we use on
$\ts\ps$ the pair of variables $(\sin\ph \, d\l, d\ph)$. Denoting by
$\bar{D}\ma$ the differential in these new variables, we get from
(\ref{dmap})
\begin{equation}
	\bar{D}\ma_{\z_i} = \left[ 
	\begin{array}{cc} 
		-1 - \eta_i \tau_i & 
		\tau_i \\
		\eta_i + \eta_{i+1} + \eta_i \eta_{i+1} \tau_i \ \ &
		-1 - \eta_{i+1} \tau_i
	\end{array}
	\right] , 
\end{equation}
with $\eta_i := k_i / \sin\ph_i$. Notice that $\det \bar{D}\ma_{\z_i}
= 1$. Hence, upon imposing
\begin{equation}
	\bar{D}\ma_{\z_1} \bar{D}\ma_{\z_0} =: - \left[ 
	\begin{array}{cc} 
		-1 - \hat{\eta}_0 \hat{\tau} & 
		\hat{\tau} \\
		\hat{\eta}_0 + \hat{\eta}_2 + \hat{\eta}_0
		\hat{\eta}_2 \hat{\tau} \ \ &
		-1 - \hat{\eta}_2 \hat{\tau}
	\end{array}
	\right] , 
\end{equation}
and $\hat{k}_i = \hat{\eta}_i \sin\ph_i$ $(i=0,2)$, we get the three
parameters as in the statement of the lemma.  
\qed

\end{document}